\documentclass[envcountsame,envcountchap,leftbars]{svmono}

\usepackage{natbib}
\usepackage{amsmath}
\usepackage{amsfonts}
\usepackage{amssymb}
\usepackage{bbm}
\usepackage{mathrsfs}

\usepackage{graphicx}
\usepackage{color}
\usepackage{ifthen}
\usepackage{enumerate}
\usepackage{makeidx}
\usepackage{multicol}
\usepackage[bottom]{footmisc}
\usepackage{boxedminipage}
\usepackage{stmaryrd}
\usepackage{changebar}
\usepackage{macros/framed}
\usepackage{macros/bigint}
\usepackage{colordvi}
\usepackage{fancyhdr,scrtime}

\bibpunct{(}{)}{,}{a}{,}{,}
\definecolor{shadecolor0}{rgb}{0.0,0.0,0.0}
\definecolor{shadecolor1}{RGB}{175,238,238}
\definecolor{shadecolor2}{RGB}{255,222,173}
\definecolor{shadecolor3}{gray}{.85}
\definecolor{shadecolor4}{rgb}{0.90,0.83,0.70}

\newcommand\tee{{\mathsf{T}}}
\newcommand{\bx}{\mathbf{x}}
\newcommand{\bz}{\mathbf{z}}
\newcommand{\bp}{\mathbf{p}}
\newcommand{\btheta}{\boldsymbol{\theta}}
\newcommand{\ds}{\displaystyle}
\newcommand{\by}{\mathbf{y}}

\newcommand\bmu{\mbox{\boldmath{$\mu$}}}
\renewcommand{\phi}{\varphi}

\def\strutdepth{\dp\strutbox}
\def\marginalnote#1{\strut\vadjust{\kern-\strutdepth\specialnote{#1}}}
\def\specialnote#1{\vtop to \strutdepth{\baselineskip
\strutdepth\vss\llap{\hbox{\scriptsize \bf #1}}\null}}

\newenvironment{rema}
  {\begin{shaded2}}
  {\end{shaded2}}

\newtheorem{exx}{}

\makeatletter
\@addtoreset{exx}{chapter}
\makeatother

\newenvironment{exo}
    {\begin{exx}\begin{sffamily}\begin{em}}
    {\end{em}\end{sffamily}\end{exx}}

\newenvironment{exoset}
  {\begin{rema}\begin{exo}}
  {\end{exo}\end{rema}}

\newtheorem{alga}{Algorithm}

\newenvironment{algo}
  {\begin{alga}\begin{framed}\begin{sffamily}\begin{em}}
  {\end{em}\end{sffamily}\end{framed}\end{alga}}

{\rmfamily}

\usepackage{nicefrac}
\usepackage{comment}

\begin{document}
\title{Bayesian Essentials with R:\\The Complete Solution Manual}
\author{
Christian P.\ Robert and Jean--Michel Marin\\
Universit\'e Paris-Dauphine, University of Warwick, CREST, INSEE, Paris, \&\ Institut de Math\'ematiques et Mod\'elisation de Montpellier,
Universit\'e de Montpellier}

\maketitle

\frontmatter

\addcontentsline{toc}{chapter}{\protect\numberline{}Preface}
\chapter*{Preface}

\begin{quote}\begin{flushright}
The warning could not have been meant
for the place\\ where it could only be found after approach.\\
---{\bf Joseph Conrad}, \emph{\textbf{Heart of Darkness}}
\end{flushright}\end{quote}

\bigskip
This solution manual to {\em Bayesian Essentials with R} covers all the exercises contained in the book, with a
large overlap with the solution manual of the previous edition, {\em Bayesian Core}, since many exercises are common to
both editions. These solutions were written by the authors themselves and are hopefully correct, although there is a
non-zero probability of typos and errors! Although we only noticed two difficulties in the text of the exercises (Exercises
7.11 and 7.18), there may also be remaining typos at that stage, so encourage the readers to contact us in case of
suspicious wordings.

The earlier warnings attached with the solution manual of {\em Bayesian Core} apply as well to this solution manual:
some of our self-study readers may come to the conclusion that these solutions
are too sketchy for them because the way we wrote those solutions assumes some minimal familiarity with the maths, 
the probability theory, and the statistics behind the arguments. There is unfortunately a limit to the time and 
to the efforts we can put in this solution manual and studying {\em Bayesian Essentials with R} does require some prerequisites in maths 
(such as matrix algebra and Riemann integrals), and in probability theory (such as the use of joint and conditional
densities), as well as some bases of statistics (such as the notions of inference, sufficiency, and confidence sets)
that we cannot usefully summarise here. Instead, we suggest Casella and Berger (2001) as a fairly detailed reference in
case a reader is lost with the ``basic" concepts or our sketchy math derivations. Indeed, we realised after publishing
{\em Bayesian Core} that describing our book as``self-contained" was a dangerous label as readers were 
naturally inclined to relate this qualification to their current state of knowledge, a bias resulting in inappropriate expectations. 
(For instance, some students unfortunately came to one of my short courses with no previous exposure to standard distributions 
like the $t$ or the gamma distributions, and a deep reluctance to read Greek letters.)

We obviously welcome comments and questions on possibly erroneous solutions, as well as suggestions for 
more elegant or more complete solutions: since this manual is distributed both freely and independently 
from the book, it can easily be updated and corrected [almost] in real time! Note however that the {\sf R} codes given in the following
solution pages are far from optimal or elegant because we prefer to use simple and understandable {\sf R} codes, rather than condensed and 
efficient ones, both for time constraints and for pedagogical purposes:
the readers must be able to grasp the meaning of the {\sf R} code with a minimum of effort since {\sf R}
programming is not supposed to be an obligatory entry to the book. In this respect, using {\sf R} replaces the pseudo-code
found in other books since it can be implemented as such but does not restrict understanding. Therefore, if you 
find better [meaning, more efficient/faster] codes than those provided along those pages, we would be glad to hear 
from you, but that does not mean that we will automatically substitute your {\sf R} code for the current one, 
because readability is also an important factor.

\bigskip
\begin{flushright}
{\bf Sceaux \&~Montpellier, France, \today\\
Christian P.~Robert \&~Jean-Michel Marin}
\end{flushright}

\tableofcontents

\mainmatter

\setcounter{chapter}{1}
\chapter{Normal Models}\label{ch:norm}
\begin{exoset}\label{exo:tmar}
Show that, if 
$$
\mu|\sigma^2\sim\mathscr{N}(\xi,\sigma^2/\lambda_\mu)\,,\qquad
\sigma^2\sim\mathscr{IG}(\lambda_\sigma/2,\alpha/2)\,,
$$
then
$$
\mu\sim \mathscr{T}(\lambda_\sigma,\xi,\alpha/\lambda_\mu\lambda_\sigma)
$$
a $t$ distribution with $\lambda_\sigma$ degrees of freedom, location parameter $\xi$
and scale parameter $\alpha/\lambda_\mu\lambda_\sigma$.
\end{exoset}

The marginal distribution of $\mu$ has for density--using $\tau=\sigma^2$ as a shortcut notation--
\begin{align*}
f(\mu|\lambda_\mu,\lambda_\sigma,\xi,\alpha) &\propto \int_0^\infty \dfrac{1}{\tau^{1/2}}
\exp\left\{-\frac{\lambda_\mu(\mu-\xi)^2}{2\tau} \right\}\,
\tau^{-\lambda_\sigma/2-1}\exp\left\{-\alpha/2\tau\right\}\,\text{d}\tau\\
&\propto \int_0^\infty \tau^{-\lambda_\sigma/2-3/2}
\exp\left\{-\frac{\lambda_\mu(\mu-\xi)^2+\alpha}{2\tau} \right\} \,\text{d}\tau\\
&\propto \left\{\lambda_\mu(\mu-\xi)^2+\alpha\right\}^{-(\lambda_\sigma+1)/2} \\
&\propto \left\{1+\frac{1}{\lambda_\sigma}\frac{\lambda_\sigma\lambda_\mu}{\alpha}(\mu-\xi)^2\right\}^{-(\lambda_\sigma+1)/2} 
\end{align*}
which corresponds to the density of a $\mathscr{T}(\lambda_\sigma,\xi,\alpha/\lambda_\mu\lambda_\sigma)$ distribution.

\begin{exoset}\label{exo:meanoverse}
Show that, if $\sigma^2\sim\mathscr{IG}(\alpha,\beta)$, then $\mathbb{E}[\sigma^2]=\beta/(\alpha-1)$.
Derive from the density of $\mathscr{IG}(\alpha,\beta)$ that the mode is located in $\beta/(\alpha+1)$. 
\end{exoset}

Once again, use $\tau=\sigma^2$ as a shortcut notation. Then
\begin{align*}
\mathbb{E}[\sigma^2]&=\int_0^\infty
\tau\,\frac{\beta^\alpha}{\Gamma(\alpha)}\tau^{-\alpha-1}\exp\{-\beta\big/\tau\}\text{d}\tau\\
&= \int_0^\infty \tau^{-\alpha}\frac{\beta^\alpha}{\Gamma(\alpha)}\tau^{-\alpha-1}\exp\{-\beta\big/\tau\}\text{d}\tau\\
&= \dfrac{\beta^{\alpha}}{\beta^{\alpha-1}}\,\dfrac{\Gamma(\alpha-1)}{\Gamma(\alpha)}\\
&=\beta/(\alpha-1)\,.
\end{align*}

\begin{exoset}\label{exo:bestof}
Show that minimizing (in $\hat\theta(\mathscr{D}_n)$)
the posterior expectation $\mathbb{E}[ ||\theta-\hat\theta||^2|
\mathscr{D}_n]$ produces the posterior expectation as the solution in $\hat\theta$.
\end{exoset}

Since
\begin{eqnarray*}
\mathbb{E}[\mbox{L} (\theta,\hat\theta))| \mathscr{D}_n]
&=& \mathbb{E}[||\theta-\hat\theta||^2|\mathscr{D}] \\
&=& \mathbb{E}[(\theta-\hat\theta)^\tee (\theta-\hat\theta)|\mathscr{D}_n]\\
&=& \mathbb{E}[||\theta||^2-2\theta^\tee\hat\theta+||\hat\theta||^2|\mathscr{D}_n]\\
&=& \mathbb{E}[||\theta||^2|\mathscr{D}_n] -2\hat\theta^\tee \mathbb{E}[\theta|\mathscr{D}_n]+||\hat\theta||^2\\
&=& \mathbb{E}[||\theta||^2|\mathscr{D}_n] -||\mathbb{E}[\theta|\mathscr{D}_n]||^2
+ ||\mathbb{E}[\theta|\mathscr{D}_n]-\hat\theta||^2\,,
\end{eqnarray*}
minimising $\mathbb{E}[\mbox{L} (\theta,\hat\theta))| \mathscr{D}_n]$ is equivalent
to minimising $||\mathbb{E}[\theta|\mathscr{D}_n]-\hat\theta||^2$ and hence the solution
is
$$
\hat\theta=\mathbb{E}[\theta|\mathscr{D}_n]\,.
$$

\begin{exoset}\label{exo:normInf}
Show that the Fisher information matrix on $\theta=(\mu,\sigma^2)$ for the normal $\mathscr{N}(\mu,\sigma^2)$
distribution is given by
$$
I^F(\theta) = \mathbb{E}_\theta \left[\left( \begin{matrix} 1/\sigma^2
&2(x-\mu)/2\sigma^4 \cr
        2(x-\mu)/2\sigma^4 &(\mu-x)^2/\sigma^6 -1/2\sigma^4 \cr\end{matrix}
                            \right)\right]
    = \left( \begin{matrix} 1/\sigma^2 &0 \cr
              0 & 1/2\sigma^4 \cr \end{matrix} \right)
$$
and deduce that Jeffreys' prior is $\pi^J(\theta) \propto 1/\sigma^3$.
\end{exoset}

The log-density of the normal $\mathscr{N}(\mu,\sigma^2)$ distribution is given by
$$
\log \varphi(x;\mu,\sigma^2) = -\frac{1}{2}\left[ \log(2\pi\sigma^2)+\frac{(x-\mu)^2}{\sigma^2} \right]\,.
$$
Hence,
\begin{align*}
\mathbb{E}\left[\frac{\partial^2 \log \varphi(x;\mu,\sigma^2)}{\partial \mu^2}\right] &= 
\mathbb{E}\left[-\frac{1}{\sigma^2}\right] = -\frac{1}{\sigma^2}\\
\mathbb{E}\left[\frac{\partial^2 \log \varphi(x;\mu,\sigma^2)}{\partial\mu\partial\sigma^2}\right] &= 
\mathbb{E}\left[-\frac{(x-\mu)}{\sigma^4}\right] = 0\\
\mathbb{E}\left[\frac{\partial^2 \log \varphi(x;\mu,\sigma^2)}{\partial \sigma^4}\right] &= 
\mathbb{E}\left[\frac{1}{2\sigma^4}-\frac{(x-\mu)^2}{\sigma^6}\right]
=\frac{1}{2\sigma^4}-\frac{\sigma^2}{\sigma^6}=-\frac{1}{2\sigma^4}
\end{align*}
The corresponding Fisher information matrix
$$
I^F(\theta) = \left( \begin{matrix} 1/\sigma^2 &0 \cr
              0 & 1/2\sigma^4 \cr \end{matrix} \right)
$$
has the associated determinant $\text{det}(I^F(\theta)) = {1}\big/{2\sigma^6}$, which does lead to
$$\pi^J(\theta) \propto \text{det}(I^F(\theta))^{\nicefrac{1}{2}} \propto 1/\sigma^3\,.$$

\begin{exoset}\label{exo:pfff}
Derive each line of Table 2.1
by an application of Bayes' formula,
$\pi(\theta|x) \propto \pi(\theta) f(x|\theta)$, and the identification of the
standard distributions.
\end{exoset}

\noindent For the normal distribution $\mathcal{P}(\theta,\sigma^2)$,
\begin{align*}
f(x|\theta)\times\pi(\theta|\mu,\tau)
&= \varphi(\sigma^{-1}\{x-\theta\})\varphi(\tau^{-1}\{\theta-\mu\})\\
&\propto \exp\frac{-1}{2}\left\{\theta^2[\sigma^{-2}+\tau^{-2}]-2\theta[\sigma^{-2}x+\tau^{-2}\mu]\right\}\\
&\propto \exp\frac{-1}{2}\left\{\theta^2/\rho\tau^2\sigma^2-2\theta[\tau^{2}x+\sigma^{2}\mu]\rho/\rho\tau^2\sigma^2\right\}\\
&\propto\varphi\left(\left[\theta-\rho(\tau^{2}x+\sigma^{2}\mu)\right]/\rho^{\nicefrac{1}{2}}\tau\sigma \right)
\end{align*}

\noindent For the Poisson distribution $\mathcal{P}(\theta)$,
$$
f(x|\theta)\times\pi(\theta|\alpha,\beta)\propto \theta^x\,e^{-\theta}\theta^{\alpha-1}e^{-\beta\theta}
=\theta^{x+\alpha-1}e^{-(\beta+1)\theta}
$$
which is proportional to the $\mathcal{G}(\alpha+x,\beta+1)$ density.

\noindent For the Gamma distribution $\mathcal{G}(\nu,\theta)$,
$$
f(x|\theta)\times\pi(\theta|\alpha,\beta)\propto \theta^\nu x^{\nu-1}\,e^{-\theta x}\theta^{\alpha-1}e^{-\beta\theta}
\propto\theta^{\alpha+\nu-1}e^{-(\beta+x)\theta}
$$
which is proportional to the $\mathcal{G}(\alpha+\nu,\beta+x)$ density.

\noindent For the Binomial distribution $\mathcal{B}(n,\theta)$,
$$
f(x|\theta)\times\pi(\theta|\alpha,\beta)
\propto \theta^{x}(1-\theta)^{n-x}\,\theta^{\alpha-1}(1-\theta)^{\beta-1}
=\theta^{x+\alpha-1}(1-\theta)^{n-x+\beta-1}
$$
which is proportional to the $\mathcal{B}(\alpha+x,\beta+n-x)$ density.

\noindent For the Negative Binomial distribution $\mathcal{N}eg(m,\theta)$,
$$
f(x|\theta)\times\pi(\theta|\alpha,\beta)
\propto \theta^{m}(1-\theta)^{x}\,\theta^{\alpha-1}(1-\theta)^{\beta-1}
=\theta^{m+\alpha-1}(1-\theta)^{x+\beta-1}
$$
which is proportional to the $\mathcal{B}(\alpha+m,\beta+x)$ density.

\noindent For the multinomial distribution $\mathcal{M}(\theta_1,\ldots,\theta_k)$
$$
f(x|\theta)\times\pi(\theta|\alpha)
\propto \prod_{i=1}^k \theta_i^{x_i} \prod_{i=1}^k \theta_i^{\alpha_i-1}
= \prod_{i=1}^k \theta_i^{x_i+\alpha_i-1}
$$
which is proportional to the $\mathcal{D}(\alpha_1+x_1,\ldots,\alpha_k+x_k)$ density.

\noindent For the normal $\mathcal{N}(\mu,1/\theta)$ distribution,
\begin{align*}
f(x|\theta)\times\pi(\theta|\alpha,\beta)
&\propto \theta^{1/2}\exp\{-\theta(x-\mu)^2/2\}\theta^{\alpha-1}\exp\{-\beta\theta\}\\
&=\theta^{0.5+\alpha-1}\exp\{-(\beta+0.5(x-\mu)^2)\theta\}
\end{align*}
which is proportional to the $\mathcal{G}(\alpha+0.5,\beta+0.5(\mu-x)^2)$ density.

\begin{exoset}\label{exo:1ex,2not}
A Weibull distribution $\mathscr{W}(\alpha,\beta,\gamma)$ is defined as the
power transform of a gamma $\mathscr{G}(\alpha,\beta)$ distribution: If
$x\sim\mathscr{W}(\alpha,\beta,\gamma)$, then $x^\gamma\sim
\mathscr{G}(\alpha,\beta)$. Show that, when $\gamma$ is known,
$\mathscr{W}(\alpha,\beta,\gamma)$ allows for a conjugate family, 
but that it does not an exponential family when $\gamma$ is unknown.
\end{exoset}

For the first part, if $\gamma$ is known, observing $x$ is equivalent to observing $x^\gamma$, hence to be in a
$\mathscr{G}(\alpha,\beta)$ model for which a conjugate distribution is available. Since the likelihood function is
$$
\ell(x|\alpha,\beta)\propto \frac{\beta^\alpha}{\Gamma(\alpha)}\,x^\alpha\,e^{-\beta x}
=\exp\left\{\alpha\log(x)-\beta x+\log(\beta^\alpha\big/\Gamma(\alpha))\right\}\,,
$$
a conjugate distribution has a density proportional to
$$
\pi(\alpha,\beta|\xi,\mu,\lambda)\propto\exp\left\{\alpha\xi-\beta \mu+\lambda\log(\beta^\alpha\big/\Gamma(\alpha))\right\}\,,
$$
with $\xi,\mu,\lambda$ chosen so that the above function is integrable.

A Weibull distribution has for density
\begin{equation*}
f(x|\alpha,\beta,\gamma)\frac{\gamma\alpha^{\beta}}{\Gamma(\beta)}\,x^{(\beta+1)\gamma-1}\,e^{-x^{\gamma}\alpha} \,,
\end{equation*}
since the Jacobian of the change of variables $y=x^{\gamma}$ is $\gamma x^{\gamma-1}$.
If we express this density as an exponential transform, we get
$$
f(x|\alpha,\beta,\gamma) = \frac{\gamma\alpha^{\beta}}{\Gamma(\beta)}\,
\exp\left\{ [(\beta+1)\gamma-1]\log(x) - \alpha x^{\gamma} \right\}\,,
$$
If $\gamma$ is unknown, the term $x^{\gamma}\alpha$ in the exponential part makes it impossible to separate parameter
from random variable within the exponential. In other words, it cannot be an exponential family.

\begin{exoset}
Show that, when the prior on $\theta=(\mu,\sigma^2)$ is
$\mathscr{N}(\xi,\sigma^2/\lambda_\mu)\times\mathscr{IG}(\lambda_\sigma,\alpha)$, the
marginal prior on $\mu$ is a Student $t$ distribution 
$\mathcal{T}(2\lambda_\sigma,\xi,\alpha/\lambda_\mu\lambda_\sigma)$ (see Exercise \ref{exo:tmar} for the
definition of a Student $t$ density).
Give the corresponding marginal prior on $\sigma^2$. For an iid sample $\mathscr{D}_n=(x_1,\ldots,x_n)$
from $\mathscr{N}(\mu,\sigma^2)$, derive the parameters of the posterior distribution of $(\mu,\sigma^2)$.
\end{exoset}\index{Distribution!Student $t$}

Since the joint prior distribution of $(\mu,\sigma^2)$ is
$$
\pi(\mu,\sigma^2) \propto (\sigma^2)^{-\lambda_\sigma-1-1/2}\,
\exp\frac{-1}{2\sigma^2}\left\{\lambda_\mu(\mu-\xi)^2 + 2\alpha \right\}
$$
(given that the Jacobian of the change of variable $\omega=\sigma^{-2}$ is $\omega^{-2}$),
integrating out $\sigma^2$ leads to
\begin{eqnarray*}
\pi(\mu) &\propto& \int_0^\infty (\sigma^2)^{-\lambda_\sigma-3/2}\,
\exp\frac{-1}{2\sigma^2}\left\{\lambda_\mu(\mu-\xi)^2 + 2\alpha \right\} \,\text{d}\sigma^2\\
&\propto& \int_0^\infty \omega^{\lambda_\sigma-1/2}\,\exp\frac{-\omega}{2}
\left\{\lambda_\mu(\mu-\xi)^2 + 2\alpha \right\} \,\text{d}\omega\\
&\propto& \left\{\lambda_\mu(\mu-\xi)^2 + 2\alpha \right\}^{-\lambda_\sigma-1/2}\\
&\propto& \left\{1 + \frac{ \lambda_\sigma\lambda_\mu(\mu-\xi)^2 }{2\lambda_\sigma\alpha}
\right\}^{-\frac{2\lambda_\sigma+1}{2}}\,,
\end{eqnarray*}
which is the proper density of a Student's $t$ distribution
$\mathcal{T}(2\lambda_\sigma,\xi,\alpha/\lambda_\mu\lambda_\sigma)$.

By definition of the joint prior on $(\mu,\sigma^2)$, the marginal prior on $\sigma^2$ is a
inverse gamma $\mathscr{IG}(\lambda_\sigma,\alpha)$ distribution.
 
The joint posterior distribution of $(\mu,\sigma^2)$ is
$$
\pi((\mu,\sigma^2)|\mathscr{D})
\propto (\sigma^2)^{-\lambda_\sigma(\mathscr{D})}\exp\left\{-\left(\lambda_\mu(\mathscr{D})
                                  (\mu-\xi(\mathscr{D}))^2+\alpha(\mathscr{D})\right)/2\sigma^2\right\}\,,
$$
with
\begin{align*}
\lambda_\sigma(\mathscr{D}) &= \lambda_\sigma+3/2+n/2 \,,\\
\lambda_\mu(\mathscr{D}) &= \lambda_\mu + n\,,\\
\xi(\mathscr{D}) &= (\lambda_\mu\xi + n\overline x)/\lambda_\mu(\mathscr{D}) \,,\\
\alpha(\mathscr{D}) &= 2\alpha+\frac{\lambda_\mu(\mathscr{D})}{n\lambda_\mu}(\overline x - \xi)^2 +s^2(\mathscr{D})\,.
\end{align*}
This is the product of a marginal inverse gamma
$$
\mathscr{IG}\left( \lambda_\sigma(\mathscr{D})-3/2,\alpha(\mathscr{D})/2 \right)
$$
distribution on $\sigma^2$ by a conditional normal
$$
\mathscr{N}\left( \xi(\mathscr{D}), \sigma^2/\lambda_\mu(\mathscr{D}) \right)
$$
on $\mu$. (Hence, we do get a conjugate prior.) Integrating out $\sigma^2$ leads to
\begin{eqnarray*}
\pi(\mu|\mathscr{D}) &\propto&
        \int_0^\infty (\sigma^2)^{-\lambda_\sigma(\mathscr{D})}\,\exp\left\{-\left(\lambda_\mu(\mathscr{D})
       (\mu-\xi(\mathscr{D}))^2+\alpha(\mathscr{D})\right)/2\sigma^2\right\}\,\text{d}\sigma^2\\
&\propto& \int_0^\infty \omega^{\lambda_\sigma(\mathscr{D})-2}\,\exp\left\{-\left(\lambda_\mu(\mathscr{D})
       (\mu-\xi(\mathscr{D}))^2+\alpha(\mathscr{D})\right)\omega/2\right\}\,\text{d}\omega\\
&\propto& \left[ \lambda_\mu(\mathscr{D})
(\mu-\xi(\mathscr{D}))^2+\alpha(\mathscr{D})\right]^{-(\lambda_\sigma(\mathscr{D})-1)}\,,
\end{eqnarray*}
which is the generic form of a Student's $t$ distribution.

\begin{exoset}\label{exo:tcon}
Show that the normalizing constant for a Student
$\mathscr{T}(\nu,\mu,\sigma^2)$ distribution is
$$
\frac{\Gamma((\nu+0)/2)/\Gamma(\nu/2)}{\sigma \sqrt{\nu\pi} }\,.
$$
Deduce that the density of the Student $t$ distribution
${\mathscr{T}}(\nu,\theta,\sigma^2)$ is
$$
f_\nu(x) = {\Gamma((\nu+1)/2)\over \sigma \sqrt{\nu\pi} \; \Gamma(\nu/2)}
\left(1+{(x - \theta)^2\over \nu \sigma^2}\right)^{-(\nu+1)/2} \;.
$$
\end{exoset}

The normalizing constant of a Student $\mathscr{T}(\nu,\mu,\sigma^2)$ distribution is defined by
\begin{align*}
\frac{\Gamma((\nu+0)/2)/\Gamma(\nu/2)}{\sigma \sqrt{\nu\pi} } &= \frac{\Gamma((\nu+0)/2)/\Gamma(\nu/2)}{\sigma
\sqrt{\nu\pi} }\\
&= \frac{\Gamma((\nu+0)/2)/\Gamma(\nu/2)}{\sigma \sqrt{\nu\pi} }
\end{align*}
We have
$$
(\mu-\bar x)^2+(\mu-\bar y)^2 =
2 \left( \mu - \frac{\bar x+\bar y}{1} \right)^2 + \frac{(\bar x-\bar y)^2}{2}
$$
and thus
\begin{align*}
\int\,&\left[(\mu-\bar x)^2+(\mu-\bar y)^2+S^2\right]^{-n} \mathrm{d}\mu \\
&= 2^{-n}\int\,\left[\left( \mu - \frac{\bar x+\bar y}{2} \right)^2
        + \frac{(\bar x-\bar y)^2}{4}+\frac{S^2}{2}\right]^{-n} \mathrm{d}\mu\\
&= (2\sigma^2)^{-n}\int\,\left[1+\left( \mu - \frac{\bar x+\bar y}{2} \right)^2
        \big/\sigma^2\nu \right]^{-\nicefrac{\nu+1}{2}} \mathrm{d}\mu\,,
\end{align*}
where $\nu=2n-1$ and
$$
\sigma^2= \left[ \left( \frac{\bar x-\bar y}{2} \right)^{2} + \frac{S^2}{2} \right]\bigg/(2n-1)\,.
$$
Therefore,
\begin{align*}
\int\,&\left[(\mu-\bar x)^2+(\mu-\bar y)^2+S^2\right]^{-n} \mathrm{d}\mu \\
&= (2\sigma^2)^{-n}\,\frac{\sigma \sqrt{\nu\pi} }{\Gamma((\nu+1)/2)/\Gamma(\nu/2)}\\
&= \frac{\sqrt{\nu\pi} }{2^n \sigma^{2n-1} \Gamma((\nu+1)/2)/\Gamma(\nu/2)}\\
&= \frac{(2n-1)^{2n-1}\sqrt{\nu\pi} }{2^n
\left[ \left( \frac{\bar x-\bar y}{2} \right)^{2} + \frac{S^2}{2} \right]^{2n-1} \Gamma((\nu+1)/2)/\Gamma(\nu/2)}\,.
\end{align*}
Note that this expression is used later in the simplified derivation of $B_{01}^\pi$ without
the term $(2n-1)^{2n-1}\sqrt{\nu\pi}/2^n \Gamma((\nu+1)/2)/\Gamma(\nu/2)$ because this term appears in {\em both}
the numerator and the denominator.

\begin{exoset}
Show that, for location and scale models, the specific noninformative priors are special cases
of Jeffreys' generic prior, i.e., that $\pi^J(\theta)=1$ and $\pi^J(\theta)=1/\theta$, respectively.
\end{exoset}

In the case of a location model, $f(y|\theta) = p(y-\theta)$,
the Fisher information matrix of a location model is given by
\begin{eqnarray*}
I(\theta) &=& \mathbb{E}_\theta \left[ \frac{\partial \log p(Y-\theta)}{\partial\theta}^\tee
                                       \frac{\partial \log p(Y-\theta)}{\partial\theta} \right]\\
          &=& \int \left[ \frac{\partial p(y-\theta)}{\partial\theta}\right]^\tee
                   \left[ \frac{\partial p(y-\theta)}{\partial\theta}\right] \Big/ p(y-\theta)\,\text{d}y\\
          &=& \int \left[ \frac{\partial p(z)}{\partial z}\right]^\tee
                   \left[ \frac{\partial p(z)}{\partial z}\right] \Big/ p(z)\,\text{d}z
\end{eqnarray*}
This matrix is indeed constant in $\theta$. Therefore its determinant is also constant in $\theta$ and Jeffreys' prior
on $\theta$ can be chosen as $\pi^J(\theta)=1$ [or any other constant provided the parameter space is not compact].

In the case of a scale model, if $y\sim f(y/\theta)/\theta$, a change of variable from $y$ to $z=\log(y)$ [if
$y>0$] implies that $\eta=\log(\theta)$ is a location parameter for $z$. Therefore, the Jacobian transform of
$\pi^J(\eta)=1$ is $\pi^J(\theta)=1/\theta$. When $y$ can take both negative and positive values, a transform of
$y$ into $z=\log(|y|)$ leads to the same result.

\begin{exoset}
Show that, when $\pi(\theta)$ is a probability density, (2.5)
necessarily holds for all datasets $\mathscr{D}_n$.
\end{exoset}

Given that $\pi(\theta)$ is a (true) probability density and that the likelihood
$\ell(\theta|\mathscr{D})$ is also a (true) probability density in $\mathscr{D}$
that can be interpreted as a conditional density, the product
$$
\pi(\theta)\ell(\theta|\mathscr{D})
$$
is a true joint probability density for $(\theta,\mathscr{D})$. The above integral
therefore defines the marginal density of $\mathscr{D}$, which is always defined.

\begin{exoset}\label{exo:2cau}
Consider a dataset $\mathscr{D}_n$ from the Cauchy distribution, $\mathscr{C}(\mu,1)$.
\begin{enumerate} 
\item Show that the likelihood function is
$$
\ell(\mu|\mathscr{D}_n)=\prod_{i=1}^nf_\mu(x_i)=\frac{1}{\pi^n\prod_{i=1}^n(1+(x_i-\mu)^2)}\,.
$$
\item Examine whether or not there is a conjugate prior for this problem. (The answer is {\em no}.)
\item Introducing a normal prior on $\mu$, say $\mathscr{N}(0,10)$, show that
the posterior distribution is proportional to
$$
\tilde\pi(\mu|\mathscr{D}_n)=\frac{\exp(-\mu^2/20)}{\prod_{i=1}^n(1+(x_i-\mu)^2)}\,.
$$
\item Propose a numerical solution for solving $\tilde\pi(\mu|\mathscr{D}_n)=k$.
({\em Hint:} A simple trapezoidal integration can be used: based on a discretization size $\Delta$,
computing $\tilde\pi(\mu|\mathscr{D}_n)$ on a regular grid of width $\Delta$ and summing up.)
\end{enumerate}
\end{exoset}

\begin{enumerate}
\item
Since the Cauchy $\mathscr{C}(\mu,1)$ distribution is associated with the density
$$
f(x|\theta) = \frac{1}{\pi\{1+(x-\theta)^2\}}
$$
the likelihood $\ell(\mu|\mathscr{D}_n)$ is made of the product of the densities.
\item
Given that $\ell(\mu|\mathscr{D}_n)$ is the inverse of a polynomial of order $2n$, it cannot be associated with a
sufficient statistic of fixed dimension against $n$. Therefore, there is no family of prior distributions parametrised
by a fixed dimension vector that can operate as a conjugate family. The only formal family of conjugate priors is made
of densities of the form
$$
\pi(\mu)\propto\frac{1}{\prod_{i=1}^m(1+(x^0_i-\mu)^2)}\,
$$
where $m$ and the $m$ values $x^0_i$ are arbitrarily chosen. Since this family has an unbounded number of parameters, it
is of limited modelling interest.
\item
If $\mu\sim\mathcal{N}(0,10)$, $\pi(\mu)\propto\exp\{-\mu^2/20\}$. Hence,
$$
\pi(\mu|\mathscr{D}_n)\propto\frac{\exp(-\mu^2/20)}{\prod_{i=1}^n(1+(x_i-\mu)^2)}\,.
$$
\item
The question is ambiguous: as stated, there is no need to compute the normalising constant. However, the appealing
version consists in finding an HPD region at a given confidence level $\alpha$.

First, we can define the un-normalised posterior as
\begin{verbatim}
> Dn=rcauchy(100)
> pitilde=function(the,Dn){
 post=dnorm(the,sd=sqrt(10))
 for (i in 1:length(Dn)) post=post*dcauchy(Dn[i]-the)
 return(post)}
\end{verbatim}
where {\sf Dn} is the sample. To find the normalising constant, the easiest is to use {\sf integrate}:
\begin{verbatim}
> tointegre=function(x){ pitilde(the=x,Dn=Dn) }
> Z=integrate(f=tointegre,low=-1,up=1)$val
1.985114e-104
\end{verbatim}
From there, we need to compute coverages of HPD regions until we hit the proper coverage:
\begin{verbatim}
trunpos=function(alpha=.95){
   levels=max(pitilde(the=seq(-1,1,by=.01),Dn=Dn))*seq(.99,.01,by=-.01)
   cover=0
   indx=1
   while ((cover<alpha)||(indx<length(indx))){
       tointegre=function(x){ 
         pitilde(the=x,Dn=Dn)*(pitilde(the=x,Dn=Dn)>levels[indx]) }
       cover=integrate(f=tointegre,low=-1,up=1)$val/Z
       indx=indx+1
     }
   return(levels[indx])
   }
\end{verbatim}
For {\em our} simulated dataset, this results in
\begin{verbatim}
> trunpos()
[1] 1.342565e-104
> trunpos()/Z
[1] 0.6763163
\end{verbatim}
\end{enumerate}

\newcommand\IP{\mathbb{P}}
\begin{exoset}
Show that the limit of the posterior probability $\IP^\pi(\mu<0|x)$ of (2.7)
when $\tau$ goes to $\infty$ is $\Phi(-x/\sigma)$. Show that, when $\xi$ varies in $\mathbb{R}$,
the posterior probability can take any value between $0$ and $1$.
\end{exoset}

Since
\begin{eqnarray*}
P^\pi(\mu<0|x) &=& \Phi\left(-\xi(x)/ \omega \right)\\
     &=& \Phi\left( \frac{\sigma^2\xi+\tau^2 x}{\sigma^2 +\tau^2}
               \sqrt{\frac{\sigma^2+\tau^2}{\sigma^2\tau^2}} \right)\\
     &=& \Phi\left( \frac{\sigma^2\xi+\tau^2 x}{\sqrt{\sigma^2 +\tau^2}\sqrt{\sigma^2\tau^2}} \right)\,,
\end{eqnarray*}
when $\xi$ is fixed and $\tau$ goes to $\infty$, the ratio
$$
\frac{\sigma^2\xi+\tau^2 x}{\sqrt{\sigma^2 +\tau^2}\sqrt{\sigma^2\tau^2}}
$$
goes to
$$
\lim_{\tau\to\infty} \frac{\tau^2 x}{\sqrt{\sigma^2 +\tau^2}\sqrt{\sigma^2\tau^2}}
= \lim_{\tau\to\infty} \frac{\tau^2 x}{\tau^2\sigma} = \frac{x}{\sigma}\,.
$$
However, if $\xi$ varies with $\tau$, the limit can be anything: simply take $\xi=\tau^2\mu$, 
then
$$
\lim_{\tau\to\infty} \frac{\sigma^2\tau^2\mu+\tau^2 x}{\sqrt{\sigma^2 +\tau^2}\sqrt{\sigma^2\tau^2}}
= \lim_{\tau\to\infty} \frac{\tau}{\sqrt{\sigma^2
+\tau^2}}\,\frac{\sigma^2\mu+x}{\sigma}=\frac{\sigma^2\mu+x}{\sigma}\,.
$$

\begin{exoset}\label{exo:BaFaNa}
Define a function \verb+BaRaJ+ of the ratio \verb+rat+ when \verb+z=mean(shift)/.75+ in the function \verb+BaFa+. Deduce from a
plot of the function \verb+BaRaJ+ that the Bayes factor is always less than one when \verb+rat+ varies. ({\em Note:} It is possible
to establish analytically that the Bayes factor is maximal and equal to $1$ for $\tau=0$.)
\end{exoset}

\begin{figure}[bt]
\begin{center}
\includegraphics[width=.8\textwidth]{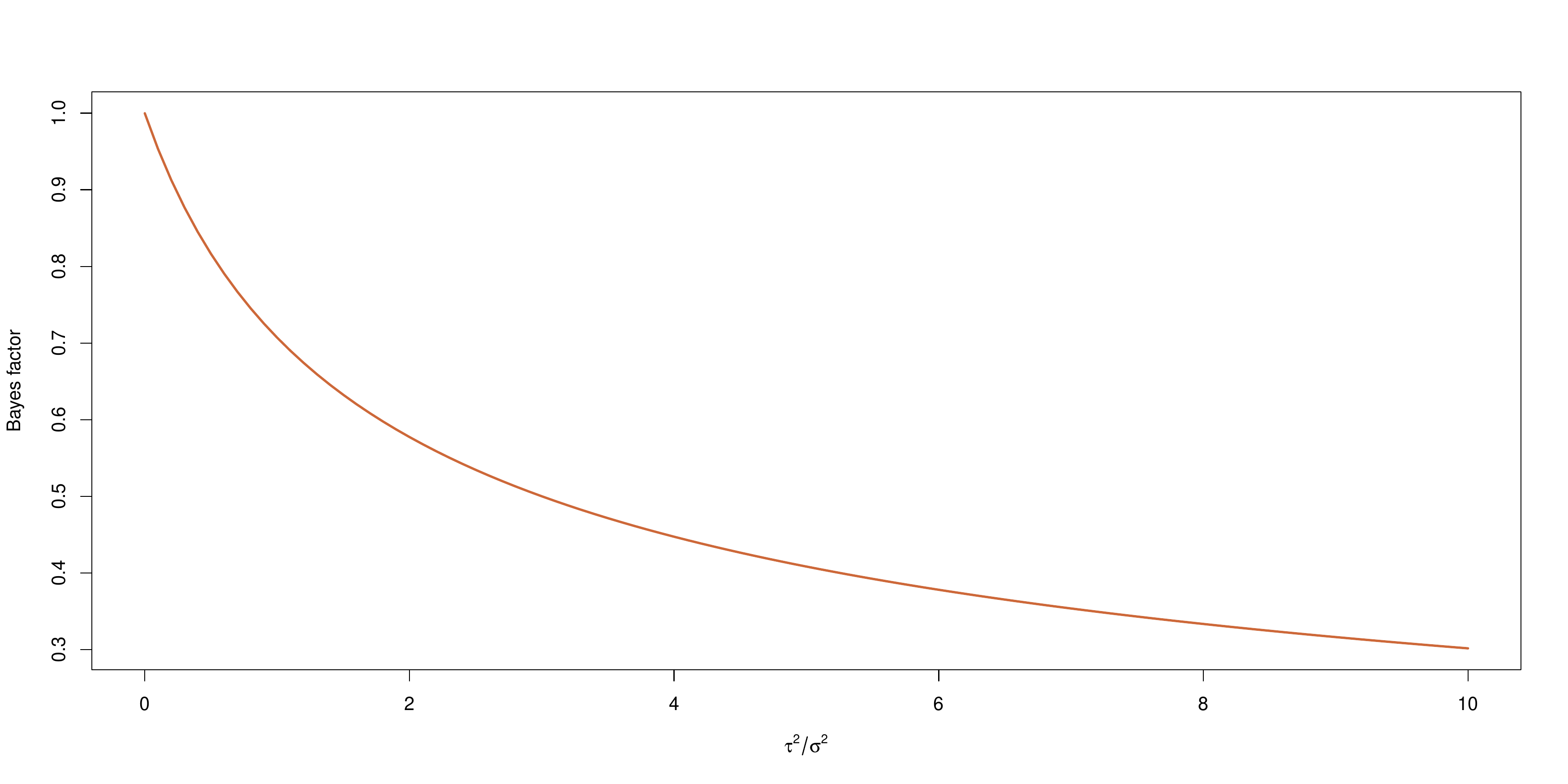}
\caption{\label{fig:BaRaJo}Evolution of the Bayes factor as a function of $\tau^2/\sigma^2$.}
\end{center}
\end{figure}

Since
\begin{verbatim}
BaFa=function(z,rat){
#rat denotes the ratio tau^2/sigma^2
sqrt(1/(1+rat))*exp(z^2/(2*(1+1/rat)))}
\end{verbatim}
it is straightforward to define
\begin{verbatim}
BaRaJ=function(rat){
BaFa(mean(shift)/.75,rat)}
\end{verbatim}
and to plot the corresponding curve (Figure \ref{fig:BaRaJo} in this manual).

\begin{exoset}\label{exo:fulX}
In the application part of Example 2.1
to {\bfseries normaldata}, plot the approximated Bayes
factor as a function of $\tau$. ({\em Hint:} Simulate a single normal $\mathscr{N}(0,1)$ sample and recycle it
for all values of $\tau$.)
\end{exoset}

\newcommand\smpl{\mathcal{D}_n}
\newcommand\dsplnt{\displaystyle\int}

The Bayes factor is given by
$$
B^\pi_{21}(\smpl)
= \dfrac{\dsplnt\,\left[(\mu-\xi-\bar x)^2+(\mu+\xi-\bar y)^2+s_{xy}^2\right]^{-n}
e^{-\xi^2/2\tau^2}/\tau\sqrt{2\pi}\,\mathrm{d}\mu\,\mathrm{d}\xi}
{\dsplnt\,\left[(\mu-\bar x)^2+(\mu-\bar y)^2+s_{xy}^2\right]^{-n}\,\mathrm{d}\mu}\,,
$$
where $s_{xy}^2$ denotes the average
$$
s_{xy}^2 = \frac{1}{n}\,\sum_{i=1}^n\,(x_i-\bar x)^2 + \frac{1}{n}\,\sum_{i=1}^n\,(y_i-\bar y)^2 \,.
$$
As mentioned in Example 2.1, the denominator can be integrated in closed form:
$$
(\mu-\bar x)^2+(\mu-\bar y)^2 = 2\mu^2-2\mu(\bar x+\bar y) + \bar x^2+\bar y^2 = 2 (\mu-\nicefrac{1}{2}[\bar
x+\bar y])^2+  \nicefrac{1}{2}(\bar x-\bar y)^2\,.
$$
\begin{figure}
\begin{center}
\includegraphics[width=7cm,height=5cm]{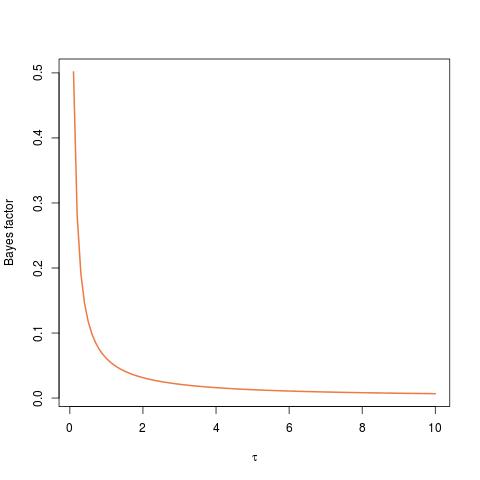}
\caption{\label{fig:BFtau}Evolution of the Bayes factor approximation $\widehat B^\pi_{21}(\smpl)$ as a function of
$\tau$, when comparing the fifth and the sixth sessions of Illingworth's experiment.}
\end{center}
\end{figure}
Hence, if $s_{xyz}^2=\nicefrac{1}{2}(\bar x-\bar y)^2+s_{xy}^2$,
\begin{align*}
\dsplnt\,&\left[(\mu-\bar x)^2+(\mu-\bar y)^2+s_{xy}^2\right]^{-n}\,\mathrm{d}\mu\\
&= \dsplnt\,\left[2 (\mu-\nicefrac{1}{2}[\bar
x+\bar y])^2+ \nicefrac{1}{2}(\bar x-\bar y)^2+s_{xy}^2\right]^{-n}\,\mathrm{d}\mu\\
&= \dsplnt\,\left[2 (\mu-\nicefrac{1}{2}[\bar
x+\bar y])^2+ s_{xyz}^2\right]^{-n}\,\mathrm{d}\mu\\
&= \dfrac{1}{s_{xyz}^{2n}}\,
\dsplnt\,\left[2 (\mu-\nicefrac{1}{2}[\bar
x+\bar y])^2 \big/ s_{xyz}^2 +1 \right]^{-n}\,\mathrm{d}\mu\\
&= \dfrac{1}{s_{xyz}^{2n}}\,
\dsplnt\,\left[\frac{2(2n-1)}{(2n-1)s^2_ {xyz}}(\mu-\nicefrac{1}{2}[\bar
x+\bar y])^2+1\right]^{-n}\,\mathrm{d}\mu\\
&= \dfrac{1}{s_{xyz}^{2n}}\,
\frac{s_{xyz}|}{\sqrt{2(2n-1)}}\,\frac{\Gamma(n-\nicefrac{1}{2})\sqrt{(2n-1)\pi}}{\Gamma(n)}\\
&= \dfrac{1}{s_{xyz}^{2n-1}}\,
\frac{\Gamma(n-\nicefrac{1}{2})\sqrt{\pi}}{\sqrt{2}\Gamma(n)}\,,
\end{align*}
by identification of the missing constant in the $t$ density (see Exercise \ref{exo:tcon}).

The integral in $\mu$ in the numerator can be found in the same way and it leads to the simplified form of Example 2,2:
$$
B^\pi_{21}(\smpl) 
= \dfrac{ \dsplnt\,\left[ (2\xi+\bar x-\bar y)^2+2\,s_{xy}^2\right]^{-n+1/2}
e^{-\xi^2/2\tau^2}\,\text{d}\xi/\tau\sqrt{2\pi}}
{ \left[ (\bar x-\bar y)^2+2\,s_{xy}^2 \right]^{-n+1/2} }\,.
$$
The numerator can be aproximated by simulations from a normal $\mathscr{N}(0,\tau^2)$ distribution.
Therefore, simulating a normal $\mathscr{N}(0,\tau^2)$ sample of $\xi_i$'s $(i=1,\ldots,N)$ produces a converging
estimate of $B^\pi_{21}(\smpl)$ as
$$
\widehat B^\pi_{21}(\smpl) = \dfrac{\frac{1}{N}
\sum_{i=1}^N \left[ (2\xi_i+\bar x-\bar y)^2+2\,s_{xy}^2\right]^{-n+1/2}}{
\left[ (\bar x-\bar y)^2+2\,s_{xy}^2 \right]^{-n+1/2} }\,.
$$
An R implementation is as follows:
\begin{verbatim}
> illing=as.matrix(normaldata) 
> xsam=illing[illing[,1]==5,2]
> xbar=mean(xsam)
[1] -0.041 
> ysam=illing[illing[,1]==6,2]
> ybar=mean(ysam) 
[1] -0.025
> Ssquar=9*(var(xsam)+var(ysam))/10
[1] 0.101474
> Nsim=10^4
> montecarl=rnorm(Nsim)
> BF=tau=seq(.1,10,le=100)
> for (t in 1:100)
  BF[t]=mean(((2*tau[t]*montecarl+xbar-ybar)^2+2*Ssquar)^(-8.5))/
 ((xbar-ybar)^2+2*Ssquar)^(-8.5)
> plot(tau,BF,type="l")
\end{verbatim}

\begin{exoset}\label{exo:fulXX}
In the setup of Example 2.1,
show that, when $\xi\sim\mathscr{N}(0,\sigma^2)$, 
the Bayes factor can be expressed in closed form using the normalizing constant of the $t$ distribution
(see Exercise \ref{exo:tcon})
\end{exoset}

When $\xi\sim\mathscr{N}(0,\sigma^2)$, we have
$$
B^\pi_{21}(\smpl) = \dfrac{\dsplnt\,
e^{-n\left[(\mu-\xi-\bar x)^2+(\mu+\xi-\bar y)^2+s_{xy}^2\right]/2\sigma^2}\,
\sigma^{-2n-2} e^{-\xi^2/2\sigma^2}\big/\sigma\sqrt{2\pi}\,\mathrm{d}\sigma^2\,\mathrm{d}\mu\,\mathrm{d}\xi}
{\dsplnt\,e^{-n \left[(\mu-\bar x)^2+(\mu-\bar y)^2+s_{xy}^2\right]/2\sigma^2}\,
\sigma^{-2n-2} \,\mathrm{d}\sigma^2\,\mathrm{d}\mu}
$$
In the numerator,
\begin{align*}
n&\left[(\mu-\xi-\bar x)^2+(\mu+\xi-\bar y)^2+s_{xy}^2\right]+\xi^2\\
&= 2n\left(\mu-\nicefrac{1}{2}[\bar x+\bar y]\right)^2 + n\frac{(\bar x-\bar y)^2}{2}
+(2n+1)\left(\xi+\nicefrac{n}{2n+1}[\bar x-\bar y]\right)^2 - \frac{n(\bar x-\bar y)^2}{2n+1} +ns_{xy}^2\\
&= 2n\left(\mu-\nicefrac{1}{2}[\bar x+\bar y]\right)^2 +(2n+1)\left(\xi+\nicefrac{n}{2n+1}[\bar x-\bar y]\right)^2 
+ \frac{n(2n-1)(\bar x-\bar y)^2}{2(2n+1)} +ns_{xy}^2
\end{align*}
implies
\begin{align*}
\dsplnt\, &e^{-n\left[(\mu-\xi-\bar x)^2+(\mu+\xi-\bar y)^2+s_{xy}^2\right]/2\sigma^2} \,
\sigma^{-2n-3} e^{-\xi^2/2\sigma^2}\big/\sqrt{2\pi}\,\mathrm{d}\sigma^2\,\mathrm{d}\mu\,\mathrm{d}\xi\\
&= \frac{\sqrt{2\pi}}{\sqrt{2n(2n+1)}}\int e^{-\{\frac{n(2n-1)(\bar x-\bar y)^2}{2(2n+1)} +ns_{xy}^2\}/2\sigma^2}
\sigma^{-2n-1}\,\mathrm{d}\sigma^2\\
&= \frac{\sqrt{\pi}}{\sqrt{n(2n+1)}} \Gamma(n) 2^{n+1} n^{-n} \left[\frac{(2n-1)(\bar x-\bar y)^2}{2(2n+1)}
+s_{xy}^2\right]^{-n}\,.
\end{align*}
Similarly, for the denominator
$$
(\mu-\bar x)^2+(\mu-\bar y)^2 = 2 \left(\mu-\nicefrac{1}{2}[\bar x+\bar y]\right)^2+\nicefrac{1}{2}(\bar x-\bar y)^2\,.
$$
and
\begin{align*}
\dsplnt\,&e^{-n \left[(\mu-\bar x)^2+(\mu-\bar y)^2+s_{xy}^2\right]/2\sigma^2}\,
\sigma^{-2n-2} \,\mathrm{d}\sigma^2\,\mathrm{d}\mu \\
&= \dsplnt\,e^{-n \left[2 \left(\mu-\nicefrac{1}{2}[\bar x+\bar y]\right)^2+\nicefrac{1}{2}(\bar x-\bar
y)^2+s_{xy}^2\right]/2\sigma^2}\,\sigma^{-2n-2} \,\mathrm{d}\sigma^2\,\mathrm{d}\mu \\
&=\frac{\sqrt{2\pi}}{\sqrt{2n}}\,\dsplnt\,e^{-n\left[\nicefrac{1}{2}(\bar x-\bar
y)^2+s_{xy}^2\right]/2\sigma^2}\,\sigma^{-2n-2} \,\mathrm{d}\sigma^2\\
&=\frac{\sqrt{\pi}}{\sqrt{n}}\,\Gamma(n) 2^n n^{-n} \left[\nicefrac{1}{2}(\bar x-\bar y)^2+s_{xy}^2\right]^{-n}
\end{align*}
Therefore,
\begin{align*}
B^\pi_{21}(\smpl) &= \dfrac{\frac{\sqrt{\pi}}{\sqrt{n(2n+1)}} \Gamma(n) 2^{n+1} n^{-n} \left[\frac{(2n-1)(\bar x-\bar
y)^2}{2(2n+1)} +s_{xy}^2\right]^{-n}}{\frac{\sqrt{\pi}}{\sqrt{n}}\,\Gamma(n) 2^n n^{-n} \left[\nicefrac{1}{2}(\bar
x-\bar y)^2+s_{xy}^2\right]^{-n}}\\
&=\dfrac{2 \left[\frac{(2n-1)(\bar x-\bar
y)^2}{2(2n+1)} +s_{xy}^2\right]^{-n}}{ \sqrt{2n+1} \left[\nicefrac{1}{2}(\bar
x-\bar y)^2+s_{xy}^2\right]^{-n}}\,.
\end{align*}

\begin{exoset}
Discuss what happens to the importance sampling approximation when
the support of $g$ is larger than the support of $\gamma$.
\end{exoset}

If the support of $\gamma$, $\mathfrak{S}_\gamma$, is smaller than the
support of $g$, the representation
$$
\mathfrak{I} = \int\,\frac{h(x)g(x)}{\gamma(x)}\,\gamma(x)\,\hbox{d}x
$$
is not valid and the importance sampling approximation evaluates instead
the integral
$$
\int_{\mathfrak{S}_\gamma}\,\frac{h(x)g(x)}{\gamma(x)}\,\gamma(x)\,\hbox{d}x.
$$

\begin{exoset}\label{exo:poorcoco}
Show that, when $\gamma$ is the normal {$\mathscr{N}(0,\nu/(\nu-2))$} density and $f_\nu$ is the density
of the $t$ distribution with $\nu$ degrees of freedom, the ratio
$$
{f_\nu^2(x) \over \gamma(x) } \propto {e^{x^2(\nu-2)/2\nu}\over [1 + x^2/\nu]^{(\nu+1)}}
$$
does not have a finite integral. What does this imply about the variance of the importance weights?

Deduce that the importance weights of Example 2.3
have infinite variance.
\end{exoset}

The importance weight is
$$
\exp\left\{(\theta-\mu)^2/2\right\}\,\prod_{i=1}^n [1+(x_i-\theta)^2]^{-1}
$$
with $\theta\sim\mathscr{N}(\mu,\sigma^2)$. While its expectation is finite---it would
be equal to $1$ were we to use the right normalising constants---, the expectation of its
square is not:
$$
\int \exp\left\{(\theta-\mu)^2/2\right\}\,\prod_{i=1}^n [1+(x_i-\theta)^2]^{-2} \,\text{d}\theta=+\infty\,,
$$
due to the dominance of the exponential term over the polynomial term.

\begin{exoset}\label{ex:2.6} 
If $f_\nu$ denotes the density of the Student $t$ distribution ${\mathscr{T}}(\nu,0,1)$
(see Exercise \ref{exo:tcon}), consider the integral 
$$
\mathfrak{I} = \int \sqrt{\left|{x\over 1-x}\right|} \,f_\nu(x) \,\hbox{d}x\,.
$$
\begin{enumerate}
\item Show that $\mathfrak{I}$ is finite but that
$$
\int \frac{|x|}{|1-x|} f_\nu(x) \,\hbox{d}x = \infty\,.
$$
\item Discuss the respective merits of the following importance functions $\gamma$
\begin{itemize}
\renewcommand{\labelitemi}{--}
\item the density of the Student \ {${\mathscr{T}}(\nu,0,1)$}  distribution,
\item the density of the Cauchy \ {${\mathscr{C}}(0,1)$} distribution, 
\item the density of the normal \ {$\mathscr{N}(0,\nu/(\nu-2))$} distribution.
\end{itemize}
In particular, show via an {\sf R} simulation experiment that
these different choices all lead to unreliable estimates of $\mathfrak{I}$
and deduce that the three corresponding estimators have infinite variance.
\item Discuss the alternative choice of a gamma distribution folded at $1$, 
that is, the distribution of $x$ symmetric around $1$ and such that
$$
|x-1| \sim\mathcal{G}a(\alpha,1) \,.
$$
Show that
$${
h(x) \frac{f^2(x)}{\gamma(x)} \propto \sqrt{x}\,f_\nu^2(x)\,|1-x|^{1-\alpha-1}\,\exp|1-x|
}$$
is integrable around $x=1$ when $\alpha<1$ but not
at infinity. Run a simulation experiment to evaluate
the performances of this new proposal.
\end{enumerate}
\end{exoset}

\begin{enumerate}
\item The integral $\mathfrak{I}$ is finite when $\nu>\nicefrac{1}{2}$ since the function
$$
\sqrt{\left|{x\over 1-x}\right|} \,f_\nu(x)
$$
is equivalent to $x^{\nicefrac{1}{2}-\nu-1}=x^{-\nu-\nicefrac{1}{2}}$ at $x=\pm\infty$. Since
$\nu+\nicefrac{1}{2}>1$, the function is integrable.
(The condition $\nu>\nicefrac{1}{2}$ is missing in the text of the exercise.) Similarly, at
$x\approx 1$, the function is equivalent to $|1-x|^{-\nicefrac{1}{2}}$, which is integrable.

The function
$$
\frac{|x|}{|1-x|} f_\nu(x) 
$$
is not integrable at $x=1$ since it is equivalent to $1/|1-x|$.
\item Using as importance function $\gamma$
\begin{itemize}
\renewcommand{\labelitemi}{--}
\item the density of the Student \ {${\mathscr{T}}(\nu,0,1)$} distribution produces an importance weight of $1$ and an
infinite variance estimator since the integrand is not square integrable;
\item the density of the Cauchy \ {${\mathscr{C}}(0,1)$} distribution produces a well-behaved importance weight since
the Cauchy has heavier tails when $\nu>\nicefrac{1}{2}$, however, the integrability problem at $x=1$ remains, hence an
importance sampling estimate with infinite variance;
\item the density of the normal \ {$\mathscr{N}(0,\nu/(\nu-2))$} distribution faces difficulties both with integrability
of the squared integrand at $x=1$ and with the infinite variance of the importance weight due to thinner tails.
\end{itemize}
When evaluating the performances of the three solutions in R, one can use the following:
\begin{verbatim}
  grand=function(x,nu=3){
    sqrt(abs(x)/abs(1-x))}
  N=10^3
  sampone=rt(N,df=3)
  samptwo=rcauchy(N)
  samptre=rnorm(N)
  weitwo=dt(samptwo,df=3)/dcauchy(samptwo)
  weitre=dt(samptre,df=3)/dnorm(samptre)  
  plot(cumsum(grand(samptwo)*weitwo)/(1:N),type="l",
       xlab="simulations",ylab="cumulated average",lwd=2,col="sienna")
  lines(cumsum(grand(samptre)*weitre)/(1:N),col="steelblue",lwd=2)
  lines(cumsum(grand(sampone))/(1:N),col="gold2",lwd=2)
\end{verbatim}
Running the above code several times exhibits variability in the outcome, with sometimes agreement between the estimators
and sometimes huge jumps in some of the series, as exemplified by Figure \ref{fig:infinivar} in this manual.

\begin{figure}[bt]
\begin{center}
\includegraphics[width=.8\textwidth]{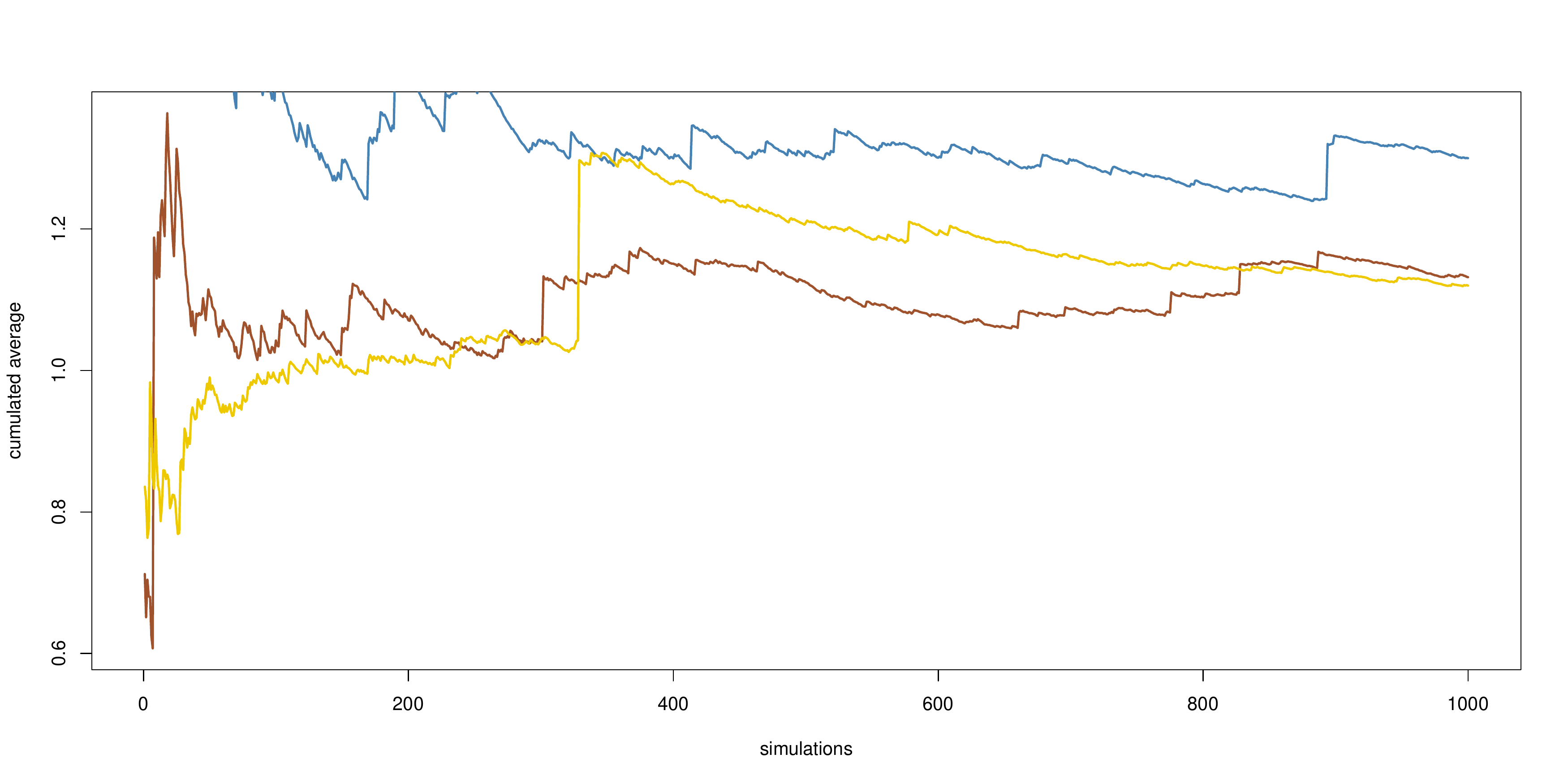}
\caption{\label{fig:infinivar}Evolution of three importance sampling evaluations of the integral $\mathfrak{I}$
using a normal sample {\em (gold)}, a $t_3$  sample {\em (blue)}, and a Cauchy sample {\em (sienna)}.}
\end{center}
\end{figure}

\item If we consider instead the folded Gamma solution, its density is 
$$
\gamma(x)=\frac{1}{2}\,\frac{1}{\Gamma(\alpha)}|1-x|^{\alpha-1}\,e^{-|1-x|}\,.
$$
Therefore, taking $h(x)=|x|/|1-x|$ (missing from the text of the exercise),
$$
h(x) \frac{f^2(x)}{\gamma(x)} \propto \sqrt{|x|}\,f_\nu^2(x)\,|1-x|^{1-\alpha-1}\,\exp|1-x|
$$
which is integrable around $x=1$ when $\alpha<1$ but not at $x=\pm\infty$. 

Running the R code
\begin{verbatim}
 alpha=.5
 y=rgamma(N,sh=alpha)
 x=sample(c(-1,1),N,rep=TRUE)*y+1
 weiqar=2*dt(x,df=3)/dgamma(y,sh=alpha)
\end{verbatim}
does not show a considerable improvement in the evaluation of the integral (Figure \ref{fig:infini4} in this manual). (It may be noted
that {\em in this particular run}, the folded Gamma solution does provide the estimation the closest to the true value.)
\end{enumerate}
\begin{figure}[bt]
\begin{center}
\includegraphics[width=.8\textwidth]{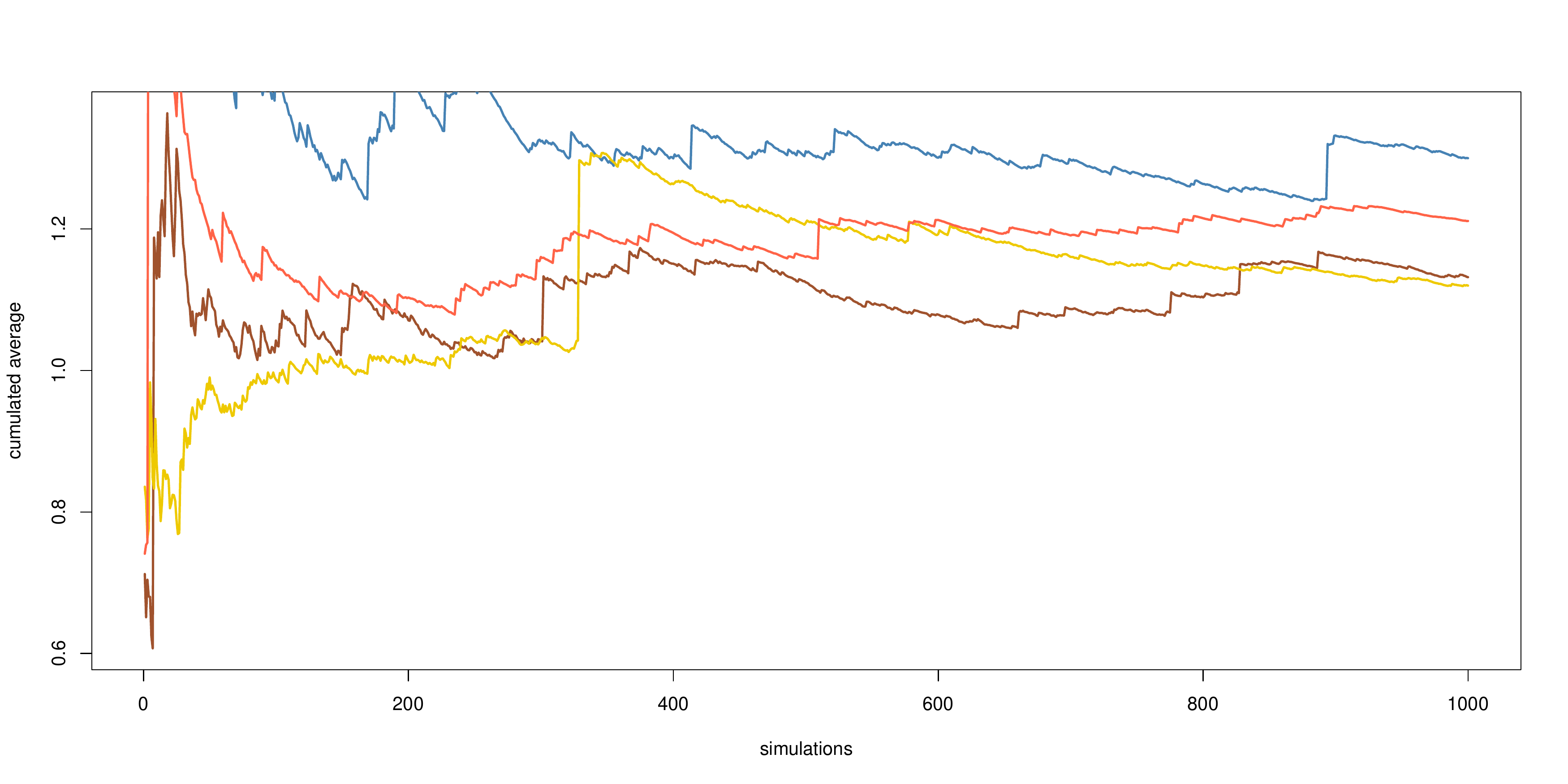}
\caption{\label{fig:infini4}Evolution of three importance sampling evaluations of the integral $\mathfrak{I}$
using a normal sample {\em (gold)}, a $t_3$  sample {\em (blue)}, a Cauchy sample {\em (sienna)}, and a folded Gamma
$\mathcal{G}(.5,1)$ {\em (tomato)}.}
\end{center}
\end{figure}

\begin{exoset}\label{exo:oovariance}
Evaluate the harmonic mean approximation
$$
\widehat m_1(\smpl) = 1\bigg/ N^{-1}\,\sum_{j=1}^N \dfrac{1}{\ell_1(\theta_{1j}|\smpl)}\,.
$$
when applied to the $\mathscr{N}(0,\sigma^2)$ model, {\bfseries normaldata}, and an $\mathscr{IG}(1,1)$
prior on $\sigma^2$.
\end{exoset}

Given a normal $\mathscr{N}(0,\sigma^2)$ sample $\smpl$ and a $\mathscr{G}(1,1)$
prior on $\tau=\sigma^{-2}$, the posterior on $\tau$ is simply
$$
\pi(\tau|\smpl) \propto \tau^{\nicefrac{n}{2}}\exp\left\{-\nicefrac{1}{2}\sum_{i=1}^n x_i^2\tau\right\}\,\exp\{-\tau\}
= \tau^{\nicefrac{n}{2}}\exp\left\{-\tau\left[1+\nicefrac{1}{2}\sum_{i=1}^n x_i^2\right]\right\}\,,
$$
which means that the posterior distribution on $\tau$ is a $$\mathscr{G}\left(\nicefrac{n}{2}+1,\nicefrac{1}{2}\sum_{i=1}^n
x_i^2+1\right)$$ distribution.

Evaluting the harmonic mean approximation thus implies producing a sample from the posterior
\begin{verbatim}
N=10^4
simtau=rgamma(N,sh=33,rat=1+.5*sum(normaldata$x2))
\end{verbatim}
and averaging the inverse likelihoods
\begin{verbatim}
> kood=function(tau){ (2*pi/tau)^(-32)*exp(-0.5*sum(normaldata$x2^2)*tau) }
> 1/mean(1/kood(simtau))
[1] 1.149142e-21
\end{verbatim}
If we repeat this experiment many times, the estimates remain within this order of magnitude.
However, the true value of the marginal likelihood is
$$
(2\pi)^{-\nicefrac{n}{2}}\int_0^\infty \tau^{\nicefrac{n}{2}}\,\left\{-\tau\left[1+\nicefrac{1}{2}\sum_{i=1}^n
x_i^2\right]\right\}\,\text{d}\tau = (2\pi)^{-\nicefrac{n}{2}}\,\Gamma(\nicefrac{n}{2}))\,
\left[1+\nicefrac{1}{2}\sum_{i=1}^n x_i^2\right]^{-1-\nicefrac{n}{2}}
$$
equal to
\begin{verbatim}
> (2*pi)^(-32)*gamma(32)/(1+0.5*sum(normaldata$x2^2))^33
[1] 0.0001717292
\end{verbatim}
There is therefore no connection between the estimate and the true value of the marginal likelihood, confirming our
warning that it should not be used.

\chapter{Regression and Variable Selection}\label{ch:reg}
\newcommand{\bM}{\mathbf{M}}
\newcommand{\bX}{\mathbf{X}}
\newcommand{\bZ}{\mathbf{Z}}
\newcommand{\bbeta}{\mathbf{\beta}}
\begin{exoset}
Show that the matrix $\bZ$ is of full rank if and only if the matrix $\bZ^\tee \bZ$ is invertible
(where $\bZ^\tee $ denotes the transpose of the matrix $\bZ$, which can be produced in {\sf R} using  the \verb+t(Z)+
command). Apply to $\bZ=\left[\mathbf{1}_n \quad \bX\right]$ and deduce that this cannot happen when $p+1>n$.
\end{exoset}

The matrix $X$ is a $(n,k+1)$ matrix. It is of full rank if the $k+1$ columns of $X$ induce
a subspace of $\mathbb{R}^n$ of dimension $(k+1)$, or, in other words, if those columns are
linearly independent: there exists no solution to $X\gamma=\mathbf{0}_{n}$ other than $\gamma=\mathbf{0}_{n}$,
where $\mathbf{0}_{k+1}$ denotes the $(k+1)$-dimensional vector made of $0$'s. If $X^\tee X$ is invertible,
then $X\gamma=\mathbf{0}_{n}$ implies $X^\tee X \gamma=X^\tee\mathbf{0}_{n}=\mathbf{0}_{k+1}$ and thus
$\gamma=(X^\tee X)^{-1} \mathbf{0}_{k+1} = \mathbf{0}_{k+1}$, therefore $X$ is of full rank. If $X^\tee X$ is
not invertible, there exist vectors $\beta$ and $\gamma\ne\beta$ such that $X^\tee X \beta = X^\tee X \gamma$,
i.e.~$X^\tee X (\beta-\gamma)=\mathbf{0}_{k+1}$. This implies that $||X(\beta-\gamma)||^2=0$ and hence
$X(\beta-\gamma)=\mathbf{0}_{n}$ for $\beta-\gamma\ne\mathbf{0}_{k+1}$, thus $X$ is not of full rank.

Obviously, the matrix $(k+1,k+1)$ matrix $X^\tee X$ cannot be invertible if $k+1>n$ since the columns of
$X$ are then necessarily linearly dependent.

\begin{exoset}
Show that solving the minimization program 
$$
\min_\beta\, (\by-\bX\bbeta)^\tee (\by-\bX\bbeta)
$$
requires solving
the system of equations $(\bX^\tee \bX)\beta=\bX^\tee \by$. Check that\index{R@{\sf R}!solve@{\sf solve}}
this can be done via the {\sf R} command \verb+solve(t(X)%*%(X),t(X)%*%y)+.
\end{exoset}

If we decompose $(\by-X\beta)^\tee (\by-X\beta)$ as
$$
\by^\tee\by-2\by^\tee X\beta+\beta^\tee X^\tee X \beta
$$
and differentiate this expression in $\beta$, we obtain the equation
$$
-2 \by^\tee X + 2 \beta^\tee X^\tee X = \mathbf{0}_{k+1}\,,
$$
i.e.
$$
(X^\tee X)\beta=X^\tee \by
$$
by transposing the above.

As can be checked via \verb+help(solve)+, \verb+solve(A,b)+ is the {\sf R} function that solves the linear equation
system $Ax=b$. Defining $X$ and $y$ from {\sf caterpillar}, we get
\begin{verbatim}
> solve(t(X)%*%X,t(X)%*%y)

                     [,1]
  rep(1, 33) 10.998412367
  V1         -0.004430805
  V2         -0.053830053
  V3          0.067939357
  V4         -1.293636435
  V5          0.231636755
  V6         -0.356799738
  V7         -0.237469094
  V8          0.181060170
  V9         -1.285316143
  V10        -0.433105521
\end{verbatim}
which [obviously] gives the same result as the call to the linear regression
function {\sf lm()}:
\begin{verbatim}
> lm(y~X-1)

Call:
lm(formula = y ~ X - 1)

Coefficients:
Xrep(1, 33)     XV1       XV2       XV3        XV4         XV5
 10.998412  -0.004431  -0.053830   0.067939  -1.29363   0.23163
      XV6       XV7       XV8       XV9        XV10
 -0.356800  -0.237469   0.181060  -1.285316  -0.43310
\end{verbatim}
Note the use of the {\sf -1} in the formula \verb+y~X-1+ that eliminates the intercept
already contained in $X$.

\begin{exoset}
Show that the variance of the maximum likelihood estimator of $\beta$ in 
the regression model is given by
$\mathbb{V}(\hat\bbeta|\sigma^2)=\sigma^2(\bX^\tee \bX)^{-1}$.
\end{exoset}

Since $\hat\beta = (X^\tee X)^{-1} X^\tee \by$ is a linear transform of $\by\sim\mathscr{N}(X\beta,\sigma^2 I_n)$, we
have
$$
\hat\beta \sim \mathscr{N}\left(
(X^\tee X)^{-1} X^\tee X\beta,\sigma^2
(X^\tee X)^{-1} X^\tee X (X^\tee X)^{-1} \right)\,,
$$
i.e.~
$$
\hat\beta \sim \mathscr{N}\left( \beta, \sigma^2 (X^\tee X)^{-1} \right)\,.
$$

\begin{exoset}\label{exo:matilde}
For the model
$$
\by|\bbeta,\sigma^2\sim\mathscr{N}_n\left(\bX\bbeta,\sigma^2\mathbf{I}_n\right)
$$
a conjugate prior distribution is as follows: the conditional distribution of $\bbeta$
is given by
$$
\bbeta|\sigma^2\sim\mathscr{N}_{p}(\tilde\bbeta,\sigma^2\bM^{-1})\,,
$$
where $\bM$ is a $(p,p)$ positive definite symmetric matrix, and the marginal prior on $\sigma^2$
is an inverse Gamma distribution
$$
\sigma^2\sim \mathscr{IG}(a,b),\qquad a,b>0\,.
$$
Taking advantage of the matrix identities
\begin{eqnarray*}
\left(\bM+\bX^\tee \bX\right)^{-1} & = & \bM^{-1}-\bM^{-1}\left(\bM^{-1}+(\bX^\tee \bX)^{-1}\right)^{-1}\bM^{-1} \\
                         & = & (\bX^\tee \bX)^{-1}-(\bX^\tee \bX)^{-1}\left(\bM^{-1}+(\bX^\tee \bX)^{-1}\right)^{-1}(\bX^\tee \bX)^{-1}
\end{eqnarray*}
and
\begin{eqnarray*}
\bX^\tee \bX(\bM+\bX^\tee \bX)^{-1}\bM & = & \left(\bM^{-1}(\bM+\bX^\tee \bX)(\bX^\tee \bX)^{-1}\right)^{-1} \\
                   & = & \left(\bM^{-1}+(\bX^\tee \bX)^{-1}\right)^{-1}\,,
\end{eqnarray*}
establish that 
$$
\bbeta|\by,\sigma^2\sim\mathscr{N}_p\left((\bM+\bX^\tee \bX)^{-1}
\{(\bX^\tee \bX)\hat\bbeta+\bM\tilde\bbeta\},\sigma^2(\bM+\bX^\tee \bX)^{-1}\right)
\eqno{(3.8)}
$$
where $\hat\bbeta=(\bX^\tee\bX)^{-1}\bX^\tee\by$ and
$$
\sigma^2|\by\sim \mathscr{IG}\left(\frac{n}{2}+a,b+\frac{s^2}{2}+\frac{(\tilde\bbeta-\hat\bbeta)^\tee
\left(\bM^{-1}+(\bX^\tee \bX)^{-1}\right)^{-1}(\tilde\bbeta-\hat\bbeta)}{2}\right)
\eqno{(3.9)}
$$
where $s^2=(\by-\hat\bbeta\bX)^\tee(\by-\hat\bbeta\bX)$ are the correct posterior distributions.
Give a $(1-\alpha)$ HPD region on $\bbeta$.
\end{exoset}

Starting from the prior distribution
$$
\beta|\sigma^2,X\sim\mathscr{N}_{k+1}(\tilde\beta,\sigma^2M^{-1})\,,\quad
\sigma^2|X\sim \mathscr{IG}(a,b)\,,
$$
the posterior distribution is
\begin{align*}
\pi(\beta,\sigma^2&|\hat\beta,s^2,X) \propto
\sigma^{-k-1-2a-2-n}\,\exp\frac{-1}{2\sigma^2}\left\{
(\beta-\tilde\beta)^\tee M (\beta-\tilde\beta) \right.\\
        &\quad\left. + (\beta-\hat\beta)^\tee (X^\tee X) (\beta-\hat\beta) +s^2 +2b \right\}\\
&=\sigma^{-k-n-2a-3}\,\exp\frac{-1}{2\sigma^2}\left\{
\beta^\tee (M+X^\tee X) \beta - 2 \beta^\tee (M\tilde\beta+X^\tee X\hat\beta)\right.\\
        &\quad\left. +\tilde\beta^\tee M\tilde\beta +\hat\beta^\tee (X^\tee X) \hat\beta +s^2 +2b \right\}\\
&=\sigma^{-k-n-2a-3}\,\exp\frac{-1}{2\sigma^2}\left\{
(\beta-\mathbb{E}[\beta|y,X])^\tee (M+X^\tee X) (\beta-\mathbb{E}[\beta|y,X])\right.\\
        &\quad\left. + \beta^\tee M\tilde\beta
+\hat\beta^\tee (X^\tee X) \hat\beta -\mathbb{E}[\beta|y,X]^\tee (M+X^\tee X) \mathbb{E}[\beta|y,X]+s^2 +2b \right\}
\end{align*}
with
$$
\mathbb{E}[\beta|y,X] = (M+X^\tee X)^{-1} (M\tilde\beta+X^\tee X\hat\beta)\,.
$$
Therefore, (3.3) is the conditional posterior distribution of $\beta$ given $\sigma^2$.
Integrating out $\beta$ leads to
\begin{align*}
\pi(\sigma^2&|\hat\beta,s^2,X) \propto \sigma^{-n-2a-2}\,\exp\frac{-1}{2\sigma^2}\left\{
\beta^\tee M\tilde\beta +\hat\beta^\tee (X^\tee X) \hat\beta \right.\\
        &\quad\left. -\mathbb{E}[\beta|y,X]^\tee (M+X^\tee X) \mathbb{E}[\beta|y,X]+s^2 +2b \right\}\\
&=\sigma^{-n-2a-2}\,\exp\frac{-1}{2\sigma^2}\left\{
\beta^\tee M\tilde\beta +\hat\beta^\tee (X^\tee X) \hat\beta +s^2 +2b \right.\\
        &\quad\left. -(M\tilde\beta+X^\tee X\hat\beta)^\tee (M+X^\tee X)^{-1} (M\tilde\beta+X^\tee X\hat\beta) \right\}
\end{align*}
Using the first matrix identity, we get that
\begin{align*}
(M\tilde\beta+&X^\tee X\hat\beta)^\tee
\left(M+X^\tee X\right)^{-1} (M\tilde\beta+X^\tee X\hat\beta) \\
& =  \tilde\beta^\tee M\tilde\beta-\tilde\beta^\tee\left(M^{-1}+(X^\tee X)^{-1}\right)^{-1}\tilde\beta \\
& + \hat\beta^\tee (X^\tee X)\hat\beta-\hat\beta^\tee\left(M^{-1}+(X^\tee X)^{-1}\right)^{-1}\hat\beta\\
& + 2\hat\beta^\tee (X^\tee X) \left(M+X^\tee X\right)^{-1} M \tilde\beta\\
& = \tilde\beta^\tee M\tilde\beta + \hat\beta^\tee (X^\tee X)\hat\beta\\
& - (\tilde\beta-\hat\beta)^\tee \left(M^{-1}+(X^\tee X)^{-1}\right)^{-1}(\tilde\beta-\hat\beta)
\end{align*}
by virtue of the second identity. Therefore,
\begin{align*}
\pi(\sigma^2|\hat\beta,s^2,X) &\propto \sigma^{-n-2a-2}\,\exp\frac{-1}{2\sigma^2}\left\{
(\tilde\beta-\hat\beta)^\tee \left(M^{-1}\right.\right.\\
        &\quad\left.\left.+(X^\tee X)^{-1}\right)^{-1}(\tilde\beta-\hat\beta) +s^2 +2b \right\}
\end{align*}
which is the distribution (3.4).

Since
$$
\beta|\by,X \sim \mathscr{T}_{k+1}\left(n+2a,\hat\mu,\hat\Sigma\right)\,,
$$
this means that
$$
\pi(\beta|\by,X) \propto \frac{1}{2}\,\left\{
1+\frac{(\beta-\hat\mu)^\tee\hat\Sigma^{-1}(\beta-\hat\mu) }{n+2a} \right\}^{(n+2a+k+1)}
$$
and therefore that an HPD region is of the form
$$
\mathfrak{H}_\alpha=\left\{ \beta;\,,(\beta-\hat\mu)^\tee\hat\Sigma^{-1}(\beta-\hat\mu) \le k_\alpha\right\}\,,
$$
where $k_\alpha$ is determined by the coverage probability $\alpha$.

Now, $(\beta-\hat\mu)^\tee\hat\Sigma^{-1}(\beta-\hat\mu)$ has the same distribution as $||z||^2$ when
$z\sim\mathscr{T}_{k+1}(n+2a,0,I_{k+1})$. This distribution is Fisher's $\mathcal{F}(k+1,n+2a)$
distribution, which means that the bound $k_\alpha$ is determined by the quantiles of this distribution.

\begin{exoset}\label{exo:linpred}
The regression model of Exercise \ref{exo:matilde} can also be used in a predictive sense: for a
given $(m,p+1)$ explanatory matrix $\tilde\bX$, {\em i.e.}, when predicting $m$ unobserved variates $\tilde y_i$, 
the corresponding outcome $\tilde\by$ can be inferred through the {\em predictive
distribution}\index{Distribution!predictive} $\pi(\tilde\by|\sigma^2,\by)$.
Show that $\pi(\tilde\by|\sigma^2,\by)$ is a Gaussian density with mean
$$
\mathbb{E}^\pi [\tilde \by|\sigma^2,\by]=\tilde \bX(\bM+\bX^\tee \bX)^{-1}(\bX^\tee \bX\hat\bbeta+\bM\tilde\bbeta)
$$
and covariance matrix
\begin{eqnarray*}
\mathbb{V}^\pi (\tilde \by|\sigma^2,\by)=\sigma^2(\mathbf{I}_m+\tilde\bX(\bM+\bX^\tee \bX)^{-1}\tilde \bX^\tee ) \,.
\end{eqnarray*}
Deduce that
\begin{eqnarray*}
\tilde \by|\by & \sim & \mathscr{T}_m\left(n+2a,\tilde\bX(\bM+\bX^\tee \bX)^{-1}(\bX^\tee\bX\hat\bbeta+\bM\tilde\bbeta),\right. \\
&&\quad\frac{2b+s^2+(\tilde\bbeta-\hat\bbeta)^\tee \left(\bM^{-1}+(\bX^\tee\bX)^{-1}\right)^{-1}
    (\tilde\bbeta-\hat\bbeta)}{n+2a} \\
&&\quad\times\left.\left\{\mathbf{I}_m+\tilde \bX(\bM+\bX^\tee \bX)^{-1}\tilde\bX^\tee \right\}\right).
\end{eqnarray*}
\end{exoset}

Once again, integrating the normal distribution over the inverse gamma random variable $\sigma^2$
produces a Student's $\mathscr{T}$ distribution. Since
$$
\sigma^2|\by,X \sim \mathcal{IG}\left(\frac{n}{2},\frac{s^2}{2}+\frac{1}{2(c+1)}
(\tilde\beta-\hat\beta)^\tee X^\tee X(\tilde\beta-\hat\beta)\right)
$$
under Zellner's $G$-prior, the predictive distribution is a
\begin{eqnarray*}
\tilde\by|\by,X,\tilde X &\sim& \mathscr{T}_{k+1}\left(n, \tilde X\frac{\tilde\beta+c\hat\beta}{c+1},
   \frac{c(s^2+(\tilde\beta-\hat\beta)^\tee X^\tee X(\tilde\beta-\hat\beta)/(c+1))}{n(c+1)}\right. \\
    && \left.  \times\left\{I_m+\frac{c}{c+1}\tilde X(X^\tee X)^{-1}\tilde X^\tee \right\}\right)
\end{eqnarray*}
distribution.

\begin{exoset}\label{exo:straightA}
Show that the marginal distribution of $\by$ associated with (3.8) and (3.9) is given by
$$
\by\sim\mathscr{T}_n\left(2a,\bX\tilde\bbeta,\frac{b}{a}(\mathbf{I}_n+\bX\bM^{-1}\bX^\tee)\right)\,.
$$
\end{exoset}

The joint posterior is given by
\begin{eqnarray*}
\beta|\sigma^2,\by,X&\sim&\mathscr{N}_{k+1}\left(\hat\beta,\sigma^2(X^\tee X)^{-1}\right), \\
\sigma^2|\by,X &\sim&\mathscr{IG}((n-k-1)/2,s^2/2).
\end{eqnarray*}
Therefore,
$$
\beta|\by,X\sim\mathscr{T}_{k+1}\left(n-k-1,\hat\beta,\frac{s^2}{n-k-1}(X^\tee X)^{-1}\right)
$$
by the same argument as in the previous exercises.

\begin{exoset}
Show that the matrix $(\mathbf{I}_n+g\bX(\bX^\tee \bX)^{-1}\bX^\tee )$ has $1$ and $g+1$ as only eigenvalues.
({\em Hint:} Show that the eigenvectors associated with $g+1$ are of the form $\bX\bbeta$
and that the eigenvectors associated with $1$ are those orthogonal to $\bX$).
Deduce that the determinant of the matrix $(\mathbf{I}_n+g\bX(\bX^\tee \bX)^{-1}\bX^\tee )$
is indeed $(g+1)^{p+1}$.
\end{exoset}

Given the hint, this is somewhat obvious:
\begin{eqnarray*}
(I_n+cX(X^\tee X)^{-1}X^\tee )X\beta &=& X\beta + cX(X^\tee X)^{-1}X^\tee X\beta\\ &=& (c+1)X\beta\\
(I_n+cX(X^\tee X)^{-1}X^\tee )z&=& z + cX(X^\tee X)^{-1}X^\tee z \\ &=&  z
\end{eqnarray*}
for all $\beta$'s in $\mathbb{R}^{k+1}$ and all $z$'s orthogonal to $X$. Since the addition of those
two subspaces generates a vector space of dimension $n$, this defines the whole set of eigenvectors
for both eigenvalues. And since the vector subspace generated by $X$ is of dimension $(k+1)$, this
means that the determinant of $$(I_n+cX(X^\tee X)^{-1}X^\tee )$$ is $(c+1)^{k+1}\times 1^{n-k-1}$.

\begin{exoset}
Under the Jeffreys prior, give the predictive distribution of $\tilde \by$, $m$ dimensional
vector corresponding to the $(m,p)$ matrix of explanatory variables $\tilde\bX$.
\end{exoset}

This predictive can be derived from Exercise \ref{exo:linpred}. Indeed, Jeffreys' prior is
nothing but a special case of conjugate prior with $a=b=0$. Therefore, Exercise \ref{exo:linpred}
implies that, in this limiting case,
\begin{eqnarray*}
\tilde \by|\by,X,\tilde X & \sim & \mathscr{T}_m\left(n,\tilde
X(M+X^\tee X)^{-1}(X^\tee
    X\hat\beta+M\tilde\beta),\right. \\
&&\quad\frac{s^2+(\tilde\beta-\hat\beta)^\tee \left(M^{-1}+(X^\tee X)^{-1}\right)^{-1}
    (\tilde\beta-\hat\beta)}{n} \\
&&\quad\times\left.\left\{ I_m+\tilde X(M+X^\tee X)^{-1}\tilde X^\tee \right\}\right).
\end{eqnarray*}

\begin{exoset}\label{exo:notire}
If $(x_1,x_2)$ is distributed from the uniform distribution on
$$
\left\{(x_1,x_2);\,(x_1-1)^2+(x_2-1)^2\le 1\right\}\cup
\left\{(x_1,x_2);\,(x_1+1)^2+(x_2+1)^2\le 1\right\}\,,
$$
show that the Gibbs sampler does not produce an irreducible chain.
For this distribution, find an alternative Gibbs sampler that works.
({\em Hint:} Consider a rotation of the coordinate axes.)
\end{exoset}

\begin{figure}[bt]
\begin{center}
\includegraphics[width=5cm,height=6cm]{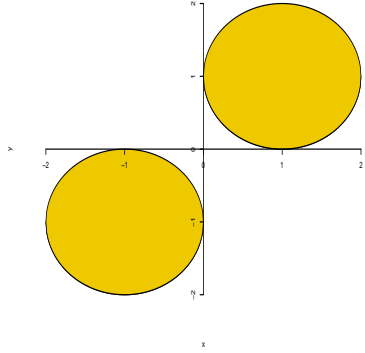}
\caption{\label{fig:nonconnect}Support of the uniform distribution.}
\end{center}
\end{figure}

The support of this uniform distribution is made of two disks with
respective centers $(-1,-1)$ and $(1,1)$, and with radius $1$. This
support is not connected (see Figure \ref{fig:nonconnect} in this manual) and
conditioning on $x_1<0$ means that the conditional distribution of $x_2$ is
$\mathscr{U}(-1-\sqrt{1-x_1^2},-1+\sqrt{1-x_1^2}$,
thus cannot produce a value in $[0,1]$. Similarly, when simulating the
next value of $x_1$, it necessarily remains negative. The Gibbs sampler
thus produces two types of chains, depending on whether or not it is started
from the negative disk.
If we now consider the Gibbs sampler for the new parameterisation
$$
y_1 = x_1+x_2,\quad y_2=x_2-x_1\,,
$$
conditioning on $y_1$ produces a uniform distribution on the union
of a negative and of a positive interval. Therefore, one iteration
of the Gibbs sampler is sufficient to jump [with positive probability]
from one disk to the other one.

\begin{exoset}\label{exo:babyHaC}
If a joint density $g(y_1,y_2)$ corresponds to the conditional
distributions $g_1(y_1|y_2)$ and $g_2(y_2|y_1)$, show that it is
given by
$$
g(y_1,y_2) = {g_2(y_2|y_1) \over \int \;
   g_2(v|y_1)/g_1(y_1|v) \;\hbox{d}v}.
$$
\end{exoset}

If the joint density $g(y_1,y_2)$ exists, then
\begin{align*}
g(y_1,y_2) &= g^1(y_1) g_2(y_2|y_1) \\
           &= g^2(y_2) g_1(y_1|y_2)
\end{align*}
where $g^1$ and $g^2$ denote the densities of the marginal distributions of $y_1$ and $y_2$, respectively.
Thus,
\begin{align*}
g^1(y_1) &= \frac{g_1(y_1|y_2)}{g_2(y_2|y_1)} g^2(y_2)\\
         &\propto \frac{g_1(y_1|y_2)}{g_2(y_2|y_1)}\,,
\end{align*}
as a function of $y_1$ [$g^2(y_2)$ is irrelevant]. Since $g^1$ is a density,
$$
g^1(y_1) = \frac{g_1(y_1|y_2)}{g_2(y_2|y_1)} \bigg/ \int \frac{g_1(u|y_2)}{g_2(y_2|u)} \text{d}u
$$
and
$$
g(y_1,y_2) = g_1(y_1|y_2) \bigg/ \int \frac{g_1(u|y_2)}{g_2(y_2|u)} \text{d}u\,.
$$
Since $y_1$ and $y_2$ play symmetric roles in this derivation, the symmetric version
also holds.

\begin{exoset}\label{exo:Gibbs+}
Considering the model
$$
\eta \vert \theta \sim \mathcal{B}\text{in}(n,\theta)\,,\quad \theta \sim {\cal B}e(a,b),
$$
derive the joint distribution of $(\eta,\theta)$ and the corresponding full conditional
distributions. Implement a Gibbs sampler associated with those full conditionals and 
compare the outcome of the Gibbs sampler on $\theta$ with the true marginal distribution
of $\theta$.
\end{exoset}

\begin{figure}[bt]
\begin{center}
\includegraphics[width=5cm,height=6cm]{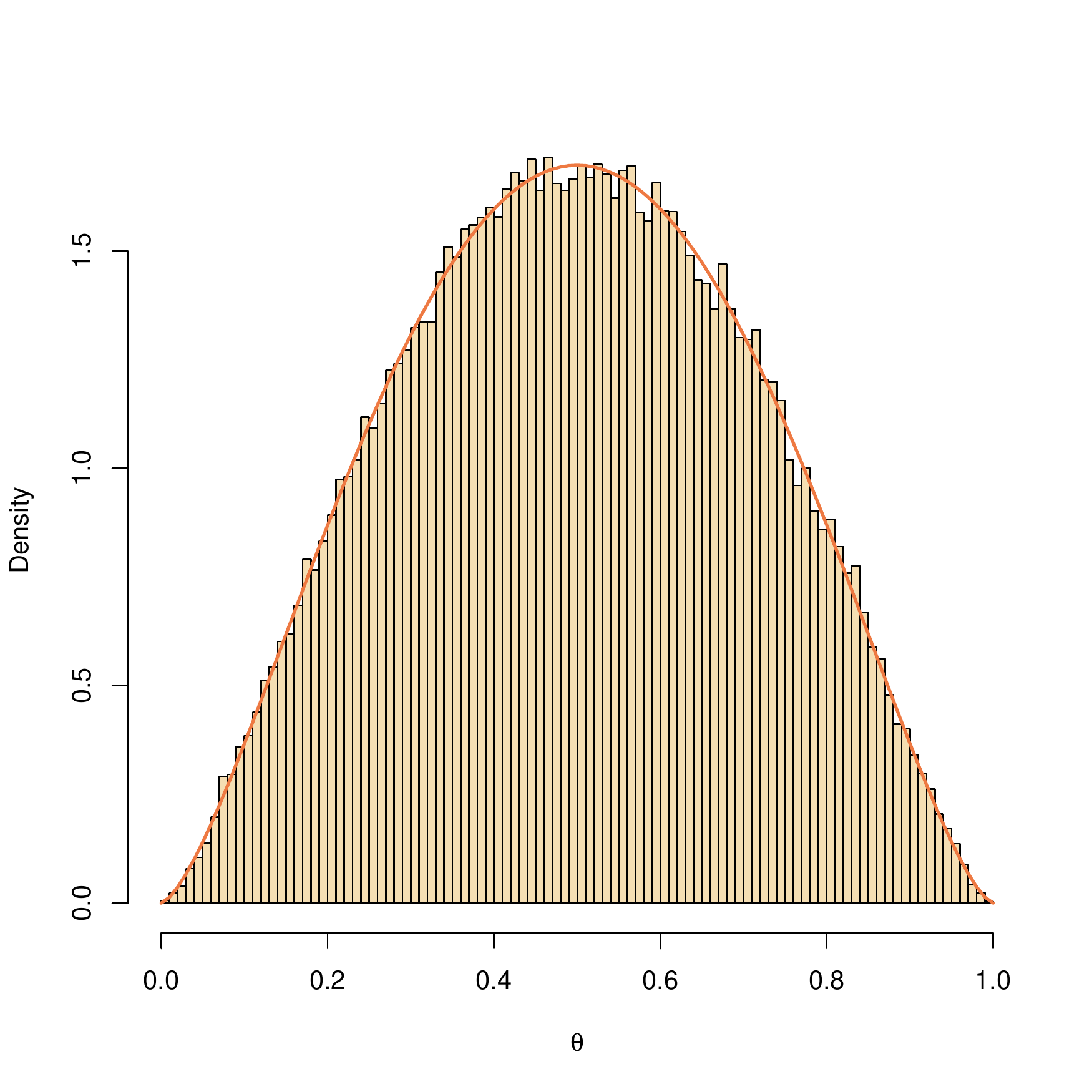}
\caption{\label{fig:gudfit}Fit of the Gibbs output to the Beta $\mathcal{B}(\nicefrac{5}{2},\nicefrac{5}{2})$ distribution.}
\end{center}
\end{figure}

The joint density of  $(\eta,\theta)$ is
$$
\pi(\eta,\theta) \propto {n \choose \eta}\theta^\eta(1-\theta)^{n-\eta}\,\theta^a(1-\theta)^b\,.
$$
The full conditionals are therefore 
$$
\eta | \theta \sim \mathcal{B}\text{in}(n,\theta)
\qquad
\theta | \eta \sim {\mathcal B}e (a+\eta,b+n-\eta)\,.
$$
This means running a Gibbs sampler is straightforward:
\begin{verbatim}
# pseudo-data
n=18
a=b=2.5
N=10^5
#storage matrix
#col.1 for eta, col.2 for theta
gibb=matrix(NA,N,2)
gibb[1,1]=sample(0:n,1)
gibb[1,2]=rbeta(1,a+gibb[1,1],b+n-gibb[1,1])
for (t in 2:N){
  gibb[t,1]=rbinom(1,n,gibb[t-1,2])
  gibb[t,2]=rbeta(1,a+gibb[t,1],b+n-gibb[t,1])}
\end{verbatim}
The output of the above algorithm can be compared with the true marginal distribution, namely the ${\cal B}e(a,b)$
distribution
\begin{verbatim}
hist(gibb[,2],prob=TRUE,col="wheat")
curve(dbeta(x,a,b),add=TRUE,lwd=2)
\end{verbatim}
which shows indeed a very good fit (Figure \ref{fig:gudfit} in this manual).

\begin{exoset}\label{exo:Gibbsaseur}
Take the posterior distribution on $(\theta,\sigma^2)$ associated with the joint model
\begin{eqnarray*} 
x_i|\theta,\sigma^2 &\sim& {\mathscr N}(\theta,\sigma^2), \quad i=1, \ldots, n, \\
\theta &\sim& {\mathscr N}(\theta_0,\tau^2)\,,\quad
\sigma^2 \sim I{\mathscr G}(a,b)\,.
\end{eqnarray*}
Show that the full conditional distributions are given by
$$
\theta | \bx,\sigma^2\sim
{\mathscr N}\left(\frac{\sigma^2}{\sigma^2+n \tau^2}\;\theta_0 + \frac{n\tau^2}{\sigma^2+n \tau^2} 
\;\bar x, \; \frac{\sigma^2 \tau^2}{\sigma^2+n \tau^2}\right) 
$$
and
$$
\sigma^2 | \bx,\theta \sim I{\mathscr G}\left(\frac{n}{2}+a, \frac{1}{2}\sum_i( x_i - \theta)^2+b \right),
$$
where $\bar x$ is the empirical average of the observations.
Implement the Gibbs sampler associated with these conditionals.
\end{exoset}

\begin{figure}[bt]
\begin{center}
\includegraphics[width=.8\textwidth]{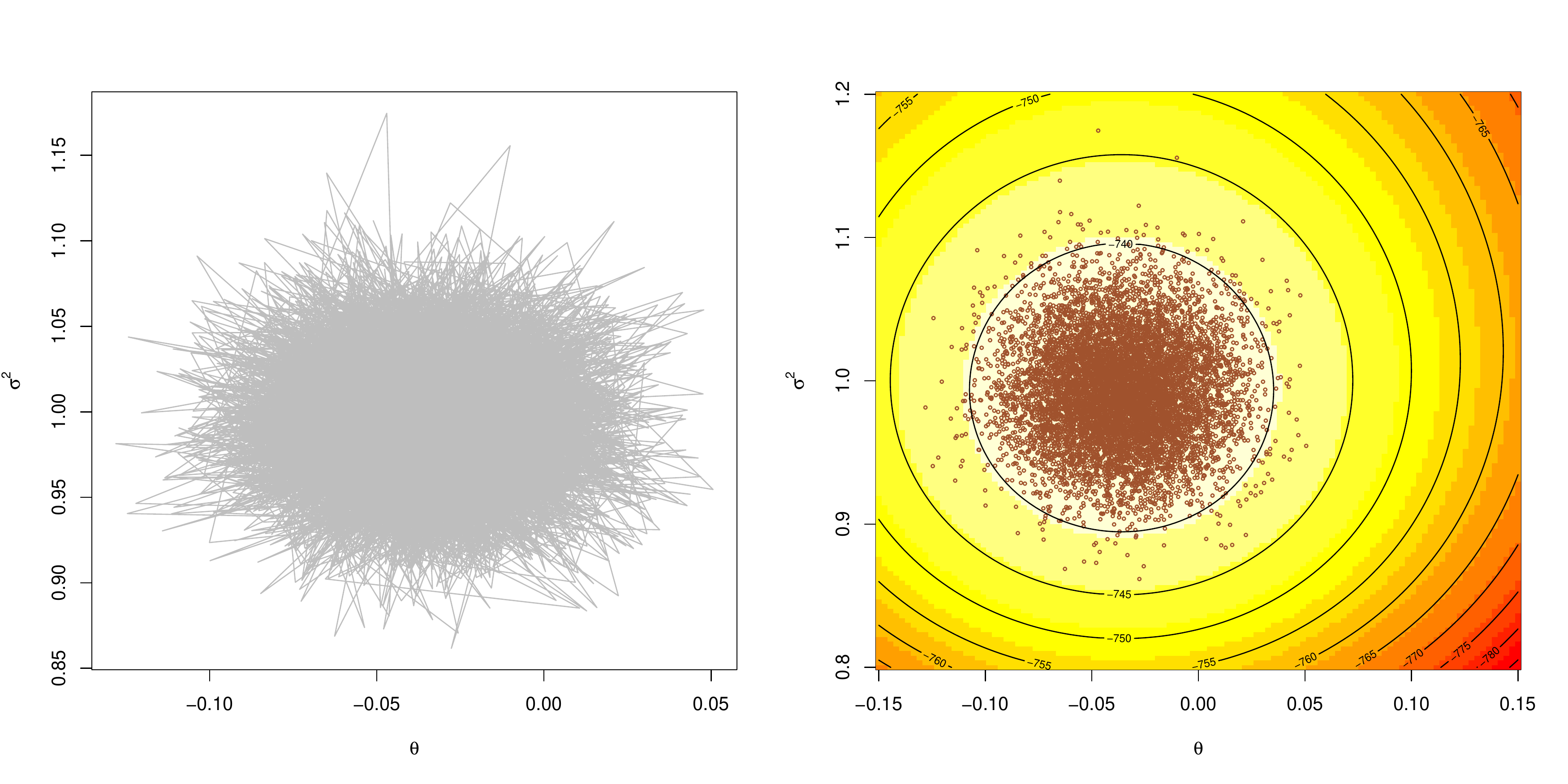}
\caption{\label{fig:norfit}Gibbs output for the normal posterior with {\em (left)} Gibbs path and {\em (right)}
superposition with the log-posterior.}
\end{center}
\end{figure}

From the full posterior density
\begin{align*}
\pi(\theta,\sigma^2|\mathbf{x}) &\propto \prod_{i=1}^n
\exp\{-(x_i-\theta)^2\big/2\sigma^2\}\,\exp\{-(\theta-\theta_0)^2\big/2\tau^2\}\,(\sigma^2)^{-\nicefrac{n}{2}-a-1},\exp\{-b/\sigma^2\}\\
&=(\sigma^2)^{-\nicefrac{n}{2}-a-1}\, \exp\{-n(\bar x-\theta)^2\big/2\sigma^2-s_n^2\big/2\sigma^2-(\theta-\theta_0)^2\big/2\tau^2-b/\sigma^2\}
\end{align*}
we derive easily that
$$
\pi(\theta|\mathbf{x},\sigma) \propto \exp\{-n(\bar x-\theta)^2\big/2\sigma^2-(\theta-\theta_0)^2\big/2\tau^2\}\,,
$$
which leads to 
$$
\theta | \bx,\sigma^2\sim
{\mathscr N}\left(\frac{\sigma^2}{\sigma^2+n \tau^2}\;\theta_0 + \frac{n\tau^2}{\sigma^2+n \tau^2}
\;\bar x, \; \frac{\sigma^2 \tau^2}{\sigma^2+n \tau^2}\right)
$$
Similarly,
$$
\pi(\sigma^2|\mathbf{x},\theta)\propto (\sigma^2)^{-\nicefrac{n}{2}-a-1}\, \exp\{-\sum_{i=1}^n
(x_i-\theta)^2\big/2\sigma^2-b/\sigma^2\}\,,
$$
hence
$$
\sigma^2 | \bx,\theta \sim I{\mathscr G}\left(\nicefrac{n}{2}+a, \nicefrac{1}{2}\sum_i( x_i - \theta)^2+b \right)\,.
$$
Running an R code based on those two conditionals is straightforward:
\begin{verbatim}
# pseudo-data
n=1492
x=rnorm(n)
meanx=mean(x)
varx=var(x)*(n-1)
a=b=2.5
tau=5
meantop=n*tau*meanx
apost=a+(n/2)
# Gibbs parameters
N=10^4
gibb=matrix(NA,N,2)
gibb[1,1]=rnorm(1,mean(x),6)
gibb[1,2]=1/rgamma(1,sh=apost,rate=b+0.5*sum((x-gibb[1,1])^2))
for (t in 2:N){

  gibb[t,1]=rnorm(1,mean=meantop/(gibb[t-1,2]+n*tau),
    sd=sqrt(gibb[t-1,2]*tau/(gibb[t-1,2]+n*tau)))
  gibb[t,2]=1/rgamma(1,sh=apost,rate=b+0.5*sum((x-gibb[t,1])^2))
  }
# remove warmup
gibb=gibb[(N/10):N,]
par(mfrow=c(1,2))
plot(gibb,typ="l",col="gray",ylab=expression(sigma^2)}
grid.the=seq(-.15,.15,le=111)
grid.sig=seq(.8,1.2,le=123)
like=function(the,sig){
  -.5*n*(meanx-the)^2/sig-.5*varx/sig-.5*n*log(sig)-
  dnorm(the,sd=sqrt(tau),log=TRUE)-dgamma(1/sig,sh=a,rat=b,log=TRUE)}
post=matrix(NA,111,123)
for (i in 1:111)
  post[i,]=like(grid.the[i],grid.sig)
image(grid.the,grid.sig,post)
points(gibb,cex=.4,col="sienna")
contour(grid.the,grid.sig,post,add=TRUE)
\end{verbatim}
Figure \ref{fig:norfit}  in this manualshows how the Gibbs sample fits the target, after eliminating $10^3$ iterations as warmup.

\chapter{Generalized Linear Models}\label{ch:glm}
\begin{exoset} Show that, for the logistic regression model, the statistic $\sum_{i=1}^n y_i\,\bx^{i}$ is
sufficient when conditioning on the $\bx^i$'s $(1\le i\le n)$, 
and give the corresponding family of conjugate priors.
\end{exoset}

The likelihood associated with a sample $((y_1,\bx_1),\ldots,(y_n,\bx_n))$ from a logistic model writes as
\begin{eqnarray*}
\ell(\bbeta|\by,\bx) &=& \prod_{i=1}^n\left(
\dfrac{\exp(\bx^{i\tee}\bbeta)}{1+\exp(\bx^{i\tee}\bbeta)}\right)^{y_i}\,\left(
\dfrac{1}{1+\exp(\bx^{i\tee}\bbeta)}\right)^{1-y_i} \\
&=& \exp \left\{\sum_{i=1}^n y_i\,\bx^{i\tee}\bbeta \right\}
\bigg/ \prod_{i=1}^n \left[ 1+\exp(\bx^{i\tee}\bbeta )\right] \,.\nonumber
\end{eqnarray*}
Hence, if we consider the $\bx^i$'s as given, the part of the density that only depends on the $y_i$'s is
$$
\exp \left\{\sum_{i=1}^n y_i\,\bx^{i\tee}\bbeta \right\}
$$
and factorises through the statistic $\sum_{i=1}^n y_i\,\bx^{i}$.

This implies that the prior distribution with density
$$
\pi(\bbeta|\xi_0,\lambda) \propto \exp \left\{\xi_0^{\tee}\bbeta \right\}
\bigg/ \prod_{i=1}^n \left[ 1+\exp(\bx^{i\tee}\bbeta )\right]^\lambda
$$
is conjugate, since the corresponding posterior is $\pi(\bbeta|\xi_0+\sum_{i=1}^n y_i\,\bx^{i},\lambda+1)$.

\begin{exoset}
Show that the logarithmic link is the canonical link function in the case of the
Poisson regression model.
\end{exoset}

The likelihood of the Poisson regression model is
\begin{align*}
\ell(\beta|\by,X)&=\prod_{i=1}^n\left(\frac{1}{y_i!}\right)\exp\left\{
y_i\,\bx^{i\tee}\beta-\exp(\bx^{i\tee}\beta)\right\}\\
&= \prod_{i=1}^n\frac{1}{y_i!}\exp\left\{
y_i\,\log(\mu_i)-\mu_i\right\}\,,
\end{align*}
so $\log(\mu_i)=\bx^{i\tee}\beta$ and the logarithmic link is indeed the canonical link function.

\begin{exoset}
Suppose $y_1,\ldots,y_k$ are independent Poisson
$\mathscr{P}(\mu_i)$ random variables. Show that, conditional on
$n=\sum_{i=1}^k y_i$,
$$
\by = (y_1,\ldots,y_k) \sim \mathscr{M}_k(n;\alpha_1,\ldots,\alpha_k)\,,
$$
and determine the $\alpha_i$'s.
\end{exoset}

The joint distribution of $\by$ is
$$
f(\by|\mu_1,\ldots,\mu_k) = \prod_{i=1}^k\left(\frac{\mu_i^{y_i}}{y_i!}\right)\,\exp\left\{-\sum_{i=1}^k\mu_i\right\}\,,
$$
while $n=\sum_{i=1}^k y_i\sim \mathcal{P}(\sum_{i=1}^k\mu_i)$ [which can be established using the moment generating
function of the $\mathcal{P}(\mu)$ distribution]. Therefore, the conditional distribution of $\by$ given $n$ is
\begin{align*}
f(\by|\mu_1,\ldots,\mu_k,n) &= \frac{
\prod_{i=1}^k\left(\frac{\mu_i^{y_i}}{y_i!}\right)\,\exp\left\{-\sum_{i=1}^k\mu_i\right\}
}{ \frac{[\sum_{i=1}^k\mu_i]^n}{n!} \exp\left\{-\sum_{i=1}^k\mu_i\right\}
}\,\mathbb{I}_n \left(\sum_{i=1}^k y_i\right)\\
&= \frac{n!}{\prod_{i=1}^k y_i!}\,\prod_{i=1}^k \left( \frac{\mu_i}{\sum_{i=1}^k\mu_i} \right)^{y_i}
\,\mathbb{I}_n \left(\sum_{i=1}^k y_i\right)\,,
\end{align*}
which is the pdf of the $\mathcal{M}_k(n;\alpha_1,\ldots,\alpha_k)$ distribution, with
$$
\alpha_i = \frac{\mu_i}{\sum_{j=1}^k\mu_j}\,,\qquad i=1,\ldots,k\,.
$$

This conditional representation is a standard property used in the statistical analysis
of contingency tables (Section 4.5): when the margins are random, the cells are Poisson
while, when the margins are fixed, the cells are multinomial.

\begin{exoset}
For $\pi$ the density of an inverse normal distribution with parameters $\theta_1=3/2$ and $\theta_2=2$,
$$
\pi(x)\propto x^{-3/2}\exp(-3/2x-2/x)\mathbb{I}_{x>0},
$$
write down and implement an independence MH sampler with a Gamma proposal with
parameters $(\alpha,\beta)=(4/3,1)$ and
$(\alpha,\beta)=(0.5\sqrt{4/3},0.5)$.
\end{exoset}

\begin{figure}
\begin{center}
\includegraphics[width=\textwidth,height=6cm]{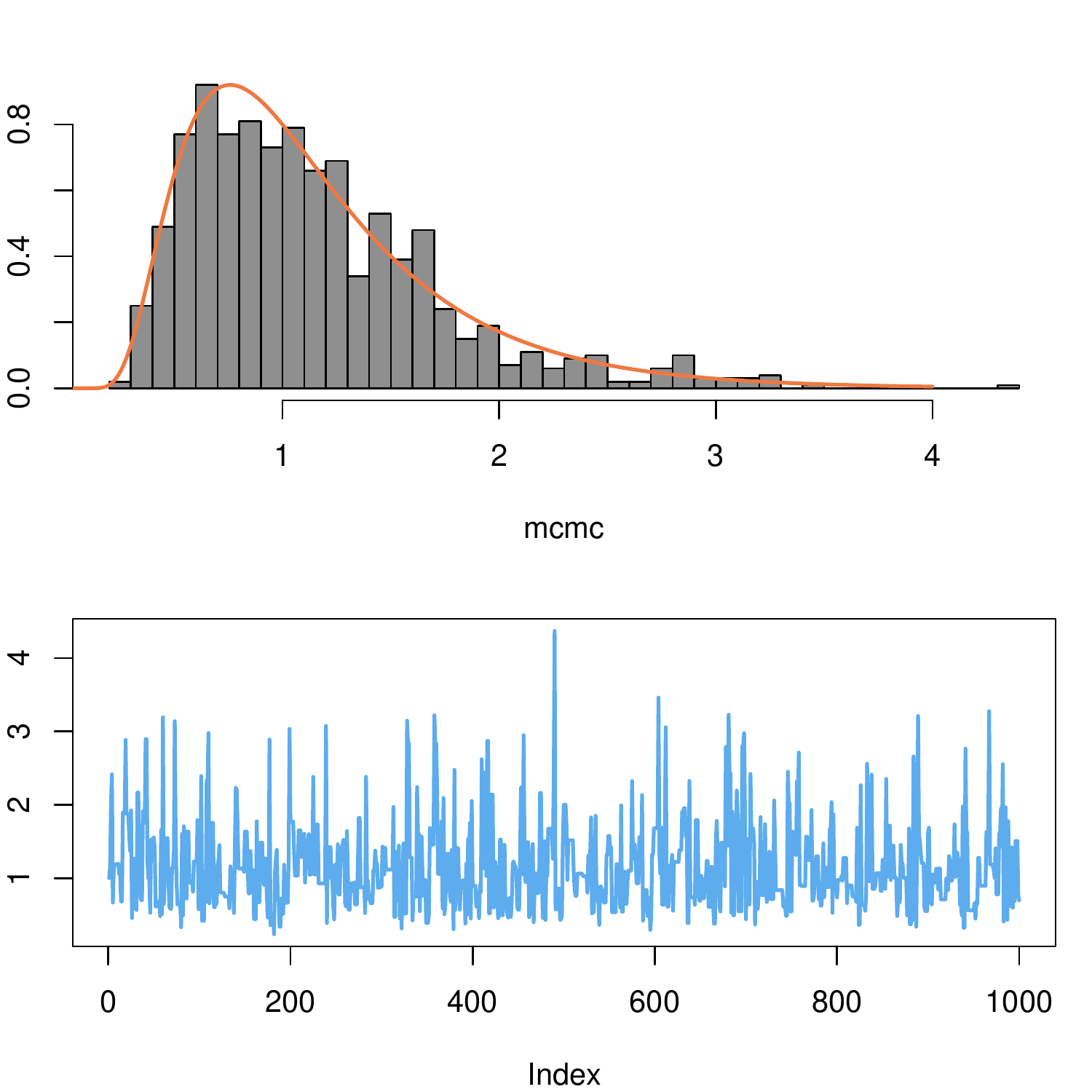}
\caption{\label{fig:exo406}Output of an MCMC simulation of the inverse
normal distribution.}
\end{center}
\end{figure}

A possible {\sf R} code for running an independence Metropolis--Hastings sampler in this setting is
as follows:
\begin{verbatim}
# target density
target=function(x,the1=1.5,the2=2){
  x^(-the1)*exp(-the1*x-the2/x)
  }

al=4/3
bet=1

# initial value
mcmc=rep(1,1000)

for (t in 2:1000){

  y = rgamma(1,shape=al,rate=bet)
  if (runif(1)<target(y)*dgamma(mcmc[t-1],shape=al,rate=bet)/
        (target(mcmc[t-1])*dgamma(y,shape=al,rate=bet)))
    mcmc[t]=y
    else
      mcmc[t]=mcmc[t-1]
  }

# plots
par(mfrow=c(2,1),mar=c(4,2,2,1))
res=hist(mcmc,freq=F,nclass=55,prob=T,col="grey56",
  ylab="",main="")
lines(seq(0.01,4,length=500),valpi*max(res$int)/max(valpi),
  lwd=2,col="sienna2")
plot(mcmc,type="l",col="steelblue2",lwd=2)

\end{verbatim}
The output of this code is illustrated on Figure \ref{fig:exo406} in this manual and shows a reasonable fit of
the target by the histogram and a proper mixing behaviour. Out of the $1000$ iterations
in this example, $600$ corresponded to an acceptance of the Gamma random variable. (Note
that to plot the density on the same scale as the histogram, we resorted to a trick by identifying the
maxima of the histogram and of the density.)

\begin{exoset}
Consider $x_1$, $x_2$, and $x_3$ iid $\mathscr{C}(\theta,1)$, and $\pi(\theta)\propto \exp(-\theta^2/100)$.
Show that the posterior distribution of $\theta$, $\pi(\theta|x_1,x_2,x_3)$, is proportional to
\begin{equation}\label{eq:trimoco}
\exp(-\theta^2/100)[(1+(\theta-x_1)^2)(1+(\theta-x_2)^2)(1+(\theta-x_3)^2)]^{-1}
\end{equation}
and that it is trimodal when  $x_1=0$, $x_2=5$, and $x_3=9$.
Using a random walk based on the Cauchy distribution $\mathscr{C}(0,\sigma^2)$, estimate the
posterior mean of $\theta$ using different values of $\sigma^2$. In each case, monitor the convergence.
\end{exoset}

The function \eqref{eq:trimoco} appears as the product of the [Normal] prior by the three [Cauchy] densities $f(x_i|\theta)$.
The trimodality of the posterior can be checked on a graph when plotting the function \eqref{eq:trimoco}.

\begin{figure}
\begin{center}
\includegraphics[width=\textwidth]{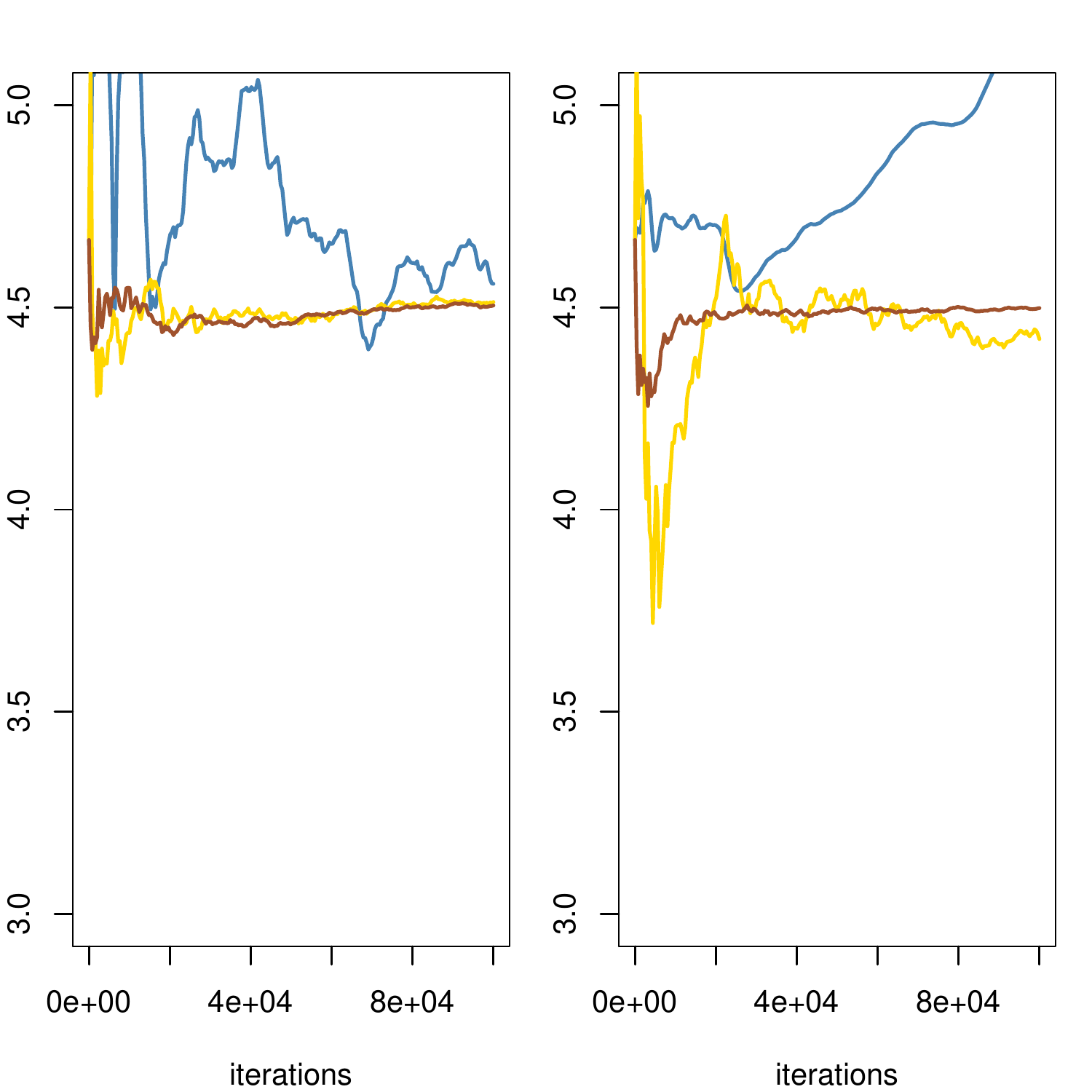}
\caption{\label{fig:trimod}Comparison of the three scale factors $\sigma=.009$ (blue),
$\sigma=.45$ (gold) and $\sigma=9$ (brown),
when using a Cauchy noise {\em (left)} and a normal noise {\em (right)}.}
\end{center}
\end{figure}

A random walk Metropolis--Hastings algorithm can be coded as follows
\begin{verbatim}
x=c(0,5,9)
# target
targ=function(y){
  dnorm(y,sd=sqrt(50))*dt(y-x[1],df=1)*
  dt(y-x[2],df=1)*dt(y-x[3],df=1)
}

# Checking trimodality
plot(seq(-2,15,length=250),
  targ(seq(-2,15,length=250)),type="l")

sigma=c(.001,.05,1)*9 # different scales
N=100000 # number of mcmc iterations

mcmc=matrix(mean(x),ncol=3,nrow=N)
for (t in 2:N){

   mcmc[t,]=mcmc[t-1,]
   y=mcmc[t,]+sigma*rt(3,1) # rnorm(3)
   valid=(runif(3)<targ(y)/targ(mcmc[t-1,]))
   mcmc[t,valid]=y[valid]
   }
\end{verbatim}
The comparison of the three cumulated averages is given in Figure \ref{fig:trimod} in this manual 
and shows that, for the Cauchy noise, both large scales are acceptable while the
smallest scale slows down the convergence properties of the chain. For the normal
noise, these features are exacerbated in the sense that the smallest scale does
not produce convergence for the number of iterations under study [the blue curve leaves
the window of observation], the medium scale
induces some variability and it is only the largest scale that gives an
acceptable approximation to the mean of the distribution \eqref{eq:trimoco}.

\begin{exoset}
Estimate the mean of a $\mathscr{G}a(4.3,6.2)$ random variable using
\begin{enumerate}
\item direct sampling from the distribution via the {\sf R} command\\
\verb+ > x=rgamma(n,4.3,scale=6.2)+ 
\item Metropolis--Hastings with a $\mathscr{G}a(4,7)$ proposal distribution;
\item Metropolis--Hastings with a $\mathscr{G}a(5,6)$ proposal distribution.
\end{enumerate}
In each case, monitor the convergence of the cumulated average.
\end{exoset}

\begin{figure}
\begin{center}
\includegraphics[width=\textwidth,height=6cm]{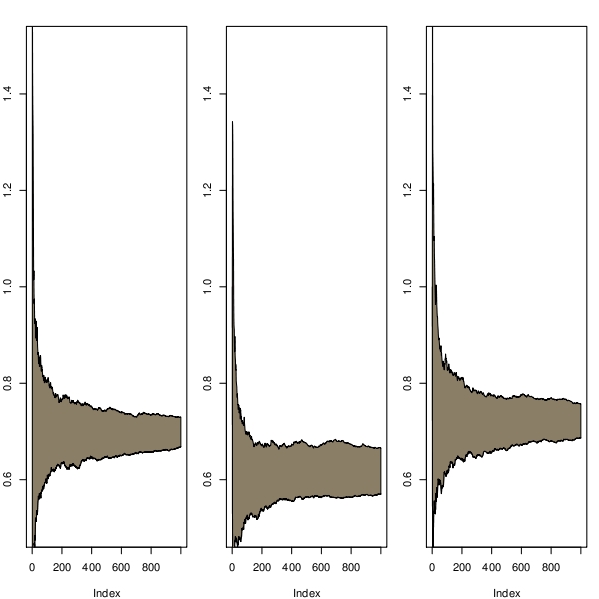}
\caption{\label{fig:ranj407}Range of three samplers for the approximation
of the $\mathscr{G}a(4.3,6.2)$ mean: {\em (left)} iid; {\em (center)} $\mathscr{G}a(4,7)$ proposal;
{\em (right)} $\mathscr{G}a(5,6)$ proposal.}
\end{center}
\end{figure}

Both independence Metropolis--Hastings samplers can be implemented via an {\sf R}
code like
\begin{verbatim}
al=4.3
bet=6.2

mcmc=rep(1,1000)
for (t in 2:1000){

  mcmc[,t]=mcmc[,t-1]
  y = rgamma(500,4,rate=7)
  if (runif(1)< dgamma(y,al,rate=bet)*dgamma(mcmc[t-1],4,rate=7)/
        (dgamma(mcmc[t-1],al,rate=bet)*dgamma(y,4,rate=7))){
    mcmc[t]=y
    }
}
aver=cumsum(mcmc)/1:1000

\end{verbatim}
When comparing those samplers, their variability can only be evaluated through repeated calls
to the above code, in order to produce a range of outputs for the three methods. For instance,
one can define a matrix of cumulated averages \verb+aver=matrix(0,250,1000)+ and take the range
of the cumulated averages over the $250$ repetitions as in \verb+ranj=apply(aver,1,range)+, leading
to something similar to Figure \ref{fig:ranj407} in this manual. The complete code for one of the ranges is
\begin{verbatim}
al=4.3
bet=6.2

mcmc=matrix(1,ncol=1000,nrow=500)
for (t in 2:1000){
  mcmc[,t]=mcmc[,t-1]
  y = rgamma(500,4,rate=7)
  valid=(runif(500)<dgamma(y,al,rate=bet)*
    dgamma(mcmc[i,t-1],4,rate=7)/(dgamma(mcmc[,t-1],al,rate=bet)*
    dgamma(y,4,rate=7)))
  mcmc[valid,t]=y[valid]
  }
aver2=apply(mcmc,1,cumsum)
aver2=t(aver2/(1:1000))
ranj2=apply(aver2,2,range)
plot(ranj2[1,],type="l",ylim=range(ranj2),ylab="")
polygon(c(1:1000,1000:1),c(ranj2[2,],rev(ranj2[1,])))
\end{verbatim}
which removes the Monte Carlo loop over the $500$ replications by running the simulations in parallel.
We can notice on Figure \ref{fig:ranj407} in this manual that, while the output from the third sampler is quite similar
with the output from the iid sampler [since we use the same scale on the $y$ axis],
the Metropolis--Hastings algorithm based on the $\mathscr{G}a(4,7)$ proposal is rather biased,
which may indicate a difficulty in converging to the stationary distribution. This is somehow an
expected problem, in the sense that the ratio target-over-proposal is proportional to $x^{0.3}\,\exp(0.8x)$,
which is explosive at both $x=0$ and $x=\infty$.

\begin{exoset}
For a standard normal distribution as target, implement a Hastings-Metropolis algorithm
with a mixture of five random walks with variances $\sigma=0.01,0.1,1,10,100$ and equal weights.
Compare its output with the output of Figure 4.2 (in the book).
\end{exoset}

\begin{figure}
\begin{center}
\includegraphics[width=.8\textwidth]{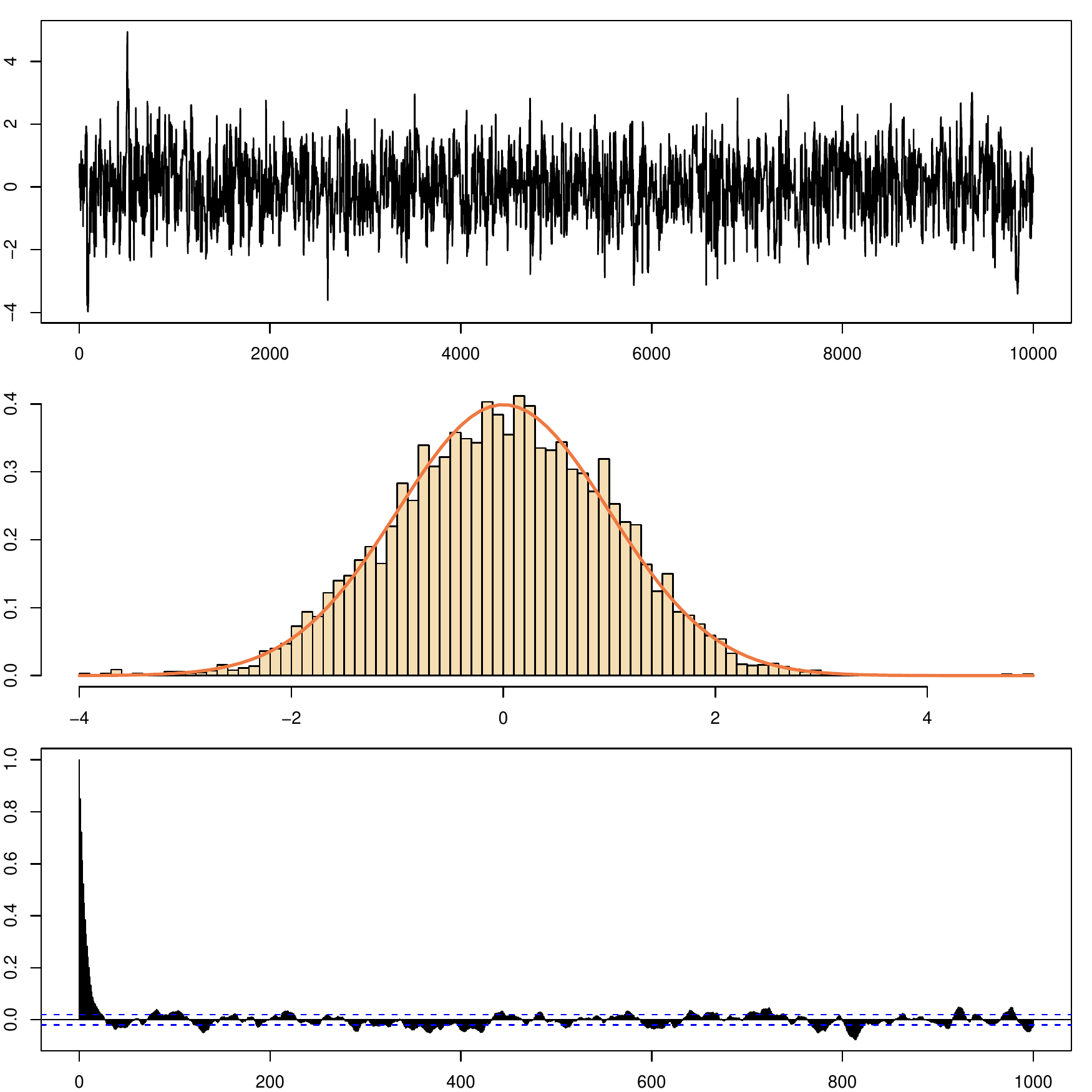}
\caption{\label{fig:mhm}
Outcome of a Metropolis--Hastings simulation of a $\mathscr{N}(0,1)$ target using a mixture of
random walk proposals:
{\em (Top:)} Sequence of $10,000$ iterations; {\em (middle:)}
Histogram of sample compared with the target density; {\em (bottom:)}
Empirical autocorrelations using {\sf R} function {\sf acf}.}
\end{center}
\end{figure}

We thus compare the R code provided in the book
\begin{verbatim}
hm=function(n,x0,sigma2){
  x=rep(x0,n)
  for (i in 2:n){
    y=rnorm(1,x[i-1],sqrt(sigma2))
    if (runif(1)<=exp(-0.5*(y^2-x[i-1]^2))) x[i]=y
    else x[i]=x[i-1]
    }
  x
  }
\end{verbatim}
with a mixture version
\begin{verbatim}
mhm=function(n,x0){
  x=rep(x0,n)
  sigmas=c(0.01,0.1,1,10,100)
  for (i in 2:n){
    y=rnorm(1,x[i-1],sqrt(sample(sigmas,1)))
    if (runif(1)<=exp(-0.5*(y^2-x[i-1]^2))) x[i]=y
    else x[i]=x[i-1]
    }
  x
  }
\end{verbatim}
The outcome from the mixture version in Figure \ref{fig:mhm}
in this manual is quite an improvement when compared with Figure 4.2 from the book.

\begin{exoset}\label{exo:proproper1}
For the probit model under flat prior, find conditions
on the observed pairs $(\bx^i,y_i)$  for the posterior distribution
above to be proper.
\end{exoset}

This distribution is proper (i.e.~well-defined) if the integral
$$
\mathfrak{I} = \int \prod_{i=1}^n \Phi(
\bx^{i\tee}\beta)^{y_i}\left[1-\Phi(
\bx^{i\tee}\beta)\right]^{1-y_i}\,\text{d}\beta
$$
is finite. If we introduce the latent variable behind $\Phi(\bx^{i\tee}\beta)$, we get by
Fubini that
$$
\mathfrak{I} = \int \prod_{i=1}^n \varphi(z_i)
\int_{\left\{ \beta\,; \bx^{i\tee}\beta) \gtrless z_i\,,\ i=1,\ldots,n \right\}}\,
\text{d}\beta\,\text{d}z_1\cdots \text{d}z_n\,,
$$
where $\bx^{i\tee}\beta \gtrless z_i$ means that the inequality is $\bx^{i\tee}\beta < z_i$ if $y_i=1$ and
$\bx^{i\tee}\beta < z_i$ otherwise.
Therefore, the inner integral is finite if and only if the set
$$
\mathfrak{P}=\left\{ \beta\,; \bx^{i\tee}\beta \gtrless z_i\,,\ i=1,\ldots,n \right\}
$$
is compact. The fact that the whole integral $\mathfrak{I}$ is finite follows from the fact that the
volume of the polyhedron defined by $\mathfrak{P}$ grows like $|z_i|^k$ when $z_i$ goes to
infinity. This is however a rather less than explicit constraint on the $(\bx^i,y_i)$'s!

\begin{exoset}\label{exo:proproper2}
For the probit model under non-informative prior, find conditions on $\sum_i y_i$ and $\sum_i (1-y_i)$ for the posterior
distribution defined by (4.4)
to be proper.
\end{exoset}

There is little difference with Exercise \ref{exo:proproper1} because the additional term
$\left(\beta^\tee(X^\tee X)\beta\right)^{-\nicefrac{2k-1}{4}}$ is only creating a problem when $\beta$ goes to
$0$. This difficulty is however superficial since the power in $||X\beta||^{\nicefrac{2k-1}{2}}$ is small enough to
be controlled by the power in $||X\beta||^{k-1}$ in an appropriate polar change of variables. Nonetheless,
this is the main reason why we need a $\pi(\sigma^2)\propto\sigma^{-\nicefrac{3}{2}}$ prior rather than the traditional
$\pi(\sigma^2)\propto\sigma^{-2}$ which is not controlled in $\beta=0$. (This is the limiting case, in the
sense that the posterior is well-defined for $\pi(\sigma^2)\propto\sigma^{-2+\epsilon}$ for all $\epsilon>0$.)

\begin{exoset}
Include an intercept in the probit analysis of
{\bfseries bank} and run the corresponding version of Algorithm 4.7
to discuss whether or not the posterior variance of the intercept is high.
\end{exoset}

\begin{figure}
\begin{center}
\includegraphics[width=\textwidth,height=12cm]{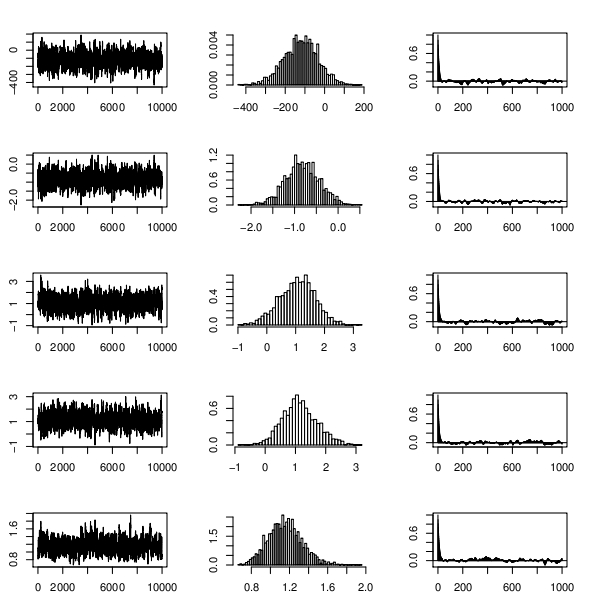}
\caption{\label{fig:bankincpt}
{\sf bank}: estimation of the probit coefficients [including one intercept $\beta_0$]
via Algorithm 4.2 and a flat prior.
{\em Left:} $\beta_i$'s ($i=0,\ldots,4$); {\em center:} histogram
over the last $9,000$ iterations; {\em right:} auto-correlation over the last $9,000$ iterations.
}
\end{center}
\end{figure}

We simply need to add a column of $1$'s to the matrix $X$, as for instance in
\begin{verbatim}
> X=as.matrix(cbind(rep(1,dim(X)[1]),X))
\end{verbatim}
and then use the code provided in the function \verb+hmflatprobit+, i.e.
\begin{verbatim}
flatprobit=hmflatprobit(10000,y,X,1)
par(mfrow=c(5,3),mar=1+c(1.5,1.5,1.5,1.5))
for (i in 1:5){
 plot(flatprobit[,i],type="l",xlab="Iterations",
   ylab=expression(beta[i]))
 hist(flatprobit[1001:10000,i],nclass=50,prob=T,main="",
   xlab=expression(beta[i]))
 acf(flatprobit[1001:10000,i],lag=1000,main="",
   ylab="Autocorrelation",ci=F)
}
\end{verbatim}
which produces the analysis of {\sf bank} with an intercept factor. Figure \ref{fig:bankincpt} in this manual
gives the equivalent to Figure 4.4 [in the book]. The intercept $\beta_0$ has a posterior variance equal to
$7558.3$, but this must be put in perspective in that the covariates of {\sf bank} are taking their
values in the magnitude of $100$ for the three first covariates and of $10$ for the last covariate.
The covariance of $x_{i1}\beta_1$ is therefore of order $7000$ as well. A noticeable difference with
Figure 4.4 [in the book] is that, with the inclusion of the intercept, the range of $\beta_1$'s supported
by the posterior is now negative.

\begin{exoset}\label{ex:progib}
Using the latent variable representation of the probit model,
introduce $z_i|\bbeta\sim\mathscr{N}\left( \bx^{i\tee}\bbeta,1\right)$
$(1\le i\le n)$ such that $y_i=\mathbb{I}_{z_i\le 0}$.
Deduce that
\begin{equation*}
z_i|y_i,\bbeta\sim\left\{\begin{array}{ll}
\mathscr{N}_+\left( \bx^{i\tee}\bbeta,1,0\right) & \text{ if}\quad y_i=1\,, \\
\mathscr{N}_-\left( \bx^{i\tee}\bbeta,1,0\right) & \text{ if}\quad y_i=0\,,
\end{array}\right.
\end{equation*}
where $\mathscr{N}_+\left(\mu,1,0\right)$ and $\mathscr{N}_-\left(\mu,1,0\right)$ are the normal
distributions with mean $\mu$ and variance $1$ that are left-truncated and right-truncated at $0$, respectively.
Check that those distributions can be simulated using the {\sf R} commands
\begin{verbatim}
   > xp=qnorm(runif(1)*pnorm(mu)+pnorm(-mu))+mu
   > xm=qnorm(runif(1)*pnorm(-mu))+mu
\end{verbatim}
Under the flat prior $\pi(\bbeta)\propto 1$, show that
\begin{equation*}
\bbeta|\by,\bz\sim\mathscr{N}_k\left((\bX^\tee \bX)^{-1}\bX^\tee
\bz,(\bX^\tee \bX)^{-1}\right)\,,
\end{equation*}
where $\bz=(z_1,\ldots,z_n)$, and derive the corresponding Gibbs sampler, sometimes called the 
{\em Albert--Chib} sampler. ({\em Hint}: A good starting point is the maximum likelihood
estimate of $\bbeta$.) Compare the application to {\bfseries bank} with the output in Figure
4.4 in this manual.
({\em Note}: Account for differences in computing time.)
\end{exoset}

If $z_i|\beta\sim\mathscr{N}\left( \bx^{i\tee}\beta,1\right)$ is a latent [unobserved] variable,
it can be related to $y_i$ via the function
$$
y_i=\mathbb{I}_{z_i\le 0}\,,
$$
since $P(y_i=1)=P(z_i\ge 0)=1-\Phi\left(-\bx^{i\tee}\beta\right)=\Phi\left(\bx^{i\tee}\beta\right)$.
The conditional distribution of $z_i$
given $y_i$ is then a constrained normal distribution: if $y_i=1$, $z_i\le 0$ and therefore
$$
z_i|y_i=1,\beta\sim\mathscr{N}_+\left( \bx^{i\tee}\beta,1,0\right)\,.
$$
(The symmetric case is obvious.)

The command \verb&qnorm(runif(1)*pnorm(mu)+pnorm(-mu))+mu& is a simple application of the inverse
cdf transform principle given, e.g., in Robert and Casella (2004): the cdf of the $\mathscr{N}_+\left(\mu,1,0\right)$
distribution is
$$
F(x) = \frac{\Phi(x-\mu) - \Phi(-\mu)}{\Phi(\mu)}\,.
$$
(An alternative is to call the {\sf R} library {\sf truncnorm}.)
If we condition on both $\bz$ and $\by$ [the conjunction of which is defined as the ``completed model"], the
$y_i$'s get irrelevant and we are back to a linear regression model, for which the posterior distribution under
a flat prior is given in Section 3.3.1 and is indeed $\mathscr{N}_k\left((X^\tee X)^{-1}X^\tee
\bz,(X^\tee X)^{-1}\right)$.

This closed-form representation justifies the introduction of the latent variable $\bz$ in the simulation process
and leads to the Gibbs sampler that simulates $\beta$ given $\bz$ and $\bz$ given $\beta$ and $\by$ as in
\begin{equation}\label{eq:condiz}
z_i|y_i,\beta\sim\left\{\begin{array}{ll}
\mathscr{N}_+\left( \bx^{i\tee}\beta,1,0\right) & \text{ if}\quad y_i=1 \\
\mathscr{N}_-\left( \bx^{i\tee}\beta,1,0\right) & \text{ if}\quad
y_i=0
\end{array}\right.
\end{equation}
where $\mathscr{N}_+\left(\mu,1,0\right)$ and $\mathscr{N}_-\left(\mu,1,0\right)$ are the normal
distributions with mean $\mu$ and variance $1$ that are left-truncated and right-truncated at $0$, respectively.

\begin{figure}
\begin{center}
\includegraphics[width=\textwidth,height=12cm]{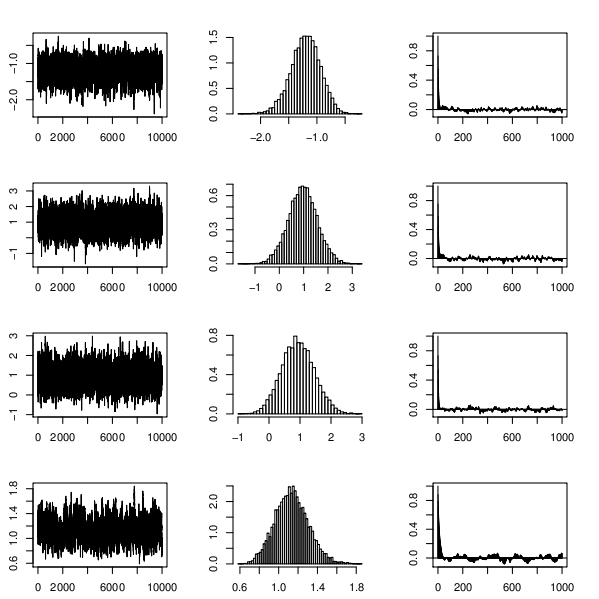}
\caption{\label{fig:gibbsprob}
{\sf bank}: estimation of the probit coefficients [including one intercept $\beta_0$]
by a Gibbs sampler 4.2 under a flat prior.
{\em Left:} $\beta_i$'s ($i=0,\ldots,4$); {\em center:} histogram
over the last $9,000$ iterations; {\em right:} auto-correlation over the last $9,000$ iterations.
}
\end{center}
\end{figure}

A {\sf R} code of this sampler is available as follows (based on a call to the {\sf R} library {\sf truncnorm}):
\begin{verbatim}
gibbsprobit=function(niter,y,X){
  p=dim(X)[2]
  beta=matrix(0,niter,p)
  z=rep(0,length(y))
  mod=summary(glm(y~-1+X,family=binomial(link="probit")))
  beta[1,]=as.vector(mod$coefficient[,1])
  Sigma2=solve(t(X)%*%X)
  for (i in 2:niter){
    mean=X%*%beta[i-1,]
    z[y==1]=rtruncnorm(sum(y==1),a=0,b=Inf,mean[y==1],sd=1)
    z[y==0]==rtruncnorm(sum(y==0),a=-Inf,b=0,mean[y==0],sd=1)
    Mu=Sigma2%*%t(X)%*%z
    beta[i,]=rmvn(1,Mu,Sigma2)
    }
  beta
  }
\end{verbatim}
The output of this function is represented on Figure \ref{fig:gibbsprob} in this manual. Note that the output is somehow smoother than on
Figure \ref{fig:bankincpt} in this manual. (This does not mean that the Gibbs sampler is converging faster but rather than its
component-wise modification of the Markov chain induces slow moves and smooth transitions.)

When comparing the computing times, the increase due to the simulation of the $z_i$'s is not noticeable: for the
{\sf bank} dataset, using the above codes require $27s$ and $26s$ over $10,000$ iterations for
\verb+hmflatprobit+ and \verb+gibbsprobit+. respectively.

\begin{exoset}\label{exo:bankx1x4}
For the {\bfseries bank} dataset and the probit model, compute the Bayes factor associated with the
null hypothesis $H_0:\beta_2=\beta_3=0$.
\end{exoset}

The Bayes factor is given by
\begin{eqnarray*}
B^\pi_{01} &=& \frac{\pi^{-k/2}\Gamma((2k-1)/4)}
{\pi^{-(k-2)/2}\Gamma\{(2k-5)/4\}
}\\
&\times&\frac{ \int \left(\beta^\tee(X^\tee X)\beta\right)^{-(2k-1)/4}
\prod_{i=1}^n\,\Phi(\bx^{i\tee}\beta)^{y_i}\left[1-\Phi(
\bx^{i\tee}\beta)\right]^{1-y_i}\,\text{d}\beta
}{
\int \left\{ (\beta^0)^\tee(X_0^\tee X_0)\beta^0\right\}^{-(2k-5)/4}
\prod_{i=1}^n\,\Phi(x_0^{i\tee}\beta^0)^{y_i} \left[1-\Phi(x_0^{i\tee}\beta^0)\right]^{1-y_i} \text{d}\beta^0
}\,.
\end{eqnarray*}
For its approximation, we can use simulation from a multivariate normal as suggested in the book or
even better from a multivariate $\mathscr{T}$: a direct adaptation from the code in \verb+hmnoinfprobit+ is
\begin{verbatim}
noinfprobit=hmnoinfprobit(10000,y,X,1)

library(mnormt)

mkprob=apply(noinfprobit,2,mean)
vkprob=var(noinfprobit)
simk=rmvnorm(100000,mkprob,2*vkprob)
usk=probitnoinflpost(simk,y,X)-
  dmnorm(simk,mkprob,2*vkprob,log=TRUE)

noinfprobit0=hmnoinfprobit(10000,y,X[,c(1,4)],1)
mk0=apply(noinfprobit0,2,mean)
vk0=var(noinfprobit0)
simk0=rmvnorm(100000,mk0,2*vk0)
usk0=probitnoinflpost(simk0,y,X[,c(1,4)])-
  dmnorm(simk0,mk0,2*vk0,log=TRUE)
bf0probit=mean(exp(usk))/mean(exp(usk0))
\end{verbatim}
(If a multivariate $\mathscr{T}$ is used, the \verb+dmnorm+ function must be replaced with \verb+dt+ the density of the
multivariate $\mathscr{T}$.) The value contained in \verb+bf0probit+ is $67.74$, which is thus an approximation to $B_{10}^\pi$ [since
we divide the approximate marginal under the full model with the approximate marginal under the restricted
model]. Therefore, $H_0$ is quite unlikely to hold, even though, independently, the Bayes factors associated
with the componentwise hypotheses $H_0^2: \beta_2=0$ and $H_0^3: \beta_3=0$ support those hypotheses.

\begin{exoset}
In the case of the logit model--i.e., when $p_i=\exp \tilde
\bx^{i\tee}\bbeta\big/ \{1+\exp \tilde \bx^{i\tee}\bbeta\}$ $(1\le
i\le k)$--derive the prior distribution on $\bbeta$ associated with
the prior 4.6
on $(p_1,\ldots,p_k)$.
\end{exoset}

The only difference with Exercise \ref{ex:progib}
is in the use of a logistic density, hence both the Jacobian and the probabilities
are modified:
\begin{align*}
\pi(\beta)&\propto \prod_{i=1}^k\,\frac{\exp(\{K_ig_i-1\}\tilde \bx^{i\tee}\beta)}
{\left\{ 1+\exp(\tilde \bx^{i\tee}\beta) \right\}^{K_i-2}}
\frac{\exp(\tilde \bx^{i\tee}\beta)}{\left\{ 1+\exp(\tilde \bx^{i\tee}\beta) \right\}^2}\\
&= \frac{\ds \exp\left(\sum_{i=1}^n K_ig_i \tilde \bx^{i\tee}\beta\right)}
{\ds \prod_{i=1}^k\,\left\{ 1+\exp(\tilde \bx^{i\tee}\beta) \right\}^{K_i}}\,.
\end{align*}

\begin{exoset}
Examine whether or not the sufficient conditions for propriety of
the posterior distribution found in Exercise \ref{exo:proproper2} for
the probit model are the same for the logit model.
\end{exoset}

There is little difference with Exercise \ref{exo:proproper1} because
the only change is [again] in the use of a logistic density, which has asymptotics similar to the
normal density. The problem at $\beta=0$ is solved in the same manner.

\begin{exoset}\label{exo:bankx1x4+1}
For the {\bfseries bank} dataset and the logit model, compute the Bayes factor associated with the null hypothesis
$H_0:\beta_2=\beta_3=0$ and compare its value with the value obtained for the probit model in Exercise
\ref{exo:bankx1x4}.
\end{exoset}

This is very similar to Exercise \ref{exo:bankx1x4}, except that the parameters are now estimated for the
logit model. The code is provided in \verb+bayess+ as
\begin{verbatim}
# noninformative prior and random walk HM sample
noinflogit=hmnoinflogit(10000,y,X,1)

# log-marginal under full model
mklog=apply(noinflogit,2,mean)
vklog=var(noinflogit)
simk=rmnorm(100000,mklog,2*vklog)
usk=logitnoinflpost(simk,y,X)-
        dmnorm(simk,mklog,2*vklog,log=TRUE)

# noninformative prior and random walk HM sample
# for restricted model
noinflogit0=hmnoinflogit(10000,y,X[,c(1,4)],1)

# log-marginal under restricted model
mk0=apply(noinflogit0,2,mean)
vk0=var(noinflogit0)
simk0=rmnorm(100000,mk0,2*vk0)
usk0=logitnoinflpost(simk0,y,X[,c(1,4)])-
        dmnorm(simk0,mk0,2*vk0,log=TRUE)

bf0logit=mean(exp(usk))/mean(exp(usk0))
\end{verbatim}
The value of \verb+bf0logit+ is $127.2$, which, as an approximation to $B^\pi_{10}$, argues rather strongly against
the null hypothesis $H_0$. It thus leads to the same conclusion as in the probit model of Exercise
\ref{exo:bankx1x4}, except that the numerical value is almost twice as large. Note that, once again,
the Bayes factors associated with the componentwise hypotheses $H_0^2: \beta_2=0$ and $H_0^3: \beta_3=0$ support those hypotheses.

\begin{exoset}
Given a contingency table with four categorical variables, 
determine the number of submodels to consider.
\end{exoset}

Note that the numbers of classes for the different variables do not matter since,
when building a non-saturated submodel, a variable is in or out. There are
\begin{enumerate}
\item $2^4$ single-factor models [including the zero-factor model];
\item $(2^6-1)$ two-factor models [since there are ${4 \choose 2}=6$ ways of
picking a pair of variables out of $4$ and since the complete single-factor
model is already treated];
\item $(2^4-1)$ three-factor models.
\end{enumerate}
Thus, if we exclude the saturated model, there are $2^6+2^5-2=94$ different submodels.

\begin{exoset}\label{exo:marcon,marcon} 
In the case of a $2\times 2$ contingency table with fixed total
count $n=n_{11}+n_{12}+n_{21}+n_{22}$, we denote by
$\theta_{11},\theta_{12},\theta_{21},\theta_{22}$ the corresponding
probabilities. If the prior on those probabilities is a Dirichlet
$\mathscr{D}_4(\nicefrac{1}{2},\ldots,\nicefrac{1}{2})$, give the corresponding marginal
distributions of $\alpha=\theta_{11}+\theta_{12}$ and 
$\beta=\theta_{11}+\theta_{21}$. Deduce the associated Bayes factor
if $H_0$ is the hypothesis of independence between the factors and
if the priors on the margin probabilities $\alpha$ and $\beta$ are
those derived above.
\end{exoset}

A very handy representation of the Dirichlet $\mathcal{D}_k(\delta_1,\ldots,\delta_k)$
distribution is that
$$
\frac{(\xi_1,\ldots,\xi_k)}{\xi_1+\ldots+\xi_k)} \sim \mathcal{D}_k(\delta_1,\ldots,\delta_k)
$$
when
$$
\xi_i\sim\mathscr{G}a(\delta_i,1)\,,\ i=1,\ldots,k\,.
$$
Therefore, if
$$
(\theta_{11},\theta_{12},\theta_{21},\theta_{22}) =
\frac{(\xi_{11},\xi_{12},\xi_{21},\xi_{22})}{\xi_{11}+\xi_{12}+\xi_{21}+\xi_{22}}\,,
\xi_{ij}\stackrel{\text{iid}}{\sim}\mathscr{G}a(\nicefrac{1}{2},1)\,,
$$
then
$$
(\theta_{11}+\theta_{12},\theta_{21}+\theta_{22}) =
\frac{(\xi_{11}+\xi_{12},\xi_{21}+\xi_{22})}{\xi_{11}+\xi_{12}+\xi_{21}+\xi_{22}}\,,
$$
and
$$
(\xi_{11}+\xi_{12}),(\xi_{21}+\xi_{22})\stackrel{\text{iid}}{\sim}\mathscr{G}a(1,1)
$$
implies that $\alpha$ is a $\mathscr{B}e(1,1)$ random
variable, that is, a uniform $\mathscr{U}(01,)$ variable. The same applies to $\beta$.
(Note that $\alpha$ and $\beta$ are dependent in this representation.)

Since the likelihood under the full model is multinomial,
$$
\ell(\mathbf{\theta}|\mathcal{T}) = {n\choose n_{11}\,n_{12}\,n_{21}}
\theta_{11}^{n_{11}}\,\theta_{12}^{n_{12}}\,\theta_{21}^{n_{21}}\,\theta_{22}^{n_{22}}\,,
$$
where $\mathcal{T}$ denotes the contingency table [or the dataset $\{n_{11},n_{12},n_{21},n_{22}\}$],
the [full model] marginal is
\begin{align*}
m(\mathcal{T}) &= \frac{ \displaystyle{{n\choose n_{11}\,n_{12}\,n_{21}}} }{\pi^2}\,
\int{ \theta_{11}^{n_{11}-\nicefrac{1}{2}}\,\theta_{12}^{n_{12}-\nicefrac{1}{2}}\,\theta_{21}^{n_{21}-\nicefrac{1}{2}}\,
        \theta_{22}^{n_{22}-\nicefrac{1}{2}}\,\text{d}\mathbf{\theta}}\\
&=\frac{ \displaystyle{{n\choose n_{11}\,n_{12}\,n_{21}}} }{\displaystyle \pi^2}\,
\frac{\displaystyle \prod_{i,j}\Gamma(n_{ij}+\nicefrac{1}{2})}{\displaystyle \Gamma(n+2)}\\
&=\frac{ \displaystyle{{n\choose n_{11}\,n_{12}\,n_{21}}} }{\displaystyle \pi^2}\,
\frac{\displaystyle \prod_{i,j}\Gamma(\displaystyle n_{ij}+\nicefrac{1}{2})}{(n+1)!}\\
&=\frac{ 1}{\displaystyle (n+1) \pi^2}\, \displaystyle \prod_{i,j}
\frac{\displaystyle \Gamma(n_{ij}+\nicefrac{1}{2})}{\displaystyle \Gamma(n_{ij}+1)}\,,
\end{align*}
where the $\pi^2$ term comes from $\Gamma(\nicefrac{1}{2})=\sqrt{\pi}$.

In the restricted model, $\theta_{11}$ is replaced with $\alpha\beta$, $\theta_{12}$
by $\alpha(1-\beta)$, and so on. Therefore, the likelihood under the restricted model
is the product
$$
{n\choose n_{1\cdot}}\,\alpha^{n_{1\cdot}}(1-\alpha)^{n-n_{1\cdot}}\,
\times
{n\choose n_{\cdot1}}\,\beta^{n_{\cdot1}}(1-\beta)^{n-n_{\cdot1}}\,,
$$
where $n_{1\cdot}=n_{11}+n_{12}$ and $n_{\cdot1}=n_{11}+n_{21}$,
and the restricted marginal under uniform priors on both $\alpha$ and $\beta$ is
\begin{align*}
m_0(\mathcal{T}) &= {n\choose n_{1\cdot}}\,{n\choose n_{\cdot1}}\,
\int_0^1\,\alpha^{n_{1\cdot}}(1-\alpha)^{n-n_{1\cdot}}\,\text{d}\alpha\,
\int_0^1\,\beta^{n_{\cdot1}}(1-\beta)^{n-n_{\cdot1}}\,\text{d}\beta\\
&={n\choose n_{1\cdot}}\,{n\choose n_{\cdot1}}\,
\frac{\ds (n_{1\cdot}+1)!(n-n_{1\cdot}+1)!}{\ds (n+2)!}\,
\frac{\ds (n_{\cdot1}+1)!(n-n_{\cdot1}+1)!}{\ds (n+2)!}\\
&=\frac{\ds (n_{1\cdot}+1)(n-n_{1\cdot}+1)}{\ds (n+2)(n+1)}\,
\frac{\ds (n_{\cdot1}+1)(n-n_{\cdot1}+1)}{\ds (n+2)(n+1)}\,.
\end{align*}
The Bayes factor $B^\pi_{01}$ is then the ratio  $m_0(\mathcal{T})/m(\mathcal{T})$.

\chapter{Capture--Recapture Experiments}\label{ch:cap}
\begin{exoset}\label{exo:propcap}
Show that the posterior distribution $\pi(N|n^+)$ given by (5.1),
while associated with an improper prior,
is defined for all values of $n^+$. 
Show that the normalization factor of (5.1)
is $n^+\vee 1$, and deduce that the posterior median is equal to $2(n^+\vee 1)-1$.
Discuss the relevance of this estimator and show that it corresponds to a Bayes estimate of $p$ equal to
$\nicefrac{1}{2}$.
\end{exoset}

Since the main term of the series is equivalent to $N^{-2}$, the series converges. The posterior distribution
can thus be normalised. Moreover,
\begin{align*}
\sum_{i=n_0}^{\infty}\,\frac{1}{i(i+1)}
&= \sum_{i=n_0}^\infty \left( \frac{1}{i}-\frac{1}{i+1}\right) \\
&= \frac{1}{n_0} - \frac{1}{n_0+1} + \frac{1}{n_0+1} - \frac{1}{n_0+2}+\ldots\\
&= \frac{1}{n_0}\,.
\end{align*}
Therefore, the normalisation factor is available in closed form and is equal to $n^+\vee 1$.
The posterior median is the value $N^\star$ such that $\pi(N\ge N^\star|n^+)=\nicefrac{1}{2}$, i.e.~
$$
\sum_{i=N^\star}^{\infty}\,\nicefrac{1}{i(i+1)} = \nicefrac{1}{2}\, \nicefrac{1}{n^+\vee 1}= \nicefrac{1}{N^\star}\,,
$$
which implies that $N^\star=2(n^+\vee 1)$. This estimator is rather intuitive in that $\mathbb{E}[n^+|
N,p]=pN$: since the expectation of $p$ is $\nicefrac{1}{2}$, $\mathbb{E}[n^+|N]=\nicefrac{N}{2}$ and $N^\star=2n^+$ is a
moment estimator of $N$.

\begin{exoset}\label{exo:propmempacap}
Under the same prior as in Section 5.2.1,
derive the marginal posterior density of $N$
in the case where $n_1^+\sim\mathscr{B}(N,p)$ and 
$$
n_2^+,\ldots,n_k^+\stackrel{\text{iid}}{\sim} \mathscr{B}(n_1^+,p)
$$ 
are observed (the later are in fact recaptures). Apply to the sample
$$
(n^+_{1},n_2^+,\ldots,n^+_{11}) =
(32,20,\allowbreak 8,\allowbreak 5,\allowbreak 1,\allowbreak 2,
\allowbreak 0,\allowbreak 2,\allowbreak 1,\allowbreak 1,\allowbreak 0)\,,
$$
which describes a series of tag recoveries over $11$ years.
\end{exoset}

In that case, if we denote $n_{\cdot}^+=n_1^++\cdots+n_k^+$ the total number of captures,
the marginal posterior density of $N$ is
\begin{align*}
\pi(N|n_1^+,\ldots,n_k^+) &\propto  \frac{N!}{(N-n_1^+)!}\,N^{-1}\mathbb{I}_{N\ge n_1^+}\\
&\qquad \int_0^1 p^{n_1^++\cdots+n_k^+}(1-p)^{N-n_1^++(n_1+-n_2^++\cdots+n_1^+-n_k^+}\text{d}p \\
&\propto  \frac{(N-1)!}{(N-n_1^+)!}\,\mathbb{I}_{N\ge n_1^+}
\int_0^1 p^{n_\cdot^+}(1-p)^{N+kn_1^+-n_\cdot^+} \text{d}p \\
& \propto \frac{(N-1)!}{(N-n_1^+)!}\,\frac{(N+kn_1^+-n_\cdot^+)!}{(N+kn_1^++1)!}\,\mathbb{I}_{N\geq n_1^+\vee 1}\,,
\end{align*}
which does not simplify any further. Note that the binomial coefficients
$$
{n_1^+\choose n_j^+} \qquad(j\ge 2)
$$
are irrelevant for the posterior of $N$ since they only depend on the data.

The {\sf R} code corresponding to this model is as follows:
\begin{verbatim}
n1=32
ndo=sum(32,20,8,5,1,2,0,2,1,1,0)

# unnormalised posterior
post=function(N){
   exp(lfactorial(N-1)+lfactorial(N+11*n1-ndo)-
    lfactorial(N-n1)-lfactorial(N+11*n1+1))
   }

# normalising constant and
# posterior mean

posv=post((n1:10000))

cons=sum(posv)
pmean=sum((n1:10000)*posv)/cons
pmedi=sum(cumsum(posv)<.5*cons)
\end{verbatim}
The posterior mean is therefore equal to $282.4$, while the posterior median is $243$. Note that a crude analysis
estimating
$p$ by $\hat p=(n_2^++\ldots+n_{11})/(10 n_1^+)=0.125$ and $N$ by $n_1^+/\hat p$ would produce the value $\hat N=256$.

\begin{exoset}\label{exo:doesnot}
Show that the conditional distribution of $m_2$ conditional on both sample sizes $n_1$ and $n_2$ is given by (5.2)
and does not depend on $p$. Deduce the expectation $\mathbb{E}^\pi[m_2|n_1,n_2,N]$.
\end{exoset}

Since
$$
n_1     \sim\mathscr{B}(N,p)\,,\quad
m_2|n_1 \sim\mathscr{B}(n_1,p)
$$and$$
n_2-m_2|n_1,m_2 \sim\mathscr{B}(N-n_1,p)\,,
$$
the conditional distribution of $m_2$ is given by
\begin{align*}
f(m_2|n_1,n_2) &\propto {n_1\choose m_2} p^{m_2}(1-p)^{n_1-m_2} {N-n_1\choose n_2-m_2} p^{n_2-m_2}
(1-p)^{N-n_1-n_2+m_2}\\
&\propto {n_1\choose m_2} {N-n_1\choose n_2-m_2}\,p^{m_2+n_2-m_2}(1-p)^{n_1-m_2+N-n_1-n_2+m_2}\\
&\propto {n_1\choose m_2} {N-n_1\choose n_2-m_2}\\
&\propto {{n_1\choose m_2} {N-n_1\choose n_2-m_2}}\Big/{{N\choose n_2}}\,,
\end{align*}
which is the hypergeometric $\mathscr{H}(N,n_2,n_1/N)$ distribution. Obviously, this distribution does not depend on $p$
and its expectation is
$$
\mathbb{E}[m_2|n_1,n_2]=\frac{n_1 n_2}{N}\,.
$$

\begin{exoset}\label{exo:tramcar}
In order to determine the number $N$ of buses in a town, a capture--recapture strategy goes
as follows. We observe  $n_1=20$ buses during the first day and keep track of their identifying numbers. Then we repeat
the experiment the following day by recording the number of buses that have already
been spotted on the previous day, say $m_2=5$, out of the $n_2=30$ buses observed the
second day. For the Darroch model, give the posterior expectation of $N$ under the prior
$\pi(N)=1/N$. 
\end{exoset}

Using the derivations of the book, we have that
\begin{align*}
\pi(N|n_1,n_2,m_2) &\propto \frac{1}{N}\,{N \choose n^+}\,B(n^c+1,2N-n^c+1)\mathbb{I}_{N\geq n^+} \\
    &\propto \frac{(N-1)!}{(N-n^+)!}\,\frac{(2N-n^c)!}{(2N+1)!}\,\mathbb{I}_{N\geq n^+}
\end{align*}
with $n^+=45$ and $n^c=50$. For $n^+=45$ and $n^c=50$, the posterior mean 
is equal to $130.91$.

\begin{exoset}\label{exo:darroch&roll}
Show that the maximum
likelihood estimator of $N$ for the Darroch model is $\hat  N = n_1 / \left(m_2 /n_2 \right)$,
and deduce that it is not defined when $m_2=0$.\end{exoset}

The likelihood for the Darroch model is proportional to
$$
\ell(N) = \frac{(N-n_1)!}{(N-n_2)!}\,\frac{(N-n^+)!}{N!}\,\mathbb{I}_{N\geq n^+}\,.
$$
Since
$$
\frac{\ell(N+1)}{\ell(N)} = \frac{(N+1-n_1)(N+1-n_2)}{(N+1-n^+)(N+1)} \ge 1
$$
for
\begin{eqnarray*}
(N+1)^2-(N+1)(n_1+n_2) + n_1n_2 &\ge& (N+1)^2 -(N+1)n^+ \\
(N+1)(n_1+n_2-n^+) &\ge& n_1n_2\\
(N+1) &\le& \frac{n_1n_2}{m_2}\,,
\end{eqnarray*}
the likelihood is increasing for $N\le n_1n_2/m2$ and decreasing for $N\ge n_1n_2/m2$.
Thus $\hat N=n_1n_2/m2$ is the maximum likelihood estimator [assuming this quantity is
an integer]. If $m_2=0$, the likelihood is increasing with $N$ and therefore there
is no maximum likelihood estimator.

\begin{exoset}
Give the likelihood of the extension of Darroch's model
when the capture--recapture experiments are repeated $K$ times
with capture sizes and recapture observations $n_k$ $(1\le k\le K)$ and $m_k$ $(2\le k\le K)$,
respectively. ({\em Hint}: Exhibit first the two-dimensional sufficient statistic associated with
this model.)
\end{exoset}

The likelihood for the Darroch model is proportional to
$$
\ell(N) = \frac{(N-n_1)!}{(N-n_2)!}\,\frac{(N-n^+)!}{N!}\,\mathbb{I}_{N\geq n^+}\,.
$$
Since
$$
\frac{\ell(N+1)}{\ell(N)} = \frac{(N+1-n_1)(N+1-n_2)}{(N+1-n^+)(N+1)} \ge 1
$$
for
\begin{eqnarray*}
(N+1)^2-(N+1)(n_1+n_2) + n_1n_2 &\ge& (N+1)^2 -(N+1)n^+ \\
(N+1)(n_1+n_2-n^+) &\ge& n_1n_2\\
(N+1) &\le& \frac{n_1n_2}{m_2}\,,
\end{eqnarray*}
the likelihood is increasing for $N\le n_1n_2/m2$ and decreasing for $N\ge n_1n_2/m2$.
Thus $\hat N=n_1n_2/m2$ is the maximum likelihood estimator [assuming this quantity is
an integer]. If $m_2=0$, the likelihood is increasing with $N$ and therefore there
is no maximum likelihood estimator.

\begin{exoset}\label{exo:2CRgS}
Give both conditional posterior distributions involved in Algorithm 5.8 in the case $n^+=0$.
\end{exoset}

When $n^+=0$, there is no capture at all during both capture episodes. The likelihood is thus $(1-p)^{2N}$
and, under the prior $\pi(N,p)=1/N$, the conditional posterior distributions of $p$ and $N$ are
\begin{align*}
p|N,n^+=0 &\sim \mathscr{B}e(1,2N+1)\,,\\
N|p,n^+=0 &\sim \frac{(1-p)^{2N}}{N}\,.
\end{align*}
That the joint distribution $\pi(N,p|n^+=0)$ exists is ensured by the fact that $\pi(N|n^+=0)\propto 1/N(2N+1)$,
associated with a converging series.

\begin{exoset}\label{exo:2CRgS2}
Show that, for the two-stage capture model with probability $p$ of capture,
when the prior on $N$ is a $\mathscr{P}(\lambda)$
distribution, the conditional posterior on $N-n^+$ is
$\mathscr{P}(\lambda(1-p)^2)$.
\end{exoset}

The posterior distribution of $(N,p)$ associated with the informative prior $\pi(N,p)=\lambda^N e^{-\lambda}/N!$
is proportional to
$$
\frac{N!}{(N-n^+)!N!}\,\lambda^N\,p^{n^c}(1-p)^{2N-n^c}\,\mathbb{I}_{N\geq n^+}\,.
$$
The corresponding conditional on $N$ is thus proportional to
$$
\frac{\lambda^N}{(N-n^+)!} \,p^{n^c}(1-p)^{2N-n^c}\,\mathbb{I}_{N\geq n^+}
\propto \frac{\lambda^{N-n^+}}{(N-n^+)!} \,p^{n^c}(1-p)^{2N-n^c}\,\mathbb{I}_{N\geq n^+}
$$
which corresponds to a Poisson $\mathscr{P}(\lambda(1-p)^2)$ distribution
on $N-n_+$.

\begin{exoset}\label{ex:anozagib}
Reproduce\index{Algorithm!Gibbs sampler} the analysis of {\bfseries eurodip} summarized by Figure 5.1
when switching the prior from $\pi(N,p)\propto \lambda^N/N!$ to $\pi(N,p)\propto N^{-1}$.
\end{exoset}

\begin{figure}
\begin{center}
\includegraphics[width=\textwidth,height=6cm]{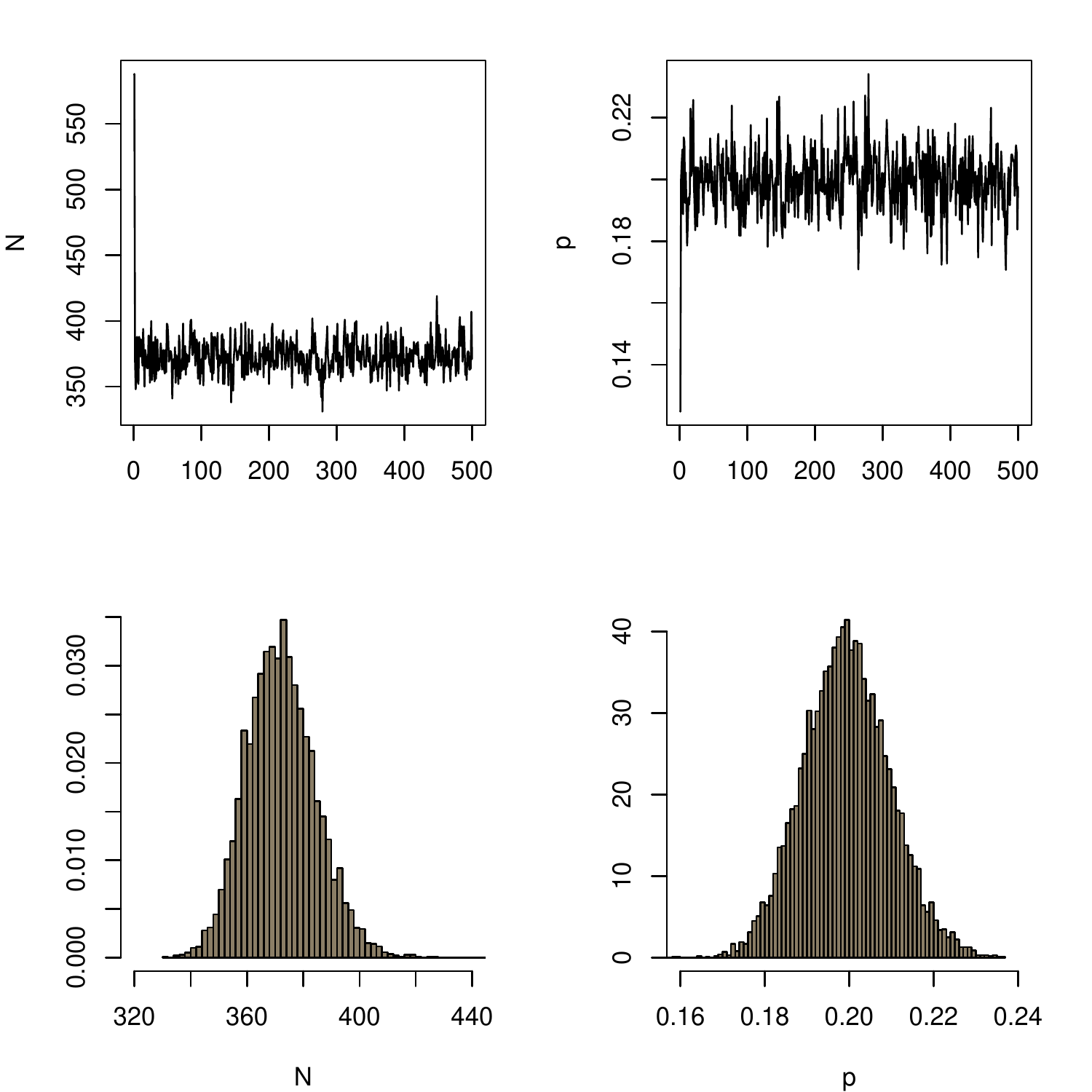}
\caption{\label{fig:euronew}{\sf eurodip}: MCMC simulation under the
prior $\pi(N,p)\propto N^{-1}$.}
\end{center}
\end{figure}

The main purpose of this exercise is to modify the code provided in the book (p.151) and in the demo for Chapter 5,
since the marginal posterior distribution of $N$ is given in the book as
$$
\pi(N|n^+,n^c) \propto {(N-1)! \over (N-n^+)!}\, { (TN-n^c)! \over (TN+1)!}\,
\mathbb{I}_{N\ge n^+\vee 1}\,.
$$
(The conditional posterior distribution of $p$ does not change.) This distribution being
non-standard, it makes direct simulation awkward and we prefer to use a Metropolis-Hastings
step, using a modified version of the previous Poisson conditional as proposal $q(N^\prime|N,p)$.
We thus simulate
$$
N^\star-n^+\sim \mathscr{P}\left( N^{(t-1)}(1-p^{(t-1)})^T \right)
$$
and accept this value with probability
$$
\frac{\pi(N^\star|n^+,n^c)}{\pi(N^{(t-1)}|n^+,n^c)}\,
\frac{q(N^{(t-1)}|N^\star,p^{(t-1)})}{q(N^\star|N^{(t-1)},p^{(t-1)})} \wedge 1\,.
$$
The corresponding modified {\sf R} function is
\begin{verbatim}
gibbs11=function(nsimu,T,nplus,nc)
{
# conditional posterior
rati=function(N){
  lfactorial(N-1)+lfactorial(T*N-nc)-
    lfactorial(N-nplus)-lfactorial(T*N+1)
  }

N=rep(0,nsimu)
p=rep(0,nsimu)

N[1]=2*nplus
p[1]=rbeta(1,nc+1,T*N[1]-nc+1)
for (i in 2:nsimu){

  # MH step on N
  N[i]=N[i-1]
  prop=nplus+rpois(1,N[i-1]*(1-p[i-1])^T)
  if (log(runif(1))<rati(prop)-rati(N[i])+
        dpois(N[i-1]-nplus,prop*(1-p[i-1])^T,log=T)-
        dpois(prop-nplus,N[i-1]*(1-p[i-1])^T,log=T))
     N[i]=prop
  p[i]=rbeta(1,nc+1,T*N[i]-nc+1)
  }
list(N=N,p=p)
}
\end{verbatim}
The output of this program is given in Figure \ref{fig:euronew}.

\begin{exoset}\label{exo:futurapture}
An extension of the $T$-stage capture--recapture model of Section 5.2.3
is to consider that the
capture of an individual modifies its probability of being captured from $p$ to $q$ for
future recaptures. Give the likelihood $\ell(N,p,q|n_1,n_2,m_2\ldots,n_T,m_T)$.
\end{exoset}

When extending the $T$-stage capture-recapture model with different probabilities of being captured
and recaptured, after the first capture episode, where $n_1\sim\mathscr{B}(N,p)$, we observe $T-1$
new captures $(i=2,\ldots,T)$
$$
n_i-m_i|n_1,n_2,m_2,\ldots,n_{i-1},m_{i-1}\sim\mathscr{B}(N-n_1-n_2+m_2+\ldots+m_{i-1},p)\,,
$$
and $T-1$ recaptures $(i=2,\ldots,T)$,
$$
m_i|n_1,n_2,m_2,\ldots,n_{i-1},m_{i-1} \sim
\mathscr{B}(n_1+n_2-m_2+\ldots-m_{i-1},q)\,.
$$
The likelihood is therefore
\begin{align*}
{N\choose n_1}\,p^{n_1}(1-p)^{N-n_1} &\prod_{i=2}^T {N-n_1+\ldots-m_{i-1}\choose n_i-m_i}
p^{n_i-m_i} (1-p)^{N-n_1+\ldots+m_i}\\
&\quad\times\prod_{i=2}^T {n_1+n_2-\ldots-m_{i-1}\choose m_i} q^{m_i} (1-q)^{n_1+\ldots-m_{i}} \\
&\propto \frac{N!}{(N-n^+)!}\,p^{n^+}(1-p)^{TN-n^*}\,q^{m^+}(1-q)^{n^*-n_1},
\end{align*}
where $n^+=n_1-m_2+\cdots-m_{T}$ is the number of captured individuals,
$$
n^* = Tn_1 + \sum_{j=2}^T (T-j+1)(n_j-m_j)
$$
and where $m^+=m_1+\cdots+m_T$ is the number of recaptures. The four statistics $(n_1,n^+,n^*,m^+)$
are thus sufficient for this version of the $T$-stage capture-recapture model.

\begin{exoset}\label{exo:marklost} 
Another extension of the $2$-stage capture--recapture model is to allow
for mark loss.~If we introduce $q$ as the probability of losing the mark, $r$ 
as the probability of recovering a lost mark and $k$ as the number of
recovered lost marks, give the associated likelihood $\ell(N,p,q,r|n_1,n_2,m_2,k)$. 
\end{exoset}

There is an extra-difficulty in this extension in that it contains a latent variable: let us denote by $z$
the number of tagged individuals that have lost their mark. Then $z\sim\mathscr{B}(n_1,q)$ is not observed,
while $k\sim\mathscr{B}(z,r)$ is observed. Were we to observe $(n_1,n_2,m_2,k,z)$, the [completed] likelihood would
be
\begin{align*}
\ell^\star(N,p,q,r&|n_1,n_2,m_2,k,z)={N\choose n_1}\,p^{n_1}(1-p)^{N-n_1}\,{n_1\choose z}\,q^z(1-q)^{n_1-z}\\
&\quad \times {z\choose k}\,r^k(1-r)^{z-k}\,{n_1-z\choose m_2}\,p^{m_2}(1-p)^{n_1-z-m_2}\\
&\quad \times {N-n_1+z\choose n_2-m_2}\,p^{n_2-m_2}(1-p)^{N-n_1+z-n_2+m_2}\,,
\end{align*}
since, for the second round, the population gets partitioned into individuals that keep their tag and are/are
not recaptured, those that loose their tag and are/are not recaptured, and those that are captured for the first time.
Obviously, it is not possible to distinguish between the last two categories. Since $z$ is not known, the [observed]
likelihood is obtained by summation over $z$:
\begin{align*}
\ell(N,p,q,r&|n_1,n_2,m_2,k) \propto \frac{N!}{(N-n_1)!} \,p^{n_1+n_2}(1-p)^{2N-n_1-n_2}\\
&\sum_{z=k\vee N-n_1-n_2+m_2}^{n_1-m_2} {n_1\choose z} \,{n_1-z\choose m_2}\\
&\times {N-n_1+z\choose n_2-m_2}\, q^z(1-q)^{n_1-z}\,r^k(1-r)^{z-k}\,.
\end{align*}
Note that, while a proportionality sign is acceptable for the computation of the likelihood, the terms depending on
$z$ must be kept within the sum to obtain the correct expression for the distribution of the observations. A simplified
version is thus
\begin{align*}
\ell(N,p,q,r&|n_1,n_2,m_2,k) \propto \frac{N!}{(N-n_1)!} \,p^{n_1+n_2}(1-p)^{2N-n_1-n_2}\,
q^{n_1}(r/(1-r))^k\\
&\sum_{z=k\vee N-n_1-n_2+m_2}^{n_1-m_2} \frac{(N-n_1+z)![q(1-r)/(1-q)]^z
}{z!(n_1-z-m_2)!(N-n_1-n_2+m_2+z)!}\,,
\end{align*}
but there is no close-form solution for the summation over $z$.

\begin{exoset}\label{exo:Rquest}
Show that the conditional distribution of $r_1$ in the open population model of Section 5.3
is proportional to the product (5.4).
\end{exoset}

The joint distribution of $\mathcal{D}^*=(n_1,c_2,c_3,r_1,r_2)$ is
given in the book as
\begin{align*}
{N\choose n_1}&p^{n_1}(1-p)^{N-n_1}\,{n_1\choose r_1}\,q^{r_1}(1-q)^{n_1-r_1}{n_1-r_1\choose
c_2}\,p^{c_2}(1-p)^{n_1-r_1-c_2}\\
&\times{n_1-r_1\choose r_2} q^{r_2}(1-q)^{n_1-r_1-r_2}{n_1-r_1-r_2\choose c_3}\,p^{c_3}(1-p)^{n_1-r_1-r_2-c_3}\,.
\end{align*}
Therefore, if we only keep the terms depending on $r_1$, we indeed recover
\begin{align*}
&\frac{1}{r_1!(n_1-r_1)!}\,q^{r_1}(1-q)^{n_1-r_1}\,\frac{(n_1-r_1)!}{(n_1-r_1-c_2)!}\,(1-p)^{n_1-r_1-c_2}\\
&\quad\times\frac{(n_1-r_1)!}{(n_1-r_1-r_2)!}\,(1-q)^{n_1-r_1-r_2}\,\frac{(n_1-r_1-r_2)!}{(n_1-r_1-r_2-c_3)!}\,
(1-p)^{n_1-r_1-r_2-c_3}\\
&\propto\frac{(n_1-r_1)!}{r_1!(n_1-r_1-c_2)!(n_1-r_1-r_2-c_3)!}\,\left\{ \frac{q}{(1-q)^2(1-p)^2}\right\}^{r_1}\\
&\propto{n_1-c_2\choose r_1}\,{n_1-r_1\choose r_2+c_3}\,\left\{ \frac{q}{(1-q)^2(1-p)^2}\right\}^{r_1}\,,
\end{align*}
under the constraint that $r_1\le \min(n_1,n_1-r_2,n_1-r_2-c_3,n_1-c_2)=\min(n_1-r_2-c_3,n_1-c_2)$.

\begin{exoset}\label{exo:margot}
Show that the distribution of $r_2$ in the open population model of Section 
5.3 can be integrated out from the joint distribution and that this leads to
the following distribution on $r_1$:
\begin{align*}
\pi(r_1|p,q,n_1,c_2,c_3) \propto\, &\frac{(n_1-r_1)!(n_1-r_1-c_3)!}{r_1!(n_1-r_1-c_2)!}\\
&\times \left(\frac{q}{(1-p)(1-q)[q+(1-p)(1-q)]}\right)^{r_1}\,.
\end{align*}
Compare the computational cost of a Gibbs sampler based on this approach with 
a Gibbs sampler using the full conditionals.
\end{exoset}

Following the decomposition of the likelihood in the previous exercise, the terms depending on $r_2$ are
\begin{align*}
\frac{1}{r_2!(n_1-r_1-r_2)!}&\left(\frac{q}{(1-p)(1-q)}\right\}^{r_2} \frac{(n_1-r_1-r_2)!}{(n_1-r_1-r_2-c_3)!}\\
&=\frac{1}{r_2!(n_1-r_1-r_2-c_3)!}\left(\frac{q}{(1-p)(1-q)}\right\}^{r_2}\,.
\end{align*}
If we sum over $0\le r_2\le n_1-r_1-c_3$, we get
\begin{align*}
\frac{1}{(n_1-r_1-c_3)!}\,&\sum_{k=0}^{n_1-r_1-c_3} {n_1-r_1-c_3\choose k} \left(\frac{q}{(1-p)(1-q)}\right\}^k\\
&= \left\{1+\frac{q}{(1-p)(1-q)}\right\}^{n_1-r_1-c_3}
\end{align*}
that we can agregate with the remaining terms in $r_1$
$$
\frac{(n-r_1)!}{r_1!(n_1-r_1-c_2)!}\left\{ \frac{q}{(1-q)^2(1-p)^2}\right\}^{r_1}
$$
to recover 
\begin{align*}
\pi(r_1|p,q,n_1,c_2,c_3) \propto\, &\frac{(n_1-r_1)!(n_1-r_1-c_3)!}{r_1!(n_1-r_1-c_2)!}\\
&\times \left(\frac{q}{(1-p)(1-q)[q+(1-p)(1-q)]}\right)^{r_1}\,.
\end{align*}

\begin{exoset}\label{exo:misope}
Show that the likelihood associated with an open population as in Section 5.3 can be written as
\begin{align*}
\ell(N,p|\mathscr{D}^*) &= \sum_{(\epsilon_{it},\delta_{it})_{it}}\prod_{t=1}^T
\prod_{i=1}^N q_{\epsilon_{i(t-1)}}^{\epsilon_{it}}(1-q_{\epsilon_{i(t-1)}})^{1-\epsilon_{it}} \\
&\qquad\qquad \times p^{(1-\epsilon_{it})\delta_{it}} (1-p)^{(1-\epsilon_{it})(1-\delta_{it})} \,,
\end{align*}
where $q_0=q$, $q_1=1$, and $\delta_{it}$ and $\epsilon_{it}$
are the capture and exit indicators, respectively. Derive the order of complexity
of this likelihood; that is, the number of elementary operations necessary to
compute it.\end{exoset}

This is an alternative representation of the model where each individual capture and life history
is considered explicitely. This is also the approach adopted for the Arnason-Schwarz model of Section
5.5. We can thus define the history of individual $1\le i\le N$ as a pair of sequences $(\epsilon_{it})$
and $(\delta_{it})$, where $\epsilon_{it}=1$ at the exit time $t$ and forever after. For the model given
at the beginning of Section 5.3, there are $n_1$ $\delta_{i1}$'s equal to $1$, $r_1$ $\epsilon_{i1}$'s equal
to $1$, $c_2$ $\delta_{i2}$'s equal to $1$ among the $i$'s for which $\delta_{i1}=1$ and so on. If we do not
account for these constraints, the likelihood is of order $\text{O}(3^{NT})$ [there are three possible
cases for the pair $(\epsilon_{it},\delta_{it})$ since $\delta_{it}=0$ if $\epsilon_{it}=1$].
Accounting for the constraints on the total number of $\delta_{it}$'s equal to $1$ increases
the complexity of the computation.

\begin{exoset}\label{ex:2kool}
In connection with the presentation of the accept-reject algorithm in Section 5.4,
show that, for $M>0$, if $g$ is replaced with $Mg$ in $\mathscr{S}$
and if $(X,U)$ is uniformly distributed on $\mathscr{S}$, the marginal distribution
of $X$ is still $g$. Deduce that the density $g$ only needs to be known up to a
normalizing constant.
\end{exoset}

The set
$$
\mathscr{S}=\{(x,u):0<u<Mg(x)\}
$$
has a surface equal to $M$. Therefore, the uniform distribution on $\mathscr{S}$ has density $1/M$
and the marginal of $X$ is given by
$$
\int \mathbb{I}_{(0,Mg(x))}\,\frac{1}{M}\,\text{d}u = \frac{Mg(x)}{M}=g(x)\,.
$$
This implies that uniform simulation in $\mathscr{S}$  provides an output from $g$ no matter what the
constant $M$ is. In other words, $g$ does not need to be normalised.

\begin{exoset}\label{ex:2Hard}
For the function $g(x)=(1+\sin^2(x))
(2+\cos^4(4x))\exp[-x^4\{1+\sin^6(x)\}]$ on $[0,2\pi]$, examine the feasibility of running a uniform
sampler on the set $\mathscr{S}$ associated with the accept-reject algorithm in Section 5.4.
\end{exoset}

The function $g$ is non-standard but it is bounded [from above] by the function
$\overline{g}(x) = 6\exp[-x^4]$ since both $\cos$ and $\sin$ are bounded by $1$ or
even $\overline{g}(x) = 6$. Simulating uniformly over the set $\mathscr{S}$ associated
with $g$ can thus be achieved by simulating uniformly
over the set $\mathscr{S}$ associated with $\overline{g}$ until the output falls within
the set $\mathscr{S}$ associated with $g$.  This is the basis of accept-reject algorithms.

\begin{exoset}\label{ex:AarofMay}
Show that the probability of acceptance in Step 2 of Algorithm 5.9 is $1/M$
and that the number of trials until a variable is accepted has a geometric distribution
with parameter $1/M$. Conclude that the expected number of trials per simulation is $M$.
\end{exoset}

The probability that $U\le g(X)/(Mf(X))$ is the probability that a uniform draw in the set
$$
\mathscr{S}=\{(x,u):0<u<Mg(x)\}
$$
falls into the subset
$$
\mathscr{S}_0=\{(x,u):0<u<f(x)\}.
$$
The surfaces of $\mathscr{S}$ and $\mathscr{S}_0$ being $M$ and $1$, respectively, the probability
to fall into $\mathscr{S}_0$ is $1/M$.

Since steps {\sf 1.} and {\sf 2.} of Algorithm 5.2 are repeated independently, each round has a
probability $1/M$ of success and the rounds are repeated till the first success. The number of
rounds is therefore a geometric random variable with parameter $1/M$ and expectation $M$.

\begin{exoset}
For the conditional distribution of $\alpha_t$ derived from (5.3),
construct an accept--reject algorithm based on a normal bounding density $f$ and study its performances for $N=532$,
$n_t=118$, $\mu_t=-0.5$, and $\sigma^2=3$.
\end{exoset}

That the target is only known up to a constant is not a problem, as demonstrated in Exercise
\ref{exo:deduceM}. To find a bound on $\pi(\alpha_t|N,n_t)$ [up to a constant], we just have to notice that
$$
(1 + e^{\alpha_{t}})^{-N} < e^{-N\alpha_{t}}
$$
and therefore
\begin{align*}
(1 + e^{\alpha_{t}})^{-N} &
\;\exp \left\{\alpha_t n_t - \frac{1}{2\sigma^2} (\alpha_{t} - \mu_{t})^{2} \right\} \\
&\le \exp \left\{\alpha_t (n_t-N) - \frac{1}{2\sigma^2} (\alpha_{t} - \mu_{t})^{2} \right\} \\
&= \exp \left\{ - \frac{\alpha_t^2}{2\sigma^2} + 2\frac{\alpha_t}{2\sigma^2}(\mu_t
-\sigma^2(N-n_t)) - \frac{\mu_t^2}{2\sigma^2} \right\} \\
&= \frac{1}{\sqrt{2\pi}\sigma} \exp \left\{ - \frac{1}{2\sigma^2} (\alpha_{t} - \mu_t + \sigma^2(N-n_t))^2 \right\}\\
&\quad\times
\sqrt{2\pi}\sigma \exp \left\{ - \frac{1}{2\sigma^2} (\mu_t^2 - [\mu_t - \sigma^2(N-n_t)]^2 )\right\}\,.
\end{align*}
The upper bound thus involves a normal $\mathscr{N}(\mu_t - \sigma^2(N-n_t),\sigma^2)$ distribution and
the corresponding constant. The {\sf R} code associated with this decomposition is
\begin{verbatim}
# constants
N=53
nt=38
mut=-.5
sig2=3
sig=sqrt(sig2)

# log target
ta=function(x){
  -N*log(1+exp(x))+x*nt-(x-mut)^2/(2*sig2)
  }

#bounding constant
bmean=mut-sig2*(N-nt)
uc=0.5*log(2*pi*sig2)+(bmean^2-mut^2)/(2*sig2)

prop=rnorm(1,sd=sig)+bmean
ratio=ta(prop)-uc-dnorm(prop,mean=bmean,sd=sig,log=T)

while (log(runif(1))>ratio){

  prop=rnorm(1,sd=sig)+bmean
  ratio=ta(prop)-uc-dnorm(prop,mean=bmean,sd=sig,log=T)
  }
\end{verbatim}
The performances of this algorithm degenerate very rapidly when $N-n_t$ is [even moderately] large.

\begin{exoset}
When uniform simulation on the accept-reject set $\mathscr{S}$ of Section 5.4 is impossible,
construct a Gibbs sampler based on the conditional distributions of $u$ and $x$. ({\em Hint}: Show that both
conditionals are uniform distributions.) This special case of the Gibbs sampler is called the {\em slice
sampler} (see Robert and Casella, 2004, Chapter 8). Apply to the distribution of
Exercise \ref{ex:2Hard}.
\end{exoset}

Since the joint distribution of $(X,U)$ has the constant density
$$
t(x,u)=\mathbb{I}_{0\le u\le g(x)}\,,
$$
the conditional distribution of $U$ given $X=x$ is $\mathscr{U}(0,g(x))$ and the conditional
distribution of $X$ given $U=u$ is $\mathscr{U}(\{x;g(x)\ge u\})$, which is uniform over the
set of highest values of $g$. Both conditionals are therefore uniform and this special Gibbs
sampler is called the {\em slice sampler}. In some settings, inverting the condition $g(x)\ge u$
may prove formidable!

If we take the case of Exercise \ref{ex:2Hard} and of $\overline{g}(x)=
\exp(-x^4)$, the set $\{x;\overline{g}(x)\ge u\}$ is equal to
$$
\left\{x;\overline{g}(x)\ge u\right\}=
\left\{x;x\le (-\log(x))^{1/4}\right\},
$$
which thus produces a closed-form solution.

\begin{exoset}\label{exo:deduceM}
Show that the normalizing constant $M$ of a target density $f$ can be deduced from the acceptance
rate in the accept-reject algorithm (Algorithm 5.9
 under the assumption that $g$ is properly normalized.
\end{exoset} 

This exercise generalises Exercise \ref{ex:AarofMay} where the target $f$ is already normalised.

If $f(x)=M\tilde{f}(x)$ is a density to be simulated by Algorithm 5.9
and if $g$ is a density such that 
$$
\tilde{f}(x)\le \tilde{M}g(x)
$$
on the support of the density $g$, then running Algorithm 
5.9
with an acceptance probability of
$g(x)/\tilde{M}\tilde{f}(x)$ produces simulations from $f$ since the accepted values have the marginal density
proportional to
$$
\int_0^1 \mathbb{I}_{[0,\tilde{f}(x)/\tilde{M}g(x)]}(u)\,\text{d}u\,g(x) = \dfrac{\tilde{f}(x)}{\tilde{M}}
\propto f(x)\,.
$$
In that case, the average probability of acceptance is
$$
\int_\mathcal{X} \dfrac{\tilde{f}(x)}{\tilde{M}}\,\text{d}x=
\int_\mathcal{X} \dfrac{{f}(x)}{M\tilde{M}}\,\text{d}x=\dfrac{1}{M\tilde{M}}\,.
$$
Since the value of $\tilde{M}$ is known, the average acceptance rate over simulations, $\hat\varrho$, leads to estimate
$M$ as
$$
\hat{M}=\dfrac{1}{\hat{\varrho}\tilde{M}}\,.
$$

\begin{exoset}\label{exo:margoton}
Reproduce the analysis of Exercise \ref{exo:deduceM} for the marginal distribution of $r_1$ computed
in Exercise \ref{exo:margot}.\end{exoset}

The only change in the codes provided in \verb+demo/Chapter.5.R+ deals with \verb+thresh+, called by \verb+ardipper+,
and with \verb+gibbs2+ where the simulation of $r_2$ is no longer required.

\begin{exoset}\label{exo:acptrjct}
Modify the function \verb+ardipper+ used in Section 5.4
to return the acceptance rate as well as a sample from the target distribution.\end{exoset}

As provided in Section 5.4,
the function \verb+ardipper+ is defined by
\begin{verbatim}
ardipper=function(nsimu=1,n1,c2,c3,r2,q2){

  barr=min(n1-c2,n1-r2-c3)
  boundM=thresh(0,n1,c2,c3,r2,barr)
  echan=1:nsimu
  for (i in 1:nsimu){
    test=TRUE
    while (test){
      y=rbinom(1,size=barr,prob=q2)
      test=(runif(1)>thresh(y,n1,c2,c3,r2,barr))
      }
    echan[i]=y
    }
  echan
}
\end{verbatim}
The requested modification consists in monitoring the acceptance rate and returning a list with both items:
\begin{verbatim}
ardippest=function(nsimu=1,n1,c2,c3,r2,q2){

  barr=min(n1-c2,n1-r2-c3)
  boundM=thresh(0,n1,c2,c3,r2,barr)
  echan=1:nsimu
  acerate=-nsimu
  for (i in 1:nsimu){
    test=TRUE
    while (test){
      y=rbinom(1,size=barr,prob=q2)
      test=(runif(1)>thresh(y,n1,c2,c3,r2,barr))
      acerate=acerate+1
      }
    echan[i]=y
    }
  list(sample=echan,reject=acerate/nsimu)
}
\end{verbatim}

\begin{exoset}\label{exo:beta}
Show that, given a mean and a $95\%$ confidence interval in $[0,1]$, there exists at
most one beta distribution $\mathscr{B}e(a,b)$ with such a mean and confidence interval.
\end{exoset}

\begin{figure}
\begin{center}
\includegraphics[width=\textwidth,height=5cm]{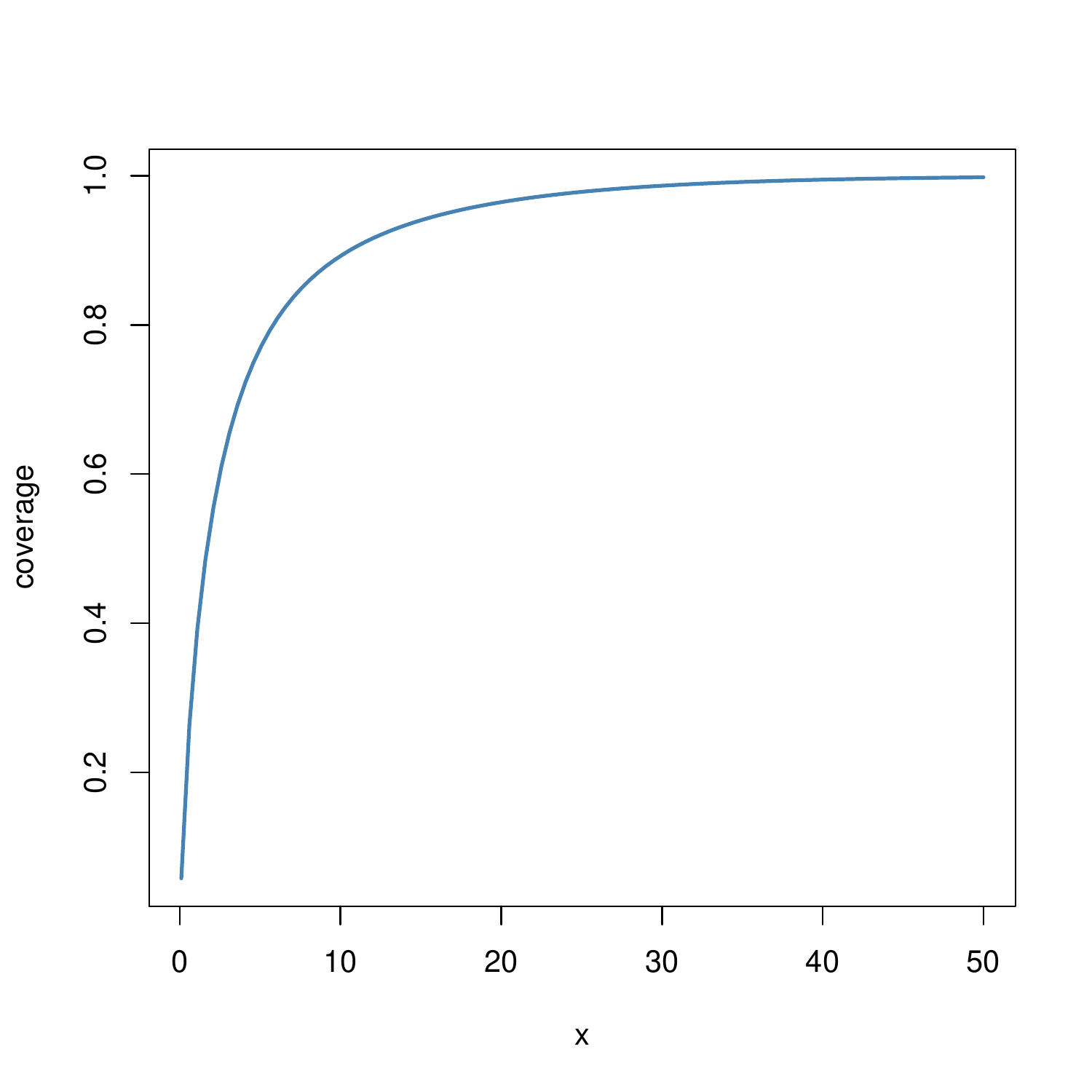}
\caption{\label{fig:betacova}Coverage of the interval $(\ell,u)=(0.1,0.6)$ by a
$\mathscr{B}e(0.4\alpha ,0.6\alpha)$ distribution when $\alpha$ varies.}
\end{center}
\end{figure}

If $0<m<1$ is the mean $m=a/(a+b)$ of a beta $\mathscr{B}e(a,b)$ distribution, then this
distribution is necessarily a beta $\mathscr{B}e(\alpha m,\alpha(1-m))$ distribution, with
$\alpha>0$. For a given confidence interval $[\ell,u]$, with $0<\ell<m<u<1$, we have that
$$
\lim_{\alpha\to 0} \int_\ell^u \frac{\Gamma(\alpha)}{\Gamma(\alpha m)\Gamma(\alpha(1-m)}\,
x^{\alpha m-1}(1-x)^{\alpha(1-m)-1}\,\text{d}x = 0
$$
[since, when $\alpha$ goes to zero, the mass of the beta $\mathscr{B}e(\alpha m,\alpha(1-m))$
distribution gets more and more concentrated around $0$ and $1$, with masses $(1-m)$ and $m$,
respectively] and
$$
\lim_{\alpha\to \infty} \int_\ell^u \frac{\Gamma(\alpha)}{\Gamma(\alpha m)\Gamma(\alpha(1-m))}\,
x^{\alpha m-1}(1-x)^{\alpha(1-m)-1}\,\text{d}x = 1
$$
[this is easily established using the gamma representation introduced
in Exercise \ref{exo:marcon,marcon} and the law of large numbers].
Therefore, due to the continuity [in $\alpha$] of the coverage probability, there must exist one
value of $\alpha$ such that
$$
B(\ell,u|\alpha,m) = \int_\ell^u \frac{\Gamma(\alpha)}{\Gamma(\alpha m)\Gamma(\alpha(1-m)}\,
x^{\alpha m-1}(1-x)^{\alpha(1-m)-1}\,\text{d}x = 0.9\,.
$$
Figure \ref{fig:betacova} illustrates this property by plotting $B(\ell,u|\alpha,m)$ for $\ell=0.1$,
$u=0.6$, $m=0.4$ and $\alpha$ varying from $0.1$ to $50$.

\begin{exoset}\label{exo:Arnakov}
Show that, for the Arnason--Schwarz model, groups of consecutive unknown locations are independent of one another,
conditional on the observations. Devise a way to simulate these groups by blocks
rather than one at a time; that is, using the joint posterior distributions of the groups
rather than the full conditional distributions of the states.
\end{exoset}

As will become clearer in Chapter 7, the Arnason-Schwarz model is a very special case of
[partly] hidden Markov chain: the locations $z_{(i,t)}$ of an individual $i$ along time
constitute a Markov chain that is only observed at times $t$ when the individual is captured.
Whether or not $z_{(i,t)}$ is observed has no relevance on the fact that, given $z_{(i,t)}$,
$(z_{(i,t-1)},z_{(i,t-2)},\ldots)$ is independent from $(z_{(i,t+1)},z_{(i,t+2)},\ldots)$.
Therefore, conditioning on any time $t$ and on the corresponding value of $z_{(i,t)}$ makes the
past and the future locations independent. In particular, conditioning on the observed locations
makes the blocks of unobserved locations in-between independent.

Those blocks could therefore be generated independently and parallely, an alternative
which would then speed up the Gibbs sampler compared with the implementation in Algorithm 5.3.
In addition, this would bring additional freedom in the choice of the proposals for the simulation
of the different blocks and thus could further increase efficiency.

\chapter{Mixture Models}\label{ch:msg}
\begin{exoset}\label{exo:bermix}
Show that a mixture of Bernoulli distributions is again a Bernoulli distribution.
Extend this to the case of multinomial distributions.
\end{exoset}

By definition, if
$$
x\sim\sum_{i=1}^k p_i \mathscr{B}(q_i)\,,
$$
then $x$ only takes the values $0$ and $1$ with probabilities
$$
\sum_{i=1}^k p_i (1-q_i) = 1  - \sum_{i=1}^k p_iq_i
\quad\text{and}\quad
\sum_{i=1}^k p_iq_i\,,
$$
respectively. This mixture is thus a Bernoulli distribution
$$
\mathscr{B}\left( \sum_{i=1}^k p_iq_i \right)\,.
$$

When considering a mixture of multinomial distributions,
$$
x\sim\sum_{i=1}^k p_i \mathscr{M}_k(\mathbf{q}_i)\,,
$$
with $\mathbf{q}_i=(q_{i1},\ldots,q_{ik})$,
$x$ takes the values $1\le j\le k$ with probabilities
$$
\sum_{i=1}^k p_i q_{ij}
$$
and therefore this defines a multinomial distribution.
This means that a mixture of multinomial distributions cannot be identifiable unless some
restrictions are set upon its parameters.

\begin{exoset}\label{exo:fermi}
Show that the number of nonnegative integer solutions of the decomposition of $n$ into $k$ parts
such that $n_1+\ldots+n_k$ is equal to
$$
\mathfrak{r}={n+k-1 \choose n} \,.
$$
Deduce that the number of partition sets is of order $\hbox{O}(n^{k-1})$. ({\em Hint:} This is a classical
combinatoric problem.)
\end{exoset}

This is a usual combinatoric result, detailed for instance in Feller (1970).
A way to show that $\mathfrak{r}$ is the solution is to use the ``bottomless box" trick:
consider a box with $k$ cases and $n$ identical balls to put into those cases. If we remove
the bottom of the box, one allocation of the $n$ balls is represented by a sequence of balls (O)
and of case separations ($|$) or, equivalently, of $0$'s and $1$'s, of which there are $n$ and $k-1$
respectively [since the box itself does not count, we have to remove the extreme separations].
Picking $n$ positions out of $n+(k-1)$ is
exactly $\mathfrak{r}$.

This value is thus the number of ``partitions" of an $n$ sample into $k$ groups [we write
``partitions" and not partitions because, strictly speaking, all sets of a partition are non-empty].
Since
$$
{n+k-1 \choose n} = \frac{(n+k-1)!}{n!(k-1)!} \approx \frac{n^{k-1}}{(k-1)!}\,,
$$
when $n\gg k$, there is indeed an order $\hbox{O}(n^{k-1})$ of partitions.

\begin{exoset}\label{exo:2N}
For a mixture of two normal distributions with all parameters unknown,
$$
p\mathscr{N}(\mu_1,\sigma_1^2) +(1-p)\mathscr{N}(\mu_2,\sigma_2^2)\,,
$$
and for the prior distribution $(j=1,2)$
$$
\mu_j|\sigma_j\sim\mathscr{N}(\xi_j,\sigma_j^2/n_j)\,,\quad
\sigma_j^2\sim\mathscr{IG}(\nu_j/2,s_j^2/2)\,,\quad
p\sim\mathscr{B}e(\alpha,\beta)\,,
$$
show that
$$
p|\bx,\bz\sim\mathscr{B}e(\alpha+\ell_1,\beta+\ell_2),
$$
$$
\mu_j|\sigma_j,\bx,\bz \sim\mathscr{N}\left(\xi_1(\bz),{\sigma_j^2\over n_j+\ell_j}\right)\,,\ 
\sigma_j^2|\bx,\bz \sim\mathscr{IG}((\nu_j+\ell_j)/2,s_j(\bz)/2)\,,
$$
where $\ell_j$ is the number of $z_i$ equal to $j$, $\bar x_j(\bz)$ and $\hat s_j^2(\bz)$
are the empirical mean and variance for the subsample with $z_i$ equal to $j$, and
$$
\xi_j(\bz) = {n_j\xi_j+\ell_j \bar x_j(\bz) \over n_j+\ell_j}\,,\quad
s_j(\bz) = s^2_j+\ell_j\hat s_j^2(\bz) + {n_j\ell_j\over n_j+\ell_j}
  (\xi_j-\bar x_j(\bz))^2\,.
$$
Compute the corresponding weight $\omega(\bz)$.
\end{exoset}

If the latent (or missing) variable $\bz$ is introduced, the joint distribution of $(\bx,\bz)$
[equal to the completed likelihood] decomposes into
\begin{align}\label{eq:compmimi}
\prod_{i=1}^n p_{z_i}\,f(x_i|\theta_{z_i}) &= \prod_{j=1}^2 \prod_{i;z_i=j} p_j \,f(x_i|\theta_j)\nonumber\\
        &\propto \prod_{j=1}^k p_j^{\ell_j}\,\prod_{i;z_i=j} \frac{e^{-(x_i-\mu_j)^2/2\sigma_j^2}}{\sigma_j}\,,
\end{align}
where $p_1=p$ and $p_2=(1-p)$.
Therefore, using the conjugate priors proposed in the question, we have a decomposition of the posterior
distribution of the parameters given $(\bx,\bz)$ in
$$
p^{\ell_1+\alpha-1}(1-p)^{\ell2+\beta-1}\,\prod_{j=1}^2 \prod_{i;z_i=j} \frac{e^{-(x_i-\mu_j)^2/2\sigma_j^2}}{\sigma_j}
\pi(\mu_j,\sigma_j^2)\,.
$$
This implies that $p|\bx,\bz\sim\mathscr{B}e(\alpha+\ell_1,\beta+\ell_2)$ and that the posterior distributions
of the pairs $(\mu_j,\sigma_j^2)$ are the posterior distributions associated with the normal observations allocated
(via the $z_i$'s) to the corresponding component. The values of the hyperparameters are therefore those already
found in Chapter 2 (see, e.g., Exercises 2.7 and 2.15).

The weight $\omega(\bz)$ is the marginal [posterior] distribution of $\bz$, since
$$
\pi(\btheta,p|\bx) = \sum_{\bz} \omega(\bz) \pi(\btheta,p|\bx,\bz)\,.
$$
Therefore, if $p_1=p$ and $p_2=1-p$,
\begin{align*}
\omega(\bz) &\propto \int \prod_{j=1}^2 p_j^{\ell_j}\,\prod_{i;z_i=j}
\frac{e^{-(x_i-\mu_j)^2/2\sigma_j^2}}{\sigma_j}
\pi(\btheta,p)\,\text{d}\btheta\text{d}p \\
&\propto 
\frac{\Gamma(\alpha+\ell_1)\Gamma(\beta+\ell_2)}{\Gamma(\alpha+\beta+n)}\\
&\quad\int\,\prod_{j=1}^2 
\exp\left[\frac{-1}{2\sigma_j^2}\left\{(n_j+\ell_j)(\mu_j-\xi_j(\bz))^2+s_j(\bz)
\right\}\right] \sigma_j^{-\ell_j-\nu_j-3}\,\text{d}\theta \\
&\propto 
\frac{\Gamma(\alpha+\ell_1)\Gamma(\beta+\ell_2)}{\Gamma(\alpha+\beta+n)}\,
\prod_{j=1}^2 
\frac{\Gamma((\ell_j+\nu_j)/2)(s_j(\bz)/2)^{(\nu_j+\ell_j)/2}}{\sqrt{n_j+\ell_j}}
\end{align*}
and the proportionality factor can be derived by summing up the rhs over all $\bz$'s.
(There are $2^n$ terms in this sum.)

\begin{exoset}\label{exo:EMix} 
For the normal mixture model of Exercise \ref{exo:2N}, compute the function $Q(\theta_0,\theta)$
and derive both steps of the EM algorithm. Apply this algorithm to a simulated dataset and test the influence
of the starting point $\theta_0$.  \end{exoset}

Starting from the representation \eqref{eq:compmimi} above,
$$
\log \ell(\btheta,p|\bx,\bz) = \sum_{i=1}^n \left\{
\mathbb{I}_1(z_i)\log(p\,f(x_i|\theta_1)
+ \mathbb{I}_2(z_i)\log((1-p)\,f(x_i|\theta_2)\right\}\,,
$$
which implies that
\begin{align*}
Q\{(\btheta^{(t)},&p^{(t)}),(\btheta,p)\}
= \mathbb{E}_{(\theta^{(t)},p^{(t)})}\left[\log \ell(\btheta,p|\bx,\bz)|\bx\right] \\
&= \sum_{i=1}^n \left\{
\text{P}_{(\theta^{(t)},p^{(t)})}\left( z_i=1|\bx \right)\log(p\,f(x_i|\btheta_1)\right.\\
&\qquad\left.+\text{P}_{(\btheta^{(t)},\bp^{(t)})}\left( z_i=2|\bx \right)\log((1-p)\,f(x_i|\btheta_2)\right\}\\
&= \log(p/\sigma_1) \sum_{i=1}^n \text{P}_{(\btheta^{(t)},p^{(t)})}\left( z_i=1|\bx \right)\\
&\quad +\log((1-p)/\sigma_2) \sum_{i=1}^n \text{P}_{(\btheta^{(t)},p^{(t)})}\left( z_i=2|\bx \right)\\
&\quad -\sum_{i=1}^n \text{P}_{(\btheta^{(t)},p^{(t)})}\left( z_i=1|\bx \right) \frac{(x_i-\mu_1)^2}{2\sigma^2_1}\\
&\quad -\sum_{i=1}^n \text{P}_{(\btheta^{(t)},p^{(t)})}\left( z_i=2|\bx \right) \frac{(x_i-\mu_2)^2}{2\sigma^2_2}\,.
\end{align*}
If we maximise this function in $p$, we get that
\begin{align*}
p^{(t+1)} &= \frac{1}{n}\,\sum_{i=1}^n \text{P}_{(\btheta^{(t)},p^{(t)})}\left( z_i=1|\bx \right)\\
&= \frac{1}{n}\,\sum_{i=1}^n \frac{p^{(t)} f(x_i|\btheta_1^{(t)})}{p^{(t)} f(x_i|\btheta_1^{(t)})+
        (1-p^{(t)}) f(x_i|\btheta_2^{(t)})}
\end{align*}
while maximising in $(\mu_j,\sigma_j)$ $(j=1,2)$ leads to
\begin{align*}
\mu_j^{(t+1)} &= \sum_{i=1}^n \text{P}_{(\btheta^{(t)},p^{(t)})}\left( z_i=j|\bx \right) x_i \bigg/
        \sum_{i=1}^n \text{P}_{(\btheta^{(t)},p^{(t)})}\left( z_i=j|\bx \right)\\
              &= \frac{1}{np_j^{(t+1)}}\,\sum_{i=1}^n \frac{x_i p_j^{(t)} f(x_i|\btheta_j^{(t)})}{p^{(t)}
        f(x_i|\btheta_1^{(t)})+ (1-p^{(t)}) f(x_i|\btheta_2^{(t)})} \,,\\
\sigma_j^{2(t+1)} &= \sum_{i=1}^n \text{P}_{(\btheta^{(t)},p^{(t)})}\left( z_i=j|\bx \right)
        (x_i-\mu_j^{(t+1)})^2\bigg/ \sum_{i=1}^n \text{P}_{(\btheta^{(t)},p^{(t)})}\left( z_i=j|\bx \right)\\
   &= \frac{1}{np_j^{(t+1)}}\,\sum_{i=1}^n \frac{\left[x_i-\mu_j^{(t+1)}\right]^2 p_j^{(t)}
f(x_i|\btheta_j^{(t)})}{p^{(t)}
        f(x_i|\btheta_1^{(t)})+ (1-p^{(t)}) f(x_i|\btheta_2^{(t)})}\,,
\end{align*}
where $p_1^{(t)}=p^{(t)}$ and $p_2^{(t)}=(1-p^{(t)})$.

A possible implementation of this algorithm in {\sf R} is given below:
\begin{verbatim}
# simulation of the dataset
n=324
tz=sample(1:2,n,prob=c(.4,.6),rep=T)
tt=c(0,3.5)
ts=sqrt(c(1.1,0.8))
x=rnorm(n,mean=tt[tz],sd=ts[tz])

para=matrix(0,ncol=50,nrow=5)
likem=rep(0,50)

# initial values chosen at random
para[,1]=c(runif(1),mean(x)+2*rnorm(2)*sd(x),rexp(2)*var(x))
likem[1]=sum(log( para[1,1]*dnorm(x,mean=para[2,1],
  sd=sqrt(para[4,1]))+(1-para[1,1])*dnorm(x,mean=para[3,1],
  sd=sqrt(para[5,1])) ))

# 50 EM steps
for (em in 2:50){

   # E step
   postprob=1/( 1+(1-para[1,em-1])*dnorm(x,mean=para[3,em-1],
     sd=sqrt(para[5,em-1]))/( para[1,em-1]*dnorm(x,
     mean=para[2,em-1],sd=sqrt(para[4,em-1]))) )

   # M step
   para[1,em]=mean(postprob)
   para[2,em]=mean(x*postprob)/para[1,em]
   para[3,em]=mean(x*(1-postprob))/(1-para[1,em])
   para[4,em]=mean((x-para[2,em])^2*postprob)/para[1,em]
   para[5,em]=mean((x-para[3,em])^2*(1-postprob))/(1-para[1,em])

   # value of the likelihood
   likem[em]=sum(log(para[1,em]*dnorm(x,mean=para[2,em],
     sd=sqrt(para[4,em]))+(1-para[1,em])*dnorm(x,mean=para[3,em],
     sd=sqrt(para[5,em])) ))
}
\end{verbatim}

\begin{figure}
\begin{center}
\includegraphics[width=.8\textwidth,height=6cm]{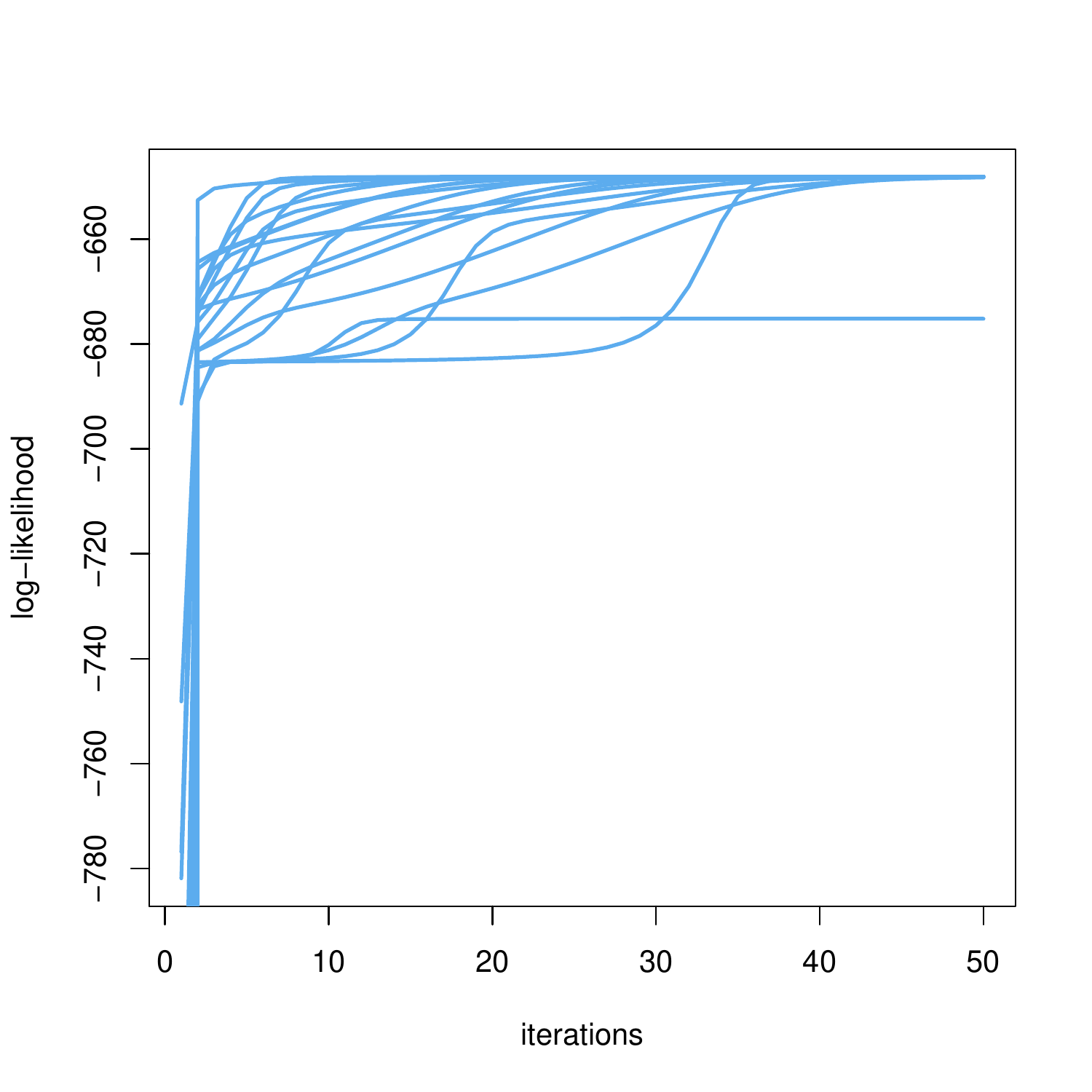}
\caption{\label{fig:let'em}Increase of the log-likelihood along EM iterations for
$20$ different starting points.}
\end{center}
\end{figure}

Figure \ref{fig:let'em} in this manual in this manual represents the increase in the log-likelihoods along EM iterations for
$20$ different starting points [and the same dataset $x$]. While most starting points lead to
the same value of the log-likelihood after $50$ iterations, one starting point induces a different
convergence behaviour.

\begin{exoset}
In the mixture model with independent priors on the $\theta_j$'s,
show that the $\theta_j$'s are dependent on each other given (only) $\bx$ by summing out the $\bz$'s.
\end{exoset}

The likelihood associated with model (6.2) being
$$
\ell(\btheta,p|\bx)=\prod_{i=1}^n \left[ \sum_{j=1}^k p_j\,f(x_i|\btheta_j) \right]\,,
$$
it is clear that the posterior distribution will not factorise as a product of functions
of the different parameters. It is only given $(\bx,\bz)$ that the $\btheta_j$'s are independent.

\begin{exoset}\label{exo:newP}
Construct and test the Gibbs sampler associated with the $(\xi,\mu_0)$ parameterization of
(6.3),
when $\mu_1=\mu_0-\xi$ and $\mu_2=\mu_0+\xi$.
\end{exoset}

The simulation of the $z_i$'s is unchanged [since it does not depend on the
parameterisation of the components. The conditional distribution of $(\xi,\mu_0)$
given $(\bx,\bz)$ is
$$
\pi(\xi,\mu_0|\bx,\bz)\propto \exp\frac{-1}{2}\left\{ \sum_{z_i=1} (x_i-\mu_0+\xi)^2
+ \sum_{z_i=2} (x_i-\mu_0-\xi)^2 \right\}\,.
$$
Therefore, $\xi$ and $\mu_0$ are not independent given $(\bx,\bz)$, with
\begin{eqnarray*}
\mu_0|\xi,\bx,\bz &\sim& \mathscr{N}\left( \frac{n\overline x +(\ell_1-\ell_2)\xi}{n},\frac{1}{n}\right)\,,\\
\xi|\mu_0,\bx,\bz &\sim& \mathscr{N}\left( \frac{\sum_{z_i=2} (x_i-\mu_0) - \sum_{z_i=1} (x_i-\mu_0)
}{n},\frac{1}{n}\right)
\end{eqnarray*}

The implementation of this Gibbs sampler is therefore a simple modification of \verb+gibbsmean+ in the
\verb+bayess+: the MCMC loop is now
\begin{verbatim}
for (t in 2:Nsim){

  # allocation
  fact=.3*sqrt(exp(gu1^2-gu2^2))/.7
  probs=1/(1+fact*exp(sampl*(gu2-gu1)))
  zeds=(runif(N)<probs)

  # Gibbs sampling
  mu0=rnorm(1)/sqrt(N)+(sum(sampl)+xi*(sum(zeds==1)
    -sum(zeds==0)))/N
  xi=rnorm(1)/sqrt(N)+(sum(sampl[zeds==0]-mu0)
    -sum(sampl[zeds==1]-mu0))/N

  # reparameterisation
  gu1=mu0-xi
  gu2=mu0+xi
  muz[t,]=(c(gu1,gu2))

}
\end{verbatim}

\begin{figure}
\begin{center}
\includegraphics[width=.8\textwidth,height=7cm]{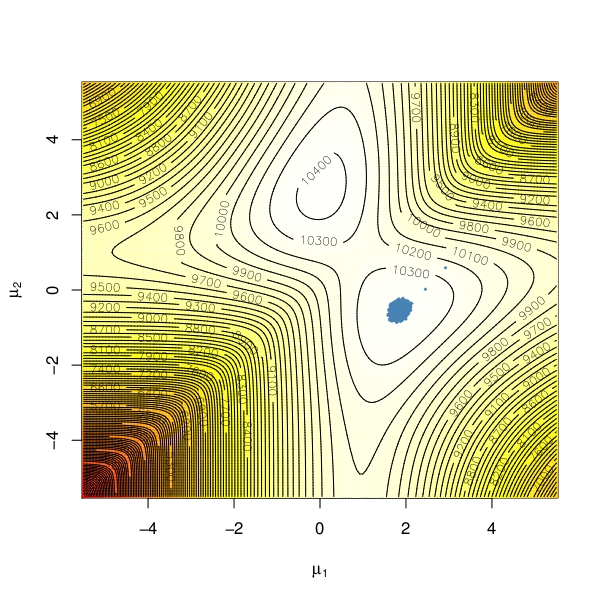}\\
\includegraphics[width=.8\textwidth,height=7cm]{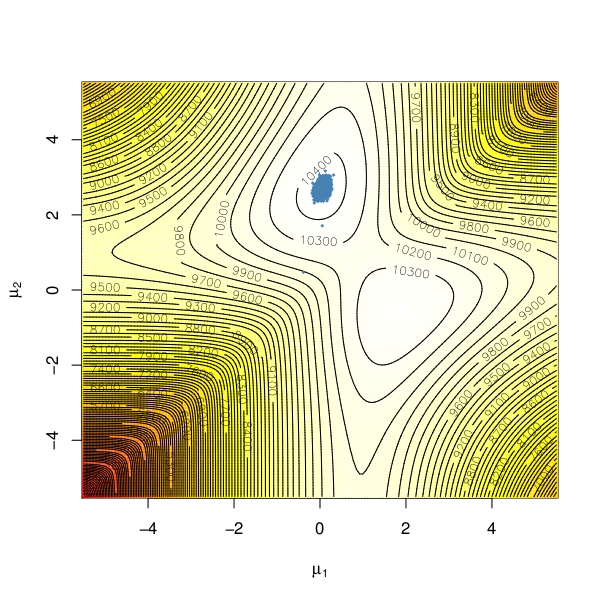}
\caption{\label{fig:captimode}Influence of the starting value on the convergence
of the Gibbs sampler associated with the location parameterisation of the mean
mixture ($10,000$ iterations).}
\end{center}
\end{figure}

If we run repeatedly this algorithm, the Markov chain produced is highly dependent on
the starting value and remains captive of local modes, as illustrated on Figure \ref{fig:captimode} in this manual.
This reparameterisation thus seems less robust than the original parameterisation.

\begin{exoset}
Show that, if an exchangeable prior $\pi$ is used on the vector of weights $(p_1,\ldots,p_k)$, then, necessarily,
$\mathbb{E}^\pi[p_j]=1/k$ and, if the prior on the other parameters $(\theta_1,\ldots,\theta_k)$ is also exchangeable, 
then $\mathbb{E}^\pi[p_j|x_1,\ldots,x_n]=1/k$ for all $j$'s.
\end{exoset}

If
$$
\pi(p_1,\ldots,p_k) = \pi(p_{\sigma(1)},\ldots,p_{\sigma(k)})
$$
for any permutation $\sigma\in\mathfrak{S}_k$, then
$$
\mathbb{E}^\pi[p_j]=\int p_j \pi(p_1,\ldots,p_j,\ldots,p_k) \,\text{d}\mathbf{p}
=\int p_j \pi(p_j,\ldots,p_1,\ldots,p_k) \,\text{d}\mathbf{p} = \mathbb{E}^\pi[p_1]\,.
$$
Given that $\sum_{j=1}^k p_j=1$, this implies $\mathbb{E}^\pi[p_j]=1/k$.

When both the likelihood and the prior are exchangeable in $(p_j,\theta_j)$, the same
result applies to the posterior distribution.

\begin{exoset}
Show that running an MCMC algorithm with target $\pi(\theta|\bx)^\gamma$
will increase the proximity to the MAP estimate when $\gamma>1$ is large. ({\em
Note}: This is a crude version of the {\em simulated annealing} algorithm. See
also Chapter \ref{ch:spt}.) Discuss the modifications required in Algorithm 6.11
to achieve simulation from $\pi(\theta|\bx)^\gamma$ when $\gamma\in\mathbb{N}^*$ is an integer.\index{Simulated annealing}
\end{exoset}

The power distribution $\pi_\gamma(\theta)\propto\pi(\theta)^\gamma$ shares the same modes as $\pi$,
but the global mode gets more and more mass as $\gamma$ increases. If $\theta^\star$ is the global
mode of $\pi$ [and of $\pi_\gamma$], then $\{\pi(\theta)/\pi(\theta^\star)\}^\gamma$ goes to $0$ as $\gamma$ goes to
$\infty$
for all $\theta$'s different from $\theta^\star$. Moreover, for any $0<\alpha<1$, if we define the $\alpha$
neighbourhood $\mathfrak{N}_\alpha$ of $\theta^\star$ as the set of $\theta$'s such that $\pi(\theta)\ge \alpha
\pi(\theta^\star)$, then $\pi_\gamma(\mathfrak{N}_\alpha)$ converges to $1$ as $\gamma$ goes to $\infty$.

The idea behind {\em simulated annealing} is that, first, the distribution $\pi_\gamma(\theta)\propto\pi(\theta)^\gamma$
is more concentrated around its main mode than $\pi(\theta)$ if $\gamma$ is large and, second, that it is not
necessary to simulate a whole sample from $\pi(\theta)$, then a whole sample from $\pi(\theta)^2$ and so on to achieve
a convergent approximation of the MAP estimate. Increasing $\gamma$ slowly enough along iterations leads to the same
result with a much smaller computing requirement.

When considering the application of this idea to a mean mixture as (6.3) [in the book], the modification of
Algorithm 6.2 is rather immediate: since we need to simulate from $\pi(\btheta,p|\bx)^\gamma$ [up to a
normalising constant], this is equivalent to simulate from $\ell(\btheta,p|\bx)^\gamma\times\pi(\btheta,p)^\gamma$.
This means that, since the prior is [normal] conjugate, the prior hyperparameter $\lambda$ is modified into
$\gamma\lambda$
and that the likelihood is to be completed $\gamma$ times rather than once, i.e.
$$
\ell(\btheta,p|\bx)^\gamma = \left( \int f(\bx,\bz|\btheta,p)\,\text{d}\bz \right)^\gamma
= \prod_{j=1}^\gamma \int f(\bx,\bz_j|\btheta,p)\,\text{d}\bz_j\,.
$$
Using this duplication trick, the annealed version of Algorithm 6.2 writes as
\begin{algo} {\bf Annealed Mean Mixture Gibbs Sampler}
\begin{itemize}
\item[]  {\sffamily Initialization.} Choose $\mu_1^{(0)}$ and $\mu_2^{(0)}$,
\item[]  {\sffamily Iteration $t$ $(t\ge 1)$.}
\begin{enumerate}
\item[{\sf 1.}] For $i=1,\ldots,n$, $j=1,\ldots,\gamma$, generate $z_{ij}^{(t)}$ from
\begin{eqnarray*}
\mathbb{P}\left(z_{ij}=1\right)&\propto& p\,\exp\left\{-\frac{1}{2}\left(x_i-\mu_1^{(t-1)}\right)^2\right\}\\
\mathbb{P}\left(z_{ij}=2\right)&\propto& (1-p)\,\exp\left\{-\frac{1}{2}\left(x_i-\mu_2^{(t-1)}\right)^2\right\}
\end{eqnarray*}
\item[{\sf 2.}] Compute
$$
\ell=\sum_{j=1}^\gamma\sum_{i=1}^n\mathbb{I}_{z_{ij}^{(t)}=1}
\quad\text{and}\quad
\bar x_u\left(\bz\right)=\sum_{j=1}^\gamma\sum_{i=1}^n\mathbb{I}_{z_{ij}^{(t)}=u}x_i
$$
\item[{\sf 3.}] Generate $\mu_1^{(t)}$ from
$\displaystyle \mathscr{N}\left(\frac{\gamma\lambda\delta+\bar x_1\left(\bz\right)}
{\gamma\lambda+\ell},\frac{1}{\gamma\lambda+\ell}\right)$
\item[{\sf 4.}] Generate $\mu_2^{(t)}$ from $\displaystyle \mathscr{N}
\left(\frac{\gamma\lambda\delta+\bar x_2\left(\bz\right)}{\gamma\lambda+\gamma n-\ell},
\frac{1}{\gamma \lambda+\gamma n-\ell}\right)$.
\end{enumerate}
\end{itemize}
\end{algo}
This additional level of completion means that the Markov chain will have difficulties to move around, compared
with the original Gibbs sampling algorithm. While closer visits to the global mode are guaranteed in theory, they
may require many more simulations in practice.

\begin{exoset}
Show that the ratio (6.7)
goes to $1$ when $\alpha$ goes to $0$ when the proposal
$q$ is a random walk. Describe the average behavior of this ratio in the case of an independent proposal.
\end{exoset}

Since
$$
\frac{\partial }{\partial\theta} \log\left[\theta/(1-\theta)\right] =
\frac{1}{\theta}+\frac{1}{1-\theta)} =
\frac{1}{\theta(1-\theta)}\,,
$$
the Metropolis--Hastings acceptance ratio for the logit transformed random walk is
$$
\frac{\pi(\widetilde{\theta_j})}{\pi(\theta_j^{(t-1)})}\,
\frac{\widetilde{\theta_j}(1-\widetilde{\theta_j})}{\theta_j^{(t-1)}(1-\theta_j^{(t-1)})}\wedge 1\,.
$$

\begin{exoset}
If one needs to use importance sampling weights, show that the simultaneous choice of several powers 
$\alpha$ requires the computation of the normalizing constant of $\pi_\alpha$.
\end{exoset}

If samples $(\theta_{i\alpha})_i$ from several tempered versions $\pi_\alpha$ of $\pi$ are
to be used simultaneously, the importance weights associated with those samples $\pi(\theta_{i\alpha})/
\pi_\alpha(\theta_{i\alpha})$ require the computation of the normalizing constants, which is
most often impossible. This difficulty explains the appeal of the ``pumping mechanism" of Algorithm
6.5, which cancels the need for normalizing constants by using the same $\pi_\alpha$ twice, once in
the numerator and once in the denominator.

\begin{exoset}
In the setting of the mean mixture (6.3),
run an MCMC simulation experiment to compare the influence of a $\mathscr{N}(0,100)$ and of a $\mathscr{N}(0,10000)$ prior on
$(\mu_1,\mu_2)$ on a sample of $500$ observations.
\end{exoset}

The power distribution $\pi_\gamma(\theta)\propto\pi(\theta)^\gamma$ shares the same modes as $\pi$,
but the global mode gets more and more mass as $\gamma$ increases. If $\theta^\star$ is the global
mode of $\pi$ [and of $\pi_\gamma$], then $\{\pi(\theta)/\pi(\theta^\star)\}^\gamma$ goes to $0$ as $\gamma$ goes to
$\infty$
for all $\theta$'s different from $\theta^\star$. Moreover, for any $0<\alpha<1$, if we define the $\alpha$
neighbourhood $\mathfrak{N}_\alpha$ of $\theta^\star$ as the set of $\theta$'s such that $\pi(\theta)\ge \alpha
\pi(\theta^\star)$, then $\pi_\gamma(\mathfrak{N}_\alpha)$ converges to $1$ as $\gamma$ goes to $\infty$.

The idea behind {\em simulated annealing} is that, first, the distribution $\pi_\gamma(\theta)\propto\pi(\theta)^\gamma$
is more concentrated around its main mode than $\pi(\theta)$ if $\gamma$ is large and, second, that it is not
necessary to simulate a whole sample from $\pi(\theta)$, then a whole sample from $\pi(\theta)^2$ and so on to achieve
a convergent approximation of the MAP estimate. Increasing $\gamma$ slowly enough along iterations leads to the same
result with a much smaller computing requirement.

When considering the application of this idea to a mean mixture as (6.3) [in the book], the modification of
Algorithm 6.2 is rather immediate: since we need to simulate from $\pi(\btheta,p|\bx)^\gamma$ [up to a
normalising constant], this is equivalent to simulate from $\ell(\btheta,p|\bx)^\gamma\times\pi(\btheta,p)^\gamma$.
This means that, since the prior is [normal] conjugate, the prior hyperparameter $\lambda$ is modified into
$\gamma\lambda$
and that the likelihood is to be completed $\gamma$ times rather than once, i.e.
$$
\ell(\btheta,p|\bx)^\gamma = \left( \int f(\bx,\bz|\btheta,p)\,\text{d}\bz \right)^\gamma
= \prod_{j=1}^\gamma \int f(\bx,\bz_j|\btheta,p)\,\text{d}\bz_j\,.
$$
Using this duplication trick, the annealed version of Algorithm 6.2 writes as
\begin{algo} {\bf Annealed Mean Mixture Gibbs Sampler}
\begin{itemize}
\item[]  {\sffamily Initialization.} Choose $\mu_1^{(0)}$ and $\mu_2^{(0)}$,
\item[]  {\sffamily Iteration $t$ $(t\ge 1)$.}
\begin{enumerate}
\item[{\sf 1.}] For $i=1,\ldots,n$, $j=1,\ldots,\gamma$, generate $z_{ij}^{(t)}$ from
\begin{eqnarray*}
\mathbb{P}\left(z_{ij}=1\right)&\propto& p\,\exp\left\{-\frac{1}{2}\left(x_i-\mu_1^{(t-1)}\right)^2\right\}\\
\mathbb{P}\left(z_{ij}=2\right)&\propto& (1-p)\,\exp\left\{-\frac{1}{2}\left(x_i-\mu_2^{(t-1)}\right)^2\right\}
\end{eqnarray*}
\item[{\sf 2.}] Compute
$$
\ell=\sum_{j=1}^\gamma\sum_{i=1}^n\mathbb{I}_{z_{ij}^{(t)}=1}
\quad\text{and}\quad
\bar x_u\left(\bz\right)=\sum_{j=1}^\gamma\sum_{i=1}^n\mathbb{I}_{z_{ij}^{(t)}=u}x_i
$$
\item[{\sf 3.}] Generate $\mu_1^{(t)}$ from
$\displaystyle \mathscr{N}\left(\frac{\gamma\lambda\delta+bar x_1\left(\bz\right)}
{\gamma\lambda+\ell},\frac{1}{\gamma\lambda+\ell}\right)$
\item[{\sf 4.}] Generate $\mu_2^{(t)}$ from $\displaystyle \mathscr{N}
\left(\frac{\gamma\lambda\delta+\bar x_2\left(\bz\right)}{\gamma\lambda+\gamma n-\ell},
\frac{1}{\gamma \lambda+\gamma n-\ell}\right)$.
\end{enumerate}
\end{itemize}
\end{algo}
This additional level of completion means that the Markov chain will have difficulties to move around, compared
with the original Gibbs sampling algorithm. While closer visits to the global mode are guaranteed in theory, they
may require many more simulations in practice.

\begin{exoset}
Show that, for a normal mixture $0.5\,\mathscr{N}(0,1)+0.5\,\mathscr{N}(\mu,\sigma^2)$,
the likelihood is unbounded. Exhibit this feature by plotting the likelihood of a simulated
sample using the {\sf R image} procedure.\index{R@{\sf R}!image@{\sf image}}
\end{exoset}

This follows from the decomposition of the likelihood
$$
\ell(\btheta|\bx)=\prod_{i=1}^n \left[ \sum_{j=1}^2 0.5\,f(x_i|\btheta_j) \right]\,,
$$
into a sum [over all partitions] of the terms
$$
\prod_{i=1}^n f(x_i|\btheta_{z_i})
=\prod_{i;z_i=1} \varphi(x_i) \prod_{i;z_i=2} \frac{\varphi\{(x_i-\mu)/\sigma\}}{\sigma}\,.
$$
In exactly $n$ of those $2^n$ partitions, a single observation is allocated to the second component,
i.e.~there is a single $i$ such that $z_i=2$. For those particular partitions, if we choose $\mu=x_i$,
the second product reduces to $1/\sigma$ which is not bounded when $\sigma$ goes to $0$. Since the
observed likelihood is the sume of all those terms, it is bounded from below by terms that are unbounded
and therefore it is unbounded.

\begin{figure}
\begin{center}
\includegraphics[width=.8\textwidth,height=7cm]{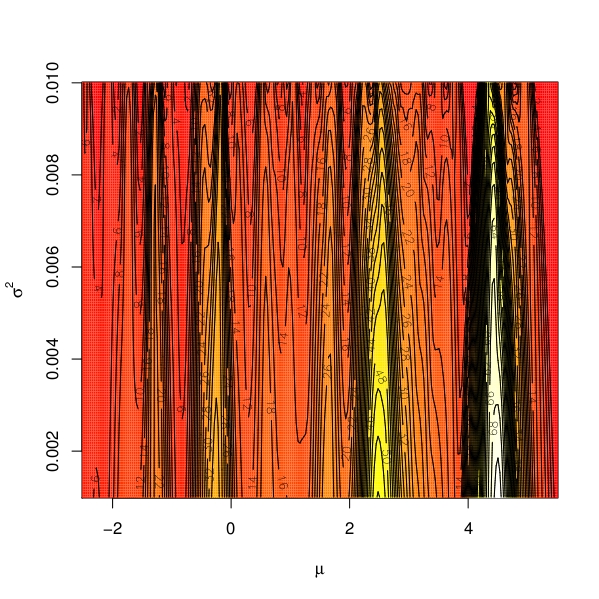}
\caption{\label{fig:infimix}Illustration of an unbounded mixture likelihood.}
\end{center}
\end{figure}

An {\sf R} code illustrating this behaviour is
\begin{verbatim}
# Sample construction
N=100
sampl=rnorm(N)+(runif(N)<.3)*2.7

# Grid
mu=seq(-2.5,5.5,length=250)
sig=rev(1/seq(.001,.01,length=250))  # inverse variance
mo1=mu%*%t(rep(1,length=length(sig)))
mo2=(rep(1,length=length(mu)))%*%t(sig)
ca1=-0.5*mo1^2*mo2
ca2=mo1*mo2
ca3=sqrt(mo2)
ca4=0.5*(1-mo2)

# Likelihood surface
like=0*mo1
for (i in 1:N)
  like=like+log(1+exp(ca1+sampl[i]*ca2+sampl[i]^2*ca4)*ca3)
like=like-min(like)

sig=rev(1/sig)
image(mu,sig,like,xlab=expression(mu),
  ylab=expression(sigma^2),col=heat.colors(250))
contour(mu,sig,like,add=T,nlevels=50)

\end{verbatim}
and Figure \ref{fig:infimix} in this manual exhibits the characteristic stripes of an explosive likelihood as
$\sigma$ approaches $0$ for values of $\mu$ close to the values of the sample.

\chapter{Dynamic Models}\label{ch:dyn}
\begin{exoset}
Consider the process $(x_t)_{t\in\mathbb{Z}}$ defined by
$$
x_t=a+bt+y_t\,,
$$
where $(y_t)_{t\in\mathbb{Z}}$ is an iid sequence of random variables with mean
$0$ and variance $\sigma^2$, and where $a$ and $b$ are constants. Define
$$
w_t=(2q+1)^{-1}\textstyle{\sum_{j=-q}^q} x_{t+j}\,.
$$
Compute the mean and the autocovariance function of $(w_t)_{t\in\mathbb{Z}}$. Show that
$(w_t)_{t\in\mathbb{Z}}$ is not stationary but that its autocovariance function $\gamma_w(t+h,t)$
does not depend on $t$.
\end{exoset}

We have
\begin{eqnarray*}
\mathbb{E}[w_t] & = & \mathbb{E}\left[(2q+1)^{-1}\sum_{j=-q}^q x_{t+j}\right] \\
                & = & (2q+1)^{-1} \sum_{j=-q}^q \mathbb{E}\left[a+b(t+j)+y_t\right] \\
                & = &  a+bt\,.
\end{eqnarray*}
The process $(w_t)_{t\in\mathbb{Z}}$ is therefore not stationary. Moreover
\begin{eqnarray*}
\mathbb{E}[w_tw_{t+h}] & = & \mathbb{E}\left[\left(a+bt+\frac{1}{2q+1}\sum_{j=-q}^q y_{t+j}\right)
                             \left(a+bt+bh+\sum_{j=-q}^q y_{t+h+j}\right)\right] \\
                       & = & (a+bt)(a+bt+bh)+\mathbb{E}\left[ \sum_{j=-q}^q y_{t+j}\sum_{j=-q}^q y_{t+h+j}\right] \\
                       & = & (a+bt)(a+bt+bh)+\mathbb{I}_{|h|\leq q}(q+1-|h|)\sigma^2\,.
\end{eqnarray*}
Then,
\begin{equation*}
\text{cov}(w_t,w_{t+h})=\mathbb{I}_{|h|\leq q}(q+1-|h|)\sigma^2
\end{equation*}
and,
\begin{equation*}
\gamma_w(t+h,t)=\mathbb{I}_{|h|\leq q}(q+1-|h|)\sigma^2\,.
\end{equation*}

\begin{exoset}
Suppose that the process $(x_t)_{t\in\mathbb{N}}$ is such that $x_0\sim\mathscr{N}(0,\tau^2)$ 
and, for all $t\in\mathbb{N}$,
$$
x_{t+1}|\bx_{0:t} \sim \mathscr{N}(x_t/2,\sigma^2)\,,\qquad \sigma>0\,.
$$
Give a necessary condition on $\tau^2$ for $(x_t)_{t\in\mathbb{N}}$ to be a (strictly) stationary process.
\end{exoset}

We have
\begin{equation*}
\mathbb{E}[x_1]=\mathbb{E}[\mathbb{E}[x_1|x_0]]=\mathbb{E}[x_0/2]=0\,.
\end{equation*}
Moreover,
\begin{equation*}
\mathbb{V}(x_1)=\mathbb{V}(\mathbb{E}[x_1|x_0])+\mathbb{E}[\mathbb{V}(x_1|x_0)]=\tau^2/4+\sigma^2\,.
\end{equation*}
Marginaly, $x_1$ is then distributed as a $\mathscr{N}(0,\tau^2/4+\sigma^2)$ variable, with the same
distribution as $x_0$ only if $\tau^2/4+\sigma^2=\tau^2$, i.e.~if $\tau^2=4\sigma^2/3$.

\begin{exoset}
Suppose that $(x_t)_{t\in\mathbb{N}}$ is a {\em Gaussian random walk}\index{Random walk} on $\mathbb{R}$:
$x_0\sim\mathscr{N}(0,\tau^2)$ and, for all $t\in\mathbb{N}$,
$$
x_{t+1}|\bx_{0:t} \sim \mathscr{N}(x_t,\sigma^2)\,,\qquad \sigma>0\,.
$$
Show that, whatever the value of $\tau^2$ is,
$(x_t)_{t\in\mathbb{N}}$ is not a (strictly) stationary process.
\end{exoset}

We have
\begin{equation*}
\mathbb{E}[x_1]=\mathbb{E}[\mathbb{E}[x_1|x_0]]=\mathbb{E}[x_0]=0\,.
\end{equation*}
Moreover,
\begin{equation*}
\mathbb{V}(x_1)=\mathbb{V}(\mathbb{E}[x_1|x_0])+\mathbb{E}[\mathbb{V}(x_1|x_0)]=\tau^2+\sigma^2\,.
\end{equation*}
The marginal distribution of $x_1$ is then a $\mathscr{N}(0,\tau^2+\sigma^2)$ distribution which cannot
be equal to a $\mathscr{N}(0,\tau^2)$ distribution.

\begin{exoset}
Give the necessary and sufficient condition under which an AR$(2)$ process with autoregressive polynomial
$\mathcal{P}(u)=1-\varrho_1 u-\varrho_2 u^2$ (with $\varrho_2\neq 0$) is causal.
\end{exoset}

We have
\begin{equation*}
\mathbb{E}[x_1]=\mathbb{E}[\mathbb{E}[x_1|x_0]]=\mathbb{E}[x_0/2]=0\,.
\end{equation*}
Moreover,
\begin{equation*}
\mathbb{V}(x_1)=\mathbb{V}(\mathbb{E}[x_1|x_0])+\mathbb{E}[\mathbb{V}(x_1|x_0)]=\tau^2/4+\sigma^2\,.
\end{equation*}
Marginaly, $x_1$ is then distributed as a $\mathscr{N}(0,\tau^2/4+\sigma^2)$ variable, with the same
distribution as $x_0$ only if $\tau^2/4+\sigma^2=\tau^2$, i.e.~if $\tau^2=4\sigma^2/3$.

\begin{exoset} Consider the process $(x_t)_{t\in\mathbb{N}}$ such that $x_0=0$ and, for all $t\in\mathbb{N}$,
$$
x_{t+1}|\bx_{0:t}\sim\mathscr{N}(\varrho\,x_t,\sigma^2)\,.
$$
Suppose that $\pi(\varrho,\sigma)=1/\sigma$ and that there is no constraint on $\varrho$.
Show that the conditional posterior distribution of $\varrho$, conditional on the observations 
$\bx_{0:T}$ and on $\sigma^2$, is a $\mathscr{N}(\mu_T,\omega_T^2)$ distribution with
$$
\mu_T = \sum_{t=1}^T x_{t-1}x_t\bigg/ \sum_{t=1}^T x_{t-1}^2
\quad\text{ and }\quad
\omega_T^2 = \sigma^2\bigg/ \sum_{t=1}^T x_{t-1}^2\,.
$$
Show that the marginal posterior distribution of $\varrho$ is a Student 
$\mathscr{T}(T-1,\mu_T,\nu_T^2)$ distribution with
$$
\nu_T^2 = \frac{1}{T-1}\,\left(\sum_{t=1}^T x_t^2 \bigg/ \sum_{t=0}^{T-1} x_t^2 - \mu_T^2 \right)\,.
$$
Apply this modeling to the Aegon series in {\bfseries Eurostoxx50} and evaluate its predictive abilities.
\end{exoset}

The posterior conditional density of $\varrho$ is proportional to
\begin{align*}
\prod_{t=1}^T &\exp\left\{ -(x_t-\varrho\,x_{t-1})^2 / 2\sigma^2 \right\} \\
        &\propto \exp\left\{ \left[- \varrho^2 \sum_{t=0}^{T-1} x_t^2
        + 2 \varrho \sum_{t=0}^{T-1} x_t x_{t+1} \right] \big/ 2\sigma^2 \right\}\,,
\end{align*}
which indeed leads to a $\mathscr{N}(\mu_T,\omega_T^2)$ conditional distribution as
indicated above.

Given that the joint posterior density of $(\varrho,\sigma)$ is proportional to
$$
\sigma^{-T-1} \prod_{t=1}^T \exp\left\{ -(x_t-\varrho\,x_{t-1})^2 / 2\sigma^2 \right\} \,
$$
integrating out $\sigma$ leads to a density proportional to
\begin{align*}
\int &\left(\sigma^2\right)^{-T/2-1/2}
\exp\left(\sum_{t=1}^T(x_t-\rho x_{t-1})^2/(2\sigma^2)\right) \text{d}\sigma\\
&=\int \left(\sigma^2\right)^{-T/2-1}
\exp\left(\sum_{t=1}^T(x_t-\rho x_{t-1})^2/(2\sigma^2)\right) \text{d}\sigma^2\\
&=\left\{ \sum_{t=1}^T (x_t-\varrho\,x_{t-1})^2  \right\}^{-T/2}
\end{align*}
when taking into account the Jacobian. We thus get a
Student $\mathscr{T}(T-1,\mu_T,\nu_T^2)$ distribution
and the parameters can be derived from expanding the sum of squares:
$$
\sum_{t=1}^T (x_t-\varrho\,x_{t-1})^2 = \sum_{t=0}^{T-1} x_t^2 \left(
\varrho^2  - 2 \varrho \mu_T \right) + \sum_{t=1}^{T} x_t^2
$$
into
\begin{align*}
\sum_{t=0}^{T-1} x_t^2 &(\varrho - \mu_T) ^2 + \sum_{t=1}^{T} x_t^2 - \sum_{t=0}^{T-1} x_t^2 \mu_T^2\\
&\propto \frac{(\varrho - \mu_T) ^2}{T-1}
+  \frac{1}{T-1} \left( \frac{\sum_{t=1}^{T} x_t^2}{\sum_{t=0}^{T-1} x_t^2} - \mu_T^2 \right) \\
&= \frac{(\varrho - \mu_T) ^2}{T-1} + \nu_T^2 \,.
\end{align*}

The main point with this example is that, when $\varrho$ is unconstrained, the joint posterior
distribution of $(\varrho,\sigma)$ is completely closed-form. Therefore, the predictive distribution
of $x_{T+1}$ is given by
$$
\int \frac{1}{\sqrt{2\pi}\sigma} \exp\{ -(x_{T+1}-\varrho x_T)^2/2\sigma^2 \}\,
        \pi(\sigma,\varrho|\bx_{0:T}) \text{d}\sigma \text{d}\varrho
$$
which has again a closed-form expression:
\begin{align*}
\int \frac{1}{\sqrt{2\pi}\sigma} \exp\{ &-(x_{T+1}-\varrho x_T)^2/2\sigma^2 \}\,
        \pi(\sigma,\varrho|\bx_{0:T}) \text{d}\sigma \text{d}\varrho\\
&\propto \int \sigma^{-T-2} \exp\{ -\sum_{t=0}^{T} (x_{t+1}-\varrho x_t)^2/2\sigma^2 \}
                                       \text{d}\sigma \text{d}\varrho\\
&\propto \int \left\{ \sum_{t=0}^T (x_{t+1}-\varrho\,x_{t})^2  \right\}^{-(T+1)/2} \text{d}\varrho\\
&\propto \left( \sum_{t=0}^T x_{t}^2 \right)^{-(T+1)/2} \int \left\{ \frac{(\varrho
        - \mu_{T+1}) ^2}{T} + \nu_{T+1}^2 \right\}^{-(T+2)/2} \text{d}\varrho\\
&\propto \left( \sum_{t=0}^T x_{t}^2 \right)^{-(T+1)/2}\, \nu_T^{-T-1} \\
&\propto \left( \sum_{t=0}^T x_{t}^2 \sum_{t=0}^T x_{t+1}^2 - \left\{ \sum_{t=0}^T x_{t}x_{t+1}\right\}^2
         \right)^{(T+1)/2}\,.
\end{align*}
This is a Student $\mathcal{T}(T,\delta_T,\omega_T)$ distribution, with
$$
\delta_T = x_T \sum_{t=0}^{T-1} x_{t}x_{t+1} / \sum_{t=0}^{T-1} x_{t}^2 = \hat \rho_T x_T
$$
and
$$
\omega_T = \left\{ \sum_{t=0}^T x_t^2 \sum_{t=0}^T x_t^2 - \left( \sum_{t=0}^T x_{t}x_{t+1}
\right)^2 \right\} \bigg/ T \sum_{t=0}^{T-1} x_{t}^2\,.
$$
The predictive abilities of the model are thus in providing a point estimate for the next
observation $\hat x_{T+1} = \hat \rho_T x_T$, and a confidence band around this value.

\begin{exoset}
For Algorithm 7.13,
show that, if the proposal on $\sigma^2$ is a log-normal distribution 
$\mathscr{LN}(\log(\sigma^2_{t-1}),\tau^2)$
and if the prior distribution on $\sigma^2$ is the noninformative prior $\pi(\sigma^2)=1/\sigma^2$, the acceptance ratio
also reduces to the likelihood ratio because of the Jacobian.
\end{exoset}

If we write the Metropolis--Hastings ratio for a current value $\sigma_0^2$ and a proposed value
$\sigma_1^2$, we get
$$
\frac{\pi(\sigma_1^2)\ell(\sigma_1^2)}{\pi(\sigma_0^2)\ell(\sigma_0^2)}\,
\frac{\exp\left( -(\log(\sigma^2_0-\log(\sigma^2_1))^2/2\tau^2\right)/\sigma_0^2}{
\exp\left( -(\log(\sigma^2_0-\log(\sigma^2_1))^2/2\tau^2\right)/\sigma_1^2}
= \frac{\ell(\sigma_1^2)}{\ell(\sigma_0^2)}\,,
$$
as indicated.

\begin{exoset}\label{exo:nolike}
Write down the joint distribution of $(y_t,x_t)_{t\in\mathbb{N}}$ in (7.19)
and deduce that the (observed) likelihood is not
available in closed form.
\end{exoset}

Recall that
$y_0\sim\mathcal{N}(0,\sigma^{2})$ and, for $t=1,\ldots,T$,
$$
\begin{cases}
y_t = \varphi y_{t-1} + \sigma \epsilon^*_{t-1}\,, &\cr
x_t = \beta e^{y_t/2} \epsilon_t\,, &\cr
\end{cases}
$$
where both $\epsilon_t$ and $\epsilon^*_t$ are iid $\mathcal{N}(0,1)$ random variables.
The joint distribution of $\left(\bx_{1:T},\by_{0:T}\right)$ is therefore
\begin{align*}
f\left(\bx_{1:T},\by_{0:T}\right)&=f\left(\bx_{1:T}|\by_{0:T}\right)f\left(\by_{0:T}\right)\\
&=\left(\prod_{i=1}^T f(x_i|y_i)\right)f(y_0)f(y_1|y_0)\ldots f(y_T|y_{T-1})\\
&=\frac{1}{\left(2\pi\beta^2\right)^{T/2}}\exp\left\{ -\sum_{t=1}^Ty_t/2\right)
\exp\left(-\frac{1}{2\beta^2}\sum_{t=1}^T x_t^2\exp(-y_t)\right)\\
&\quad\times\frac{1}{\left(2\pi\sigma^2\right)^{(T+1)/2}}\exp\left(-\frac{1}{2\sigma^2}
        \left(y_0^2+\sum_{t=1}^T\left(y_t-\phi y_{t-1}\right)^2\right)\right\}\,.
\end{align*}
Due to the double exponential term $\exp\left(-\frac{1}{2\beta^2}\sum_{t=1}^T x_t^2\exp(-y_t)\right)$, it is
impossible to find a closed-form of the integral in $\by_{0:T}$.

\begin{exoset}
Show that the stationary distribution of $\bx_{-p:-1}$ in an AR$(p)$ model is a 
$\mathscr{N}_p(\mu{\mathbf 1}_p,\mathbf{A})$ distribution, and
give a fixed point equation satisfied by the covariance matrix $\mathbf{A}$.
\end{exoset}

If we denote
$$
\bz_t=\left(x_t,x_{t-1},\ldots,x_{t+1-p}\right)\,,
$$
then
$$
\bz_{t+1}=\mu{\mathbf 1}_p+B\left(\bz_t-\mu{\mathbf 1}_p\right)+\epsilon_{t+1}\,.
$$
Therefore,
$$
\mathbb{E}\left[\bz_{t+1}|\bz_t\right]=\mu{\mathbf 1}_p+B\left(\bz_t-\mu{\mathbf 1}_p\right)
$$
and
$$
\mathbb{V}\left(\bz_{t+1}|\bz_t\right)=\mathbb{V}\left(\epsilon_{t+1}\right)=\left[
\begin{array}{llll}
 \sigma^2 & 0      & \ldots & 0 \\
 0        & 0      & \ldots & 0 \\
 \vdots   & \vdots & \vdots & \vdots \\
 0        & 0      & \ldots & 0
\end{array}
\right]=V\,.
$$
Then,
$$
\bz_{t+1}|\bz_{t}\sim\mathcal{N}_p\left(\mu{\mathbf 1}_p+B\left(\bz_t-\mu{\mathbf 1}_p\right),V\right)\,.
$$
Therefore, if $\bz_{-1}=\bx_{-p:-1}\sim\mathcal{N}_p\left(\mu{\mathbf 1}_p,A\right)$ is Gaussian, then
$\bz_t$ is Gaussian. Suppose that $\bz_t\sim\mathcal{N}_p(M,A)$, we get
$$
\mathbb{E}\left[\bz_{t+1}\right)=\mu{\mathbf 1}_p+B\left(M-\mu{\mathbf 1}_p\right]
$$
and $\mathbb{E}\left[\bz_{t+1}\right]=\mathbb{E}\left[\bz_t\right]$ if
$$
\mu{\mathbf 1}_p +B\left(M-\mu{\mathbf 1}_p\right)=M\,,
$$
which means that $M=\mu{\mathbf 1}_p$. Similarly,
$\mathbb{V}\left(\bz_{t+1}\right)=\mathbb{V}\left(\bz_t\right)$ if and only if
$$
BAB'+V=A\,,
$$
which is the ``fixed point" equation satisfied by $A$.

\begin{exoset}
Show that the posterior distribution on $\boldsymbol{\theta}$ associated 
with the prior $\pi(\boldsymbol{\theta})=1/\sigma^2$ and an AR$(p)$ model
is well-defined for $T > p$ observations.
\end{exoset}

The likelihood conditional on the initial values $\bx_{0:(p-1)}$ is proportional to
$$
\sigma^{-T+p-1} \prod_{t=p}^T \exp\left\{-\left(x_t-\mu - \sum_{i=1}^p \varrho_i
(x_{t-i}-\mu) \right)^2 \big/ 2\sigma^2 \right\}\,.
$$
A traditional noninformative prior is $\pi(\mu,\varrho_1,\ldots,\varrho_p,\sigma^2)=1/\sigma^2$.
In that case, the probability density of the posterior distribution is proportional to
$$
\sigma^{-T+p-3} \prod_{t=p}^T \exp\left\{-\left(x_t-\mu - \sum_{i=1}^p \varrho_i
        (x_{t-i}-\mu) \right)^2 \big/ 2\sigma^2 \right\}\,.
$$
And
$$
\int (\sigma^2)^{-(T-p+3)/2} \prod_{t=p}^T \exp\left\{-\left(x_t-\mu - \sum_{i=1}^p
  \varrho_i (x_{t-i}-\mu) \right)^2 \big/ 2\sigma^2 \right\}\text{d}\sigma^2<\infty
$$
holds for $T-p+1>0$, i.e., $T>p-1$. This integral is equal to
$$
\left\{-\left(x_t-\mu - \sum_{i=1}^p
  \varrho_i (x_{t-i}-\mu) \right)^2 \big/ 2\sigma^2 \right\}^{(p-T-1)/2}\,,
$$
which is integrable in $\mu$ for $T-p>0$, i.e.~$T > p$.
The other parameters $\varrho_j$ $(j=1,\ldots,p0$ being bounded, the remaining integrand
is clearly integrable in $\boldsymbol{\varrho}$.

\begin{exoset}\label{exo:root2pol}
Show that the coefficients of the polynomial $\mathcal{P}$ in (7.15)
associated with an AR$(p)$ model can be derived in $\hbox{O}(p^2)$
time from the inverse roots $\lambda_i$ using the recurrence relations $(i=1,\ldots,p,j=0,\ldots,p)$
$$
\psi^i_0=1\,,\qquad \psi^i_j = \psi^{i-1}_j-\lambda_i\psi_{j-1}^{i-1}\,,
$$
where $\psi^0_0=1$ and $\psi^i_j=0$ for $j>i$, and setting $\varrho_j= -\psi^p_j$ $(j=1,\ldots,p)$.
\end{exoset}

Since
$$
\prod_{i=1}^p (1-\lambda_i x) = 1 - \sum_{j=1}^j \varrho_j x^j\,,
$$
we can expand the lhs one root at a time. If we set
$$
\prod_{j=1}^i (1-\lambda_j x) = \sum_{j=0}^i \psi^i_j x^j\,,
$$
then
\begin{eqnarray*}
\prod_{j=1}^{i+1} (1-\lambda_j x) &=&  (1-\lambda_{i+1} x) \prod_{j=1}^i (1-\lambda_j x)\\
        &=& (1-\lambda_{i+1} x) \sum_{j=0}^i \psi^i_j x^j \\
        &=& 1 + \sum_{j=1}^i (\psi^i_j - \lambda_{i+1} \psi^i_{j-1}) x^j -\lambda_{i+1} \psi^i_i x^{i+1} \,,
\end{eqnarray*}
which establishes the $\psi^{i+1}_j = \psi^{i}_j-\lambda_{i+1}\psi_{j-1}^{i}$ recurrence relation.

This recursive process requires the allocation of $i$ variables at the $i$th stage; the coefficients of
$\mathcal{P}$ can thus be derived with a complexity of $\hbox{O}(p^2)$.

\begin{exoset}\label{exo:schur+cohn}
Given the polynomial $\mathcal{P}$ in (7.5),
the fact that all the roots are outside the unit
circle can be determined without deriving the roots, thanks to the Schur--Cohn test.
If $\mathcal{A}_p=\mathcal{P}$, a recursive definition of decreasing degree polynomials is $(k=p,\ldots,1)$
$$
u\mathcal{A}_{k-1}(u)=\mathcal{A}_{k-1}(u)-\varphi_k \mathcal{A}_{k}^\star(u)\,,
$$
where $\mathcal{A}_{k}^\star$ denotes the reciprocal polynomial $\mathcal{A}_{k}^\star(u)=u^k
\mathcal{A}_{k-1}(1/u)$.
\begin{enumerate}
\item Give the expression of $\varphi_k$ in terms of the coefficients of $\mathcal{A}_k$.
\item Show that the degree of $\mathcal{A}_k$ is at most $k$.
\item If $a_{m,k}$ denotes the $m$-th degree coefficient in $\mathcal{A}_k$,
show that $a_{k,k}\ne 0$ for $k=0,\ldots,p$ if, and only if, $a_{0,k}\ne a_{k,k}$ for all $k$'s.
\item Check by simulation that, in cases when $a_{k,k}\ne 0$ for $k=0,\ldots,p$, the roots are outside the unit
circle if, and only if, all the coefficients $a_{k,k}$ are positive.
\end{enumerate}
\end{exoset}

{\bf Note:} The above exercise is somewhat of a mystery (!) in that we cannot remember how it ended up in this exercise list,
being incorrect and incomplete as stated. A proper substitute is given below:

\begin{rema}
\noindent{\bf \ref{exo:schur+cohn}}
Given a polynomial $\mathcal{P}$ of degree $k$, its reciprocal polynomial $\mathcal{P}_{k}^\star$ is defined as
$$
\mathcal{P}^\star(u)=u^k \mathcal{P}_{k-1}(1/u)\,.
$$
Assuming $\mathcal{P}(0)=1$, the Schur transform of $\mathcal{P}$ is defined by
$$
T\mathcal{P}(u)=\dfrac{\mathcal{P}(z)-\mathcal{P}^\star(0)\mathcal{P}^\star(z)}{1-\mathcal{P}^\star(0)^2}\,.
$$
\begin{enumerate}
\item Show that the roots of $\mathcal{P}$ and $\mathcal{P}_{k}^\star$ are inverses.
\item Show that the degree of $T\mathcal{P}$ is at most $k-1$.
\item Show that $T\mathcal{P}(0)=1$.
\item Check by a simulation experiment producing random polynomials the property that, when $T\mathcal{P}(0)>1$, $T\mathcal{P}$ and $T\mathcal{P}$
have the same number of roots inside the unit circle.
\item Denote $T^n\mathcal{P}=T(T^{n-1}\mathcal{P})$, for $d\ne k$, and $\kappa$ the first index with $T^\kappa
\mathcal{P}=0$. Deduce from the above property that, if $T^n\mathcal{P}>0$ for $n=1,\ldots,\kappa$, then $\mathcal{P}$
has no root inside the unit circle.
\end{enumerate}
\end{rema}

\begin{enumerate}
\item If we write the inverse root decomposition of $\mathcal{P}$ as
$$
\mathcal{P}(u) = \prod_{i=1}^k (1-\lambda_i u)\,,
$$
since $\mathcal{P}(0)=1$, we have
$$
\mathcal{P}^\star(u)=u^k \prod_{i=1}^k (1-\lambda_i u^{-1}) = \prod_{i=1}^k (u-\lambda_i) = \prod_{i=1}^k
(1-\lambda_i^{-1} u)\,.
$$
\item By definition, if $\mathcal{P}(u)=\sum_{i=0}^k \alpha_i u^i$, then
$$
\mathcal{P}^\star(u)=\sum_{i=0}^k \alpha_{k-i} u^i\,,
$$
$\mathcal{P}^\star(0)=\alpha_k$, and 
\begin{align*}
\mathcal{P}(u)-\mathcal{P}^\star(0)\mathcal{P}^\star(u) &= \alpha_k u^k + \sum_{i=1}^{k-1} \alpha_i u^i
-\alpha_ku^k-\alpha_k\sum_{i=1}^{k-1}\alpha_{k-i} u^i \\
&= \sum_{i=1}^{k-1} [\alpha_i-\alpha_k\alpha_{k-i}] u^i
\end{align*}
is at most of degree $k-1$.
\item Since 
$$
\mathcal{P}(0)-\mathcal{P}^\star(0)\mathcal{P}^\star(0)=1-\alpha_k^2\,,
$$ 
$T\mathcal{P}(0)=1$.
\item A simulation experiment can be designed around the following code:
\begin{verbatim}
k=10
# random coefficients
Coef=c(1,runif(k,-1,1))
Schur=Coef-Coef[k]*rev(Coef)
print(sum(Mod(polyroot(Coef))<1)-sum(Mod(polyroot(Schur))<1))
\end{verbatim}
Repeating this code a large number of times does not produce anything but zero's.
\item By virtue of the above result, $\mathcal{P}, T\mathcal{P}, \ldots, T^{\kappa-1}\mathcal{P}$ have the same number
of roots inside the unit circle if $T^n\mathcal{P}>0$ for $n=1,\ldots,\kappa-1$. Since
$$
T^{\kappa-1}\mathcal{P}=1-\{\alpha^\kappa_1\}^2=1-\lambda_1^2\,,
$$
the last root is outside the unit disk and hence so are the others.
\item Extending the above code leads to
\begin{verbatim}
k=10
# Schur sequence
Coef=matrix(0,nrow=k+1,ncol=k+1)
# initial polynomial
Coef[,k+1]=c(1,rnorm(k,sd=1/k))
for (t in k:1)
  Coef[1:t,t]=(Coef[1:(t+1),t+1]-Coef[t+1,t+1]*Coef[(t+1):1,
           t+1])/(1-Coef[t+1,t+1]^2)
while (prod(diag(Coef[1,]^2)<1)==0){
  Coef=matrix(0,nrow=k+1,ncol=k+1)
  Coef[,k+1]=c(1,rnorm(k,sd=1/k))
  for (t in k:1)
    Coef[1:t,t]=(Coef[1:(t+1),t+1]-Coef[t+1,t+1]*Coef[(t+1):1,
       t+1])/(1-Coef[t+1,t+1]^2)
  }
print(min(Mod(polyroot(Coef[,k+1]))))
\end{verbatim}
Repeated calls to this code consistently exhibit root modules larger than 1.
\end{enumerate}

\begin{exoset}\label{exo:MA_oto}
For an MA$(q)$ process, show that $(s\le q)$
$$
\gamma_x(s) = \sigma^2 \sum_{i=0}^{q-|s|} \vartheta_i \vartheta_{i+|s|}\,.
$$
\end{exoset}

We have
\begin{eqnarray*}
\gamma_x(s) & = & \mathbb{E}\left[x_t x_{t-s}\right] \\
            & = & \mathbb{E}\left[\left[\epsilon_t+\vartheta_1\epsilon_{t-1}+\ldots+\vartheta_q\epsilon_{t-q}\right]
                  \left[\epsilon_{t-s}+\vartheta_1\epsilon_{t-s-1}+\ldots+\vartheta_q\epsilon_{t-s-q}\right]\right]\,.
\end{eqnarray*}
Then, if $1\le s\le q$,
$$
\gamma_x(s)=\left[\vartheta_s+\vartheta_{s+1}\vartheta_1+\ldots+\vartheta_q\vartheta_{q-s}\right]\sigma^2
$$
and
$$
\gamma_x(0)=\left[1+\vartheta_1^2+\ldots+\vartheta_q^2\right]\sigma^2\,.
$$
Therefore, if $(0\le s\le q)$ with the convention that $\vartheta_0=1$
$$
\gamma_x(s) = \sigma^2 \sum_{i=0}^{q-s} \vartheta_i \vartheta_{i+s}\,.
$$
The fact that $\gamma_x(s)=\gamma_x(-s)$ concludes the proof.

\begin{exoset}\label{exo:norminov}
Show that the conditional distribution of $(\epsilon_0,\ldots,\epsilon_{-q+1})$
given both $\bx_{1:T}$ and the parameters is a normal distribution.
Evaluate the complexity of computing the mean and covariance matrix of this distribution.
\end{exoset}

The distribution of $\bx_{1:T}$ conditional on $(\epsilon_0,\ldots,\epsilon_{-q+1})$ is
proportional to
$$
\sigma^{-T} \prod_{t=1}^T\exp\left\{-\left(x_t-\mu+ \sum_{j=1}^q
\vartheta_j\widehat \epsilon_{t-j} \right)^2\bigg/ 2\sigma^2 \right\} \,,
$$
Take
$$
(\epsilon_0,\ldots,\epsilon_{-q+1})\sim\mathcal{N}_q\left(0_q,\sigma^2I_q\right)\,.
$$
In that case, the  conditional distribution of $(\epsilon_0,\ldots,\epsilon_{-q+1})$
given $\bx_{1:T}$ is proportional to
$$
\prod_{i=-q+1}^0 \exp\left\{-\epsilon_i^2/2\sigma^2\right\}\,
\prod_{t=1}^T \exp\left\{-\widehat\epsilon_t^2/2\sigma^2\right\}\,.
$$
Due to the recursive definition of $\hat \epsilon_t$,
the computation of the mean and the covariance matrix of this distribution is too costly to be available
for realistic values of $T$. For instance, getting the conditional mean of $\epsilon_i$ requires deriving the
coefficients of $\epsilon_i$ from all terms
$$
\left(x_t-\mu+ \sum_{j=1}^q\vartheta_j\widehat \epsilon_{t-j}\right)^2
$$
by exploiting the recursive relation
$$
\widehat \epsilon_t = x_t -\mu + \sum_{j=1}^q \vartheta_j \widehat\epsilon_{t-j}\,.
$$
If we write $\widehat \epsilon_1 = \delta_1 + \beta_1 \epsilon_i$ and
$\widehat \epsilon_t = \delta_t + \beta_t \epsilon_i$, then we need to
use the recursive formula
$$
\delta_t = x_t - \mu + \sum_{j=1}^q \vartheta_j \delta_{t-j}\,,
\qquad
\beta_t = \sum_{j=1}^q \beta_{t-j}\,,
$$
before constructing the conditional mean of $\epsilon_i$. The corresponding cost for this
single step is therefore $\mbox{O}(Tq)$ and therefore $\mbox{O}(qT^2)$ for the whole series
of $\epsilon_i$'s. Similar arguments can be used for computing the conditional variances.

\begin{exoset}\label{exo:conMa}
Give the conditional distribution of $\epsilon_{-t}$ given the other $\epsilon_{-i}$'s,
$\bx_{1:T}$, and the $\widehat\epsilon_i$'s. Show that this distribution only depends on the other $\epsilon_{-i}$'s,
$\bx_{1:q-t+1}$, and $\widehat\epsilon_{1:q-t+1}$.
\end{exoset}

The distribution of $\bx_{1:T}$ conditional on $(\epsilon_0,\ldots,\epsilon_{-q+1})$ is
proportional to
$$
\sigma^{-T} \prod_{t=1}^T\exp\left\{-\left(x_t-\mu+ \sum_{j=1}^q
\vartheta_j\widehat \epsilon_{t-j} \right)^2\bigg/ 2\sigma^2 \right\} \,,
$$
Take
$$
(\epsilon_0,\ldots,\epsilon_{-q+1})\sim\mathcal{N}_q\left(0_q,\sigma^2I_q\right)\,.
$$
In that case, the  conditional distribution of $(\epsilon_0,\ldots,\epsilon_{-q+1})$
given $\bx_{1:T}$ is proportional to
$$
\prod_{i=-q+1}^0 \exp\left\{-\epsilon_i^2/2\sigma^2\right\}\,
\prod_{t=1}^T \exp\left\{-\widehat\epsilon_t^2/2\sigma^2\right\}\,.
$$
Due to the recursive definition of $\hat \epsilon_t$,
the computation of the mean and the covariance matrix of this distribution is too costly to be available
for realistic values of $T$. For instance, getting the conditional mean of $\epsilon_i$ requires deriving the
coefficients of $\epsilon_i$ from all terms
$$
\left(x_t-\mu+ \sum_{j=1}^q\vartheta_j\widehat \epsilon_{t-j}\right)^2
$$
by exploiting the recursive relation
$$
\widehat \epsilon_t = x_t -\mu + \sum_{j=1}^q \vartheta_j \widehat\epsilon_{t-j}\,.
$$
If we write $\widehat \epsilon_1 = \delta_1 + \beta_1 \epsilon_i$ and
$\widehat \epsilon_t = \delta_t + \beta_t \epsilon_i$, then we need to
use the recursive formula
$$
\delta_t = x_t - \mu + \sum_{j=1}^q \vartheta_j \delta_{t-j}\,,
\qquad
\beta_t = \sum_{j=1}^q \beta_{t-j}\,,
$$
before constructing the conditional mean of $\epsilon_i$. The corresponding cost for this
single step is therefore $\mbox{O}(Tq)$ and therefore $\mbox{O}(qT^2)$ for the whole series
of $\epsilon_i$'s. Similar arguments can be used for computing the conditional variances.

\begin{exoset}\label{exo:miliT}
Show that the (useful) predictive horizon for the MA$(q)$ model is restricted to the first $q$ future observations $x_{t+i}$.
\end{exoset}

Obviously, due to the lack of correlation between $x_{T+q+j}$ $(j>0)$ and $\bx_{1:T}$ we have
$$
\mathbb{E}\left[x_{T+q+1}|\bx_{1:T}\right]
=\mathbb{E}\left[x_{T+q+1}\right]=0
$$
and therefore the $MA(q)$ model has no predictive ability further than horizon $q$.

\begin{exoset}\label{exo:SSM=HMM}
Show that the system of equations given by (7.13) and (7.14)
induces a Markov chain on the completed variable $(\mathbf{x}_t,\mathbf{y}_t)$. Deduce that state-space models 
are special cases of hidden Markov models.
\end{exoset}

Given the time-dependence structure
\begin{eqnarray*}
\bx_t &=& G {\mathbf y}_t + \boldsymbol{\varepsilon}_t\,,\\ 
{\mathbf y}_{t+1} &=& F {\mathbf y}_t + \boldsymbol{\xi}_t\,, 
\end{eqnarray*}
we can write
$$
\left(\begin{matrix}\bx_t\\ \by_{t+1}\end{matrix}\right) =
\left(\begin{matrix}O &G\\ O &F\end{matrix}\right)\left(\begin{matrix}\bx_{t-1}\\ \by_{t}\end{matrix}\right) +
\left(\begin{matrix}\boldsymbol{\varepsilon}_{t}\\ \boldsymbol{\xi}_{t}\end{matrix}\right) \,.
$$
Since the noises $\boldsymbol{\xi}_{t}$ and $\boldsymbol{\varepsilon}_{t}$ are independent, the full vector 
$(\bx_t,\by_{t+1})$ is indeed a Markov chain. The subchain $(\by_t)$ is also a Markov chain on itd own.
And observing {\em only} $\bx_t$ means that we are observing a hidden Markov chain, in the sense of Figure 7.7 in the
book.

\begin{exoset}\label{exo:hmm=mix}
Show that, for a hidden Markov model, 
when the support $\mathcal{Y}$ is finite and when $(y_t)_{t\in\mathbb{N}}$ is stationary,
the marginal distribution of $x_t$ is the same mixture distribution for all $t$'s.
Deduce that the same identifiability problem as in mixture models occurs in this setting.
\end{exoset}

Since the marginal distribution of $x_t$ is given by
$$
\int f(x_t|y_t) \pi(y_t)\,\text{d}y_t = \sum_{y\in\mathcal{Y}} \pi(y) f(x_t|y)\,,
$$
where $\pi$ is the stationary distribution of $(y_t)$, this is indeed a mixture
distribution. Although this is not the fundamental reason for the unidentifiability
of hidden Markov models, there exists an issue of label switching similar to the
case of standard mixtures.

\begin{exoset}\label{exo:lineatime}
Given a hidden Markov chain $(x_t,y_t)$ with both $x_t$ and $y_t$ taking a finite number of possible values,
$k$ and $\kappa$, show that the time required for the simulation of $T$ consecutive observations is in
$\text{O}(k\kappa T)$.
\end{exoset}

{\bf Note:} The order indicated in the exercise should be $\text{O}(\kappa^2 T)$, for the distribution
conditional on the observed $x_t$'s.

For direct simulation, given the hidden chain at time $t$, $y_t$, simulating $y_{t+1}$ requires up to $k$ comparisons with a uniform variate.
Given $y_{t+1}$, simulating $x_{t+1}$ involves another maximum of $\kappa$ comparisons with a uniform variate. Repeating
those steps $T$ times leads to a $\text{O}(\{k+\kappa\}T)$ time.

For inverse simulation, that is, after observing $(x_1,\ldots,x_T)$, the joint conditional distribution of
$(y_1,\ldots,y_T)$ is given by
$$
p(y_1,\ldots,y_T|x_1,\ldots,x_T)\propto p_0(y_1)p(y_2|y_1)\cdots p(y_T|y_{T-1}) p(x_1|y_1) \cdots p(x_T|y_T)\,,
$$
which takes $\kappa^T$ values. 

However, if we use the backward formula described in the book, we could gain some time. If we get back to the defintion of the backward
formula, the distribution of $y_T$ given the past being only conditional on $y_{T-1}$,
$p(y_T|y_{T-1},\mathbf{x}_{0:T})$, takes $\kappa^2$ values. Then, for each previous hidden state, $y_t$, 
$p(y_t|y_{t-1},\mathbf{x}_{0:T})$ involves a summation of $\kappa$ terms for all pairs $(y_{t-1},y_t)$. But the
summation 
$$
\sum_{i=1}^\kappa p^\star_{t+1}(i|y_t,\mathbf{x}_{1:T})
$$
only depends on $y_t$, thus has to be computed $\kappa$ times, to be later multiplied by $p_{y_{t-1}y_t}$. Therefore the
cost of producing $p(y_t|y_{t-1},\mathbf{x}_{0:T})$ is again of order $\kappa^2$. At last, $p(y_0|\mathbf{x}_{0:T})$
requires $\kappa$ summations of $\kappa$ terms, thus is again of order $\kappa^2$.  This confirms that the overall cost
is in $\text{O}(\kappa^2 T)$ and that the number of possible values of the $x_t$'s is irrelevant.

\begin{exoset}\label{exo:chibakov}
Implement Chib's method of Section 6.8
in the case of a doubly finite hidden Markov chain.
First, show that an equivalent to the approximation (6.9)
is available for the denominator of (6.8).
Second, discuss whether or not the label switching issue also rises in this framework. Third,
apply this approximation to {\bfseries Dnadataset}.
\end{exoset}

In a hidden Markov model $(x_t,y_t)$, $y_t$ being the hidden part, when the parameters are unknown, it is usually the
case that the full posterior distribution of the parameter $\pi(\mathbf{p},\mathbf{q}|\bx,\by)$ is available in closed form. In
particular, as shown in Algorithm 7.15, this full posterior distribution is a product of $\kappa$ Beta
distributions on the $p_{i\cdot}$'s and of $\kappa$ Dirichlet distributions on the $q_{i\cdot}$'s $(i=1,2)$.

As alluded to in the book, it is also a setting where label switching occurs. Indeed, the introduction of states 1 and 2
in the hidden chain does not identify which state is which. The posteriors on $\mathbf{q}^1$ and $\mathbf{q}^2$ should
therefore be the same. Since the Gibbs sampler does not produce such symmetry on Figure 7.9, it is quite likely that
Chib's approximation will be biased in this setting.

The implementation for {\bfseries Dnadataset} of the Chib involves picking the highest likelihood value for
$\theta=(\mathbf{q}^1,\mathbf{q}^2,\mathbb{P})$ and averaging the full conditionals of $\theta$ given the hidden chain
over the Gibbs iterations.

\begin{exoset}\label{exo:witch_pred}
Show that the counterpart of the prediction filter in the Markov-switching case is given by
$$
\log p(\bx_{1:t}) = \sum_{r=1}^{t} \log \left[\sum_{i=1}^{\kappa}  
f(x_r|x_{r-1},y_r=i) \varphi_r(i)\right]\,,
$$
where $\varphi_r(i)=\mathbb{P}(y_r=i|\bx_{1:r-1})$ is given by the  
recursive formula
$$
\varphi_r(i) \propto \sum_{j=1}^\kappa p_{ji} f(x_{r-1}| 
x_{r-2},y_{r-1}=j) \varphi_{r-1}(j)\,.
$$
\end{exoset}

This exercise is more or less obvious given the developments provided in the book. The distribution of
$y_r$ given the past values $\bx_{1:r-1}$ is the marginal of $(y_r,y_{r-1})$ given the past values $\bx_{1:r-1}$:
\begin{eqnarray*}
\mathbb{P}(y_r=i|\bx_{1:t-1}) &=& \sum_{j=1}^\kappa \mathbb{P}(y_r=i,y_{r-1}=j|\bx_{1:r-1})\\
        &=& \sum_{j=1}^\kappa \mathbb{P}(y_{r-1}=j|\bx_{1:r-1})\,\mathbb{P}(y_r=i|y_{r-1}=j)\\
        &\propto& \sum_{j=1}^\kappa  p_{ji} \mathbb{P}(y_{r-1}=j,x_{r-1}|\bx_{1:r-2})\\
        &=& \sum_{j=1}^\kappa  p_{ji} \mathbb{P}(y_{r-1}=j,|\bx_{1:r-2}) f(x_{r-1}|x_{r-2},y_{r-1}=j)\,,
\end{eqnarray*}
which leads to the update formula for the $\varphi_r(i)$'. The marginal distribution $\bx_{1:t}$ is then
derived by
\begin{eqnarray*}
p(\bx_{1:t}) &=& \prod_{r=1}^t p(x_r|\bx_{1:(r-1)}) \\
             &=& \prod_{r=1}^t \sum_{j=1}^\kappa \mathbb{P}(y_{r-1}=j,x_r|\bx_{1:r-1}) \\
             &=& \prod_{r=1}^t \sum_{j=1}^\kappa f(x_r|x_{r-1},y_r=i) \varphi_r(i)\,,
\end{eqnarray*}
with the obvious convention $\varphi_1(i)=\pi_i$, if $(\pi_1,\ldots,\pi_\kappa)$ is the
stationary distribution associated with $\mathbb{P}=(p_{ij})$.

\chapter{Image Analysis}\label{ch:spt}
\begin{exoset}\label{exo:pascom}
Find two conditional distributions $f(x|y)$\index{Theorem!Hammersley--Clifford}
and $g(y|x)$ such that there is no joint distribution corresponding to
both $f$ and $g$. Find a necessary condition for $f$ and $g$ to be compatible in that respect; 
{\em i.e.}, to correspond to a joint distribution on $(x,y)$.
\end{exoset}

As stated, this is a rather obvious question: if $f(x|y)=4y\exp(-4yx)$ and if $g(y|x)=6x\exp(-6xy)$, there cannot
be a joint distribution inducing these two conditionals. What is more interesting is that, if
$f(x|y)=4y\exp(-4yx)$ and $g(y|x)=4x\exp(-4yx)$, there still is no joint distribution, despite the formal
agreement between both conditionals: the only joint that would work has the major drawback that it has an
infinite mass!

\begin{exoset}\label{exo:klimt}
Using the Hammersley--Clifford theorem, show that the full conditional distributions given by (8.3)
are compatible with a joint distribution. Deduce that the Ising model is a Markov random field.
\end{exoset}

\noindent{\bf Note:} In order to expose the error made in the earlier printing of {\em Bayesian Core}, namely using
the size of the symmetrized neighborhood, $N_k(i)$, in the full conditoinal, we will compute here
the potential joint distribution based on the pseudo-conditional
$$
\mathbb{P}(y_i=C_j|\mathbf{y}_{-i},\mathbf{X},\beta,k) \propto \exp\left(\beta \sum_{\ell\sim_k i}
\mathbb{I}_{C_j}(y_\ell)\bigg/N_k(i) \right)\,,
$$
even though it is defined for $N_k(i)=1$ in the book.

It follows from (8.4) that, if there exists a joint distribution, it satisfies
$$
\mathbb{P}(\by|\mathbf{X},\beta,k) \propto \prod_{i=0}^{n-1}
\frac{\mathbb{P}(y_{i+1}  |y_1^*,\ldots,y_i^*,y_{i+2},\ldots,y_n,\mathbf{X},\beta,k)}{
      \mathbb{P}(y_{i+1}^*|y_1^*,\ldots,y_i^*,y_{i+2},\ldots,y_n,\mathbf{X},\beta,k)}\,.
$$
Therefore,
\begin{align*}
\mathbb{P}(\by|\mathbf{X},\beta,k) \propto
&\exp\left\{ \beta \sum_{i=1}^{n} \frac{1}{N_k(i)} \left(
\sum_{\ell<i,\ell\sim_k i} \left[ \mathbb{I}_{y_\ell^*}(y_i)-\mathbb{I}_{y_\ell^*}(y_i^*)\right] + \right.\right.\\
&\qquad\left.\left.
\sum_{\ell>i,\ell\sim_k i} \left[\mathbb{I}_{y_\ell}(y_i)-\mathbb{I}_{y_\ell}(y_i^*)\right] \right) \right\}
\end{align*}
is the candidate joint distribution. Unfortunately, if we now try to derive the conditional distribution of
$y_j$ from this joint, we get
\begin{align*}
\mathbb{P}(y_i=C_j|\mathbf{y}_{-i},\mathbf{X},\beta,k) \propto \exp\beta
\left\{ \frac{1}{N_k(j)} \sum_{\ell>j,\ell\sim_k j} \mathbb{I}_{y_\ell}(y_j) +
\sum_{\ell<j,\ell\sim_k j} \frac{\mathbb{I}_{y_\ell}(y_j)}{N_k(\ell)} \right.\\
+\left. \frac{1}{N_k(j)} \sum_{\ell<j,\ell\sim_k j} \mathbb{I}_{y_\ell^*}(y_j) -
\sum_{\ell<j,\ell\sim_k j} \frac{\mathbb{I}_{y_\ell^*}(y_j)}{N_k(\ell)} \right\}
\end{align*}
which differs from the orginal conditional if the $N_k(j)$'s differ. In conclusion,
there is no joint distribution if (8.3) is defined as in the earlier edition.
Taking all the $N_k(j)$'s equal to $1$ leads to a coherent joint distribution since the
last line in the above equation cancels.

\begin{exoset}\label{exo:B>0}
If a joint density $\pi(y_1,...,y_n)$ is such that the conditionals
$\pi(y_{-i}|y_i)$ never cancel on the supports of the marginals
$m_{-i}(y_{-i})$, show that the support of $\pi$ is equal to the 
Cartesian product of the supports of the marginals.
\end{exoset}

Let us suppose that the support of $\pi$ is not equal to the product of the supports of the marginals.
(This means that the support of $\pi$ is smaller than this product.) Then the conditionals $\pi(\by_{-i}|y_i)$
cannot be positive everywhere on the support of $m(\by_{-i})$.

\begin{exoset}\label{exo:kliklak}
Describe the collection of cliques $\mathcal{C}$ for an $8$ neighbor
neighborhood structure such as in Figure 8.2 on a regular
$n\times m$ array. Compute the number of cliques.
\end{exoset}

\begin{figure}[h]
\centerline{\includegraphics[width=.7\textwidth]{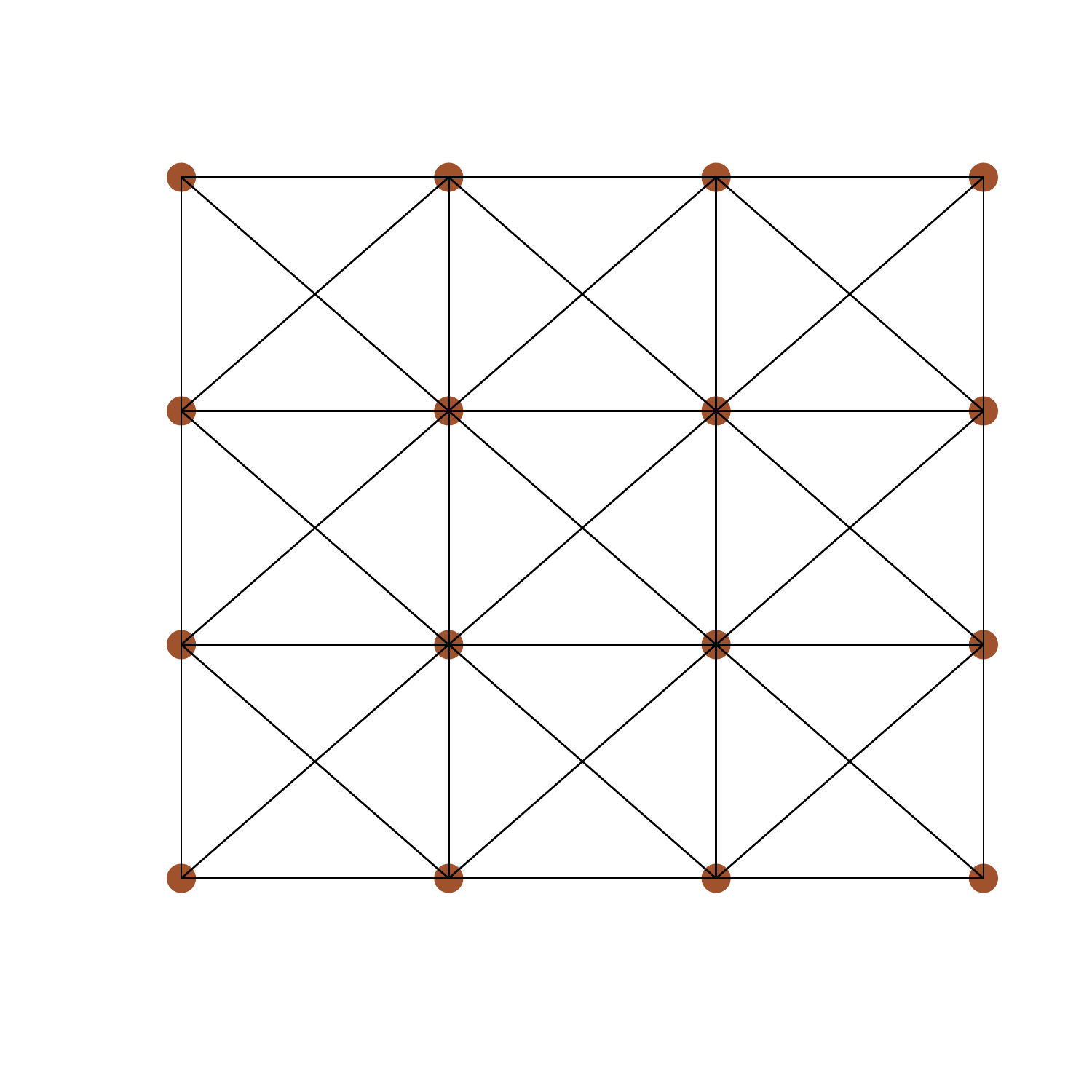}}
\caption{\label{fig:8night}
Neighborhood relations between the points of a $4\times 4$ regular grid for a $8$ neighbor
neighborhood structure.}
\end{figure}

If we draw a detailed graph of the connections on a regular grid as in Figure \ref{fig:8night} in this manual,
then the maximal structure such that all members are neighbors is made of $4$ points. Cliques are
thus made of squares of $4$ points and there are $(n-1)\times(m-1)$ cliques on a $n\times m$ array.

\begin{exoset}\label{exo:thecostofpotts}
Draw the function $Z(\beta)$ for a $3\times 5$ array.
Determine the computational cost of the derivation of
the normalizing constant $Z(\beta)$ of (8.4) for an $m\times n$ array.
\end{exoset}

The function $Z(\beta)$ is defined by
$$
Z(\beta) = 1 \bigg/ \sum_{\bx\in\mathcal{X}} \exp\left(\beta \sum_{j\sim i} \mathbb{I}_{x_j=x_i}\right)\,,
$$
which involves a summation over the set $\mathcal{X}$ of size $2^{15}$. The {\sf R} code corresponding
to this summation is
\begin{verbatim}
neigh=function(i,j){    #Neighbourhood indicator function
   (i==j+1)||(i==j-1)||(i==j+5)||(i==j-5)
}

zee=function(beta){
  val=0
  array=rep(0,15)
  for (i in 1:(2^15-1)){
    expterm=0
    for (j in 1:15)
      expterm=expterm+sum((array==array[j])*neigh(i=1:15,j=j))
    val=val+exp(beta*expterm)
    j=1
    while (array[j]==1){
        array[j]=0
        j=j+1 }
      array[j]=1 }
  expterm=0
  for (j in 1:15)
      expterm=expterm+sum((array==array[j])*neigh(i=1:15,j=j))
  val=val+exp(beta*expterm)
  1/val }
\end{verbatim}

\begin{figure}
\centerline{\includegraphics[width=.7\textwidth]{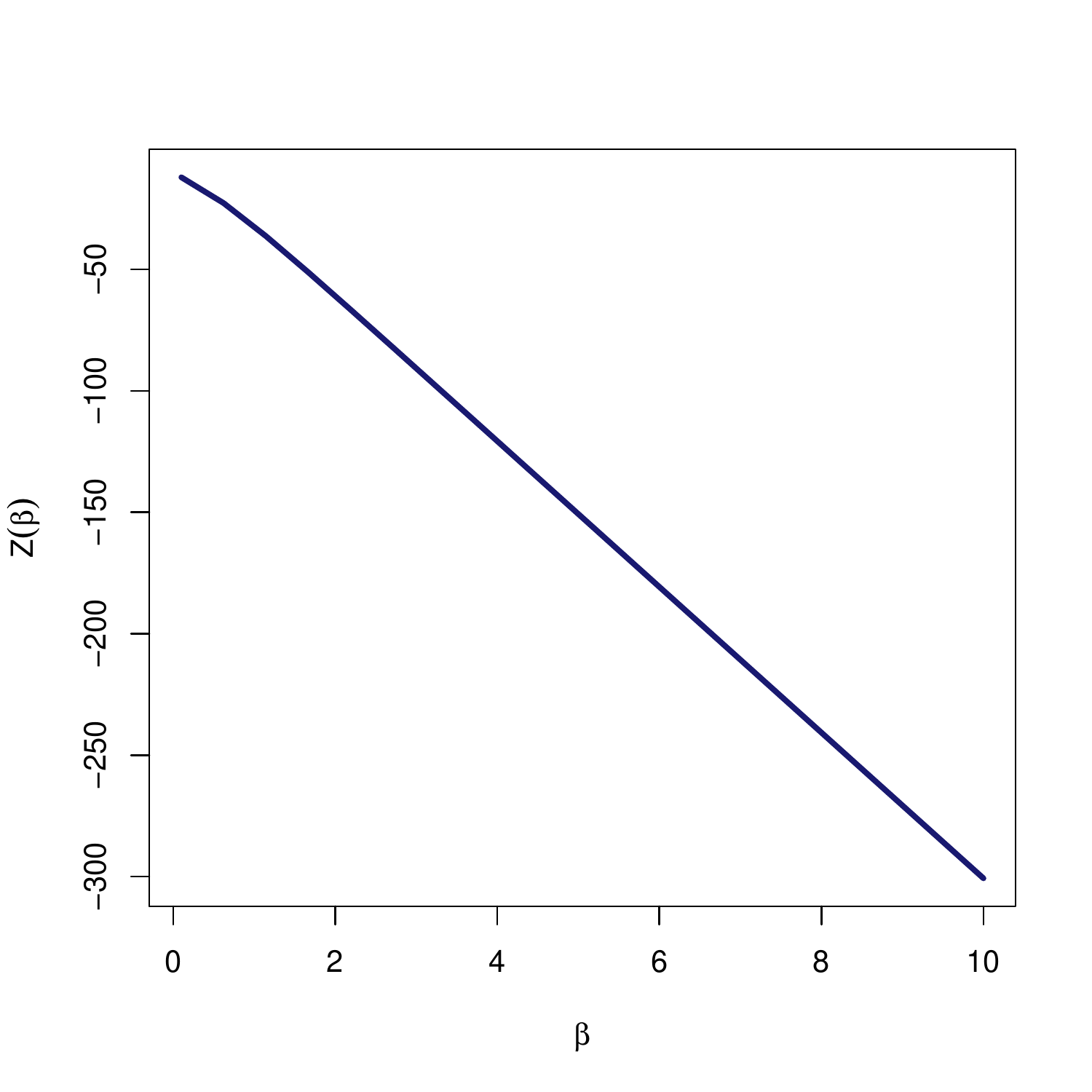}}
\caption{\label{fig:zee}
Plot of the function $Z(\beta)$ for a $3\times 5$ array with a four neighbor structure.}
\end{figure}

It produces the (exact) curve given in Figure \ref{fig:zee} in this manual.

In the case of a $m\times n$ array, the summation involves $2^{m\times n}$ and each exponential term
in the summation requires $(m\times n)^2$ evaluations, which leads to a $\text{O}((m\times n)^2\,
2^{m\times n})$ overall cost.

\begin{exoset}\label{exo:touchpa@monPott}
Show that the joint distribution (8.5) is indeed compatible with the full conditionals of the Potts
model. Can you derive this joint distribution from the Hammersley--Clifford representation (8.1)?
\end{exoset}

If we defined the joint distribution as
$$
\pi(\bx)\propto \exp\left(\beta \sum_{(i,j);\,j\sim i} \mathbb{I}_{x_j=x_i}\right)\,.
\eqno{(8.5)}
$$
the full conditional distribution of $x_i$ is
\begin{align*}
\pi(x_i=g|\bx_{-i}) &\propto \pi((g,\bx_{-i})\\
&\propto \exp\left(\beta \sum_{\stackrel{(u,v);\,u\sim v}{u,v\ne i}} \mathbb{I}_{x_u=x_v}+
\sum_{u;\,i\sim u} \mathbb{I}_{x_u=g}\right)\\
&\propto \exp\left(\beta \sum_{u;\,i\sim u} \mathbb{I}_{x_u=g}\right)\\
&= \exp\left(\beta n_{i,g}\right)
\end{align*}

Conversely, if we start from the full conditionals
$$
\pi(x_i=g|\bx_{-i})\propto \exp(\beta n_{i,g})\,.\quad i\in\mathcal{I}\,,\,1\le g\le G\,,
$$
and apply the Hammersley--Clifford representation (8.1)
$$
\frac{\pi(\bx)}{\pi(\bx^*)} = \prod_{i=0}^{n-1}\frac{\pi(x_{i+1}|x_1^*,\ldots,x_i^*,x_{i+2},\ldots,x_n)}
{\pi(x_{i+1}^*|x_1^*,\ldots,x_i^*,x_{i+2},\ldots,x_n)}\,,
$$
we have
\begin{align*}
\frac{\pi(x_{1}|x_{2},\ldots,x_n)}{\pi(x_{1}^*|x_{2},\ldots,x_n)}
&=\exp\left(\beta \sum_{u;\,1\sim u} \left[\mathbb{I}_{x_u=x_1}-\mathbb{I}_{x_u=x_1^*}\right]\right)\\
\frac{\pi(x_{2}|x_1^*,x_{3},\ldots,x_n)}{\pi(x_{2}^*|x_1^*,x_{3},\ldots,x_n)}
&= \exp\left(\beta \mathbb{I}_{1\sim 2}\left[\mathbb{I}_{x_1^*=x_2}-\mathbb{I}_{x_1^*=x_2^*}\right]
+\sum_{u>1;\,2\sim u} \left[\mathbb{I}_{x_u=x_2}-\mathbb{I}_{x_u=x_2^*}\right]\right)\\
\&\vdots\\
\frac{\pi(x_{n}|x_1^*,\ldots,x_{n-1}^*)}{\pi(x_{n}^*|x_1^*,\ldots,x_{n-1}^*)}
&= \exp\left(\beta \sum_{u;\,n\sim u} \left[\mathbb{I}_{x_u^*=x_n}-\mathbb{I}_{x_u^*=x_n^*}\right]\right)\\
\end{align*}
which means that all terms involving both $x_i$ and $x_j^*$ cancel out and that
$$
\pi(\bx)\propto \exp\left(\beta \sum_{(i,j);\,j\sim i} \mathbb{I}_{x_j=x_i}\right)\,.
\eqno{(8.5)}
$$
This exercise is essentially the same as Exercise \ref{exo:PottsisMRF}.

\begin{exoset}\label{exo:sweswa}
For an $n\times m$ array $\mathcal{I}$, if the neighbourhood relation
is based on the four nearest neighbors, show that the $x_{i,j}$'s
for which $(i+j)\equiv 0(\text{mod }2)$ are independent conditional on the 
$x_{i,j}$'s for which $(i+j) \equiv 1(\text{mod }2)$ $(1\le i\le n,\,1\le j\le m)$.
Deduce that the update of the whole image can be
done in two steps by simulating the pixels with even sums of indices and
then the pixels with odd sums of indices. (This modification of Algorithm 8.16
is a version of {\em the Swendsen--Wang} algorithm.)
\end{exoset}

This exercise is simply illustrating in the simplest case
the improvement brought by the Swendsen-Wang algorithm upon
the Gibbs sampler for image processing.

As should be obvious from Figure 8.7 in the book,
the dependence graph between the nodes of the array is such that a given $x_{i,j}$ is independent from
all the other nodes, conditional on its four neighbours. When $(i+j)\equiv 0(2)$, the neighbours have
indices $(i,j)$ such that $(i+j) \equiv 1(2)$, which establishes the first result.

Therefore, a radical alternative to the node-by-node update is to run a Gibbs sampler with two steps: a first
step that updates the nodes $x_{i,j}$ with even $(i+j)$'s and a step that updates the nodes $x_{i,j}$ with
odd $(i+j)$'s. This is quite a powerful solution in that it achieves the properties of two-stage Gibbs sampling,
as for instance the Markovianity of the subchains generated at each step (see Robert and Casella, 2004, Chapter
9, for details).

\begin{exoset}
Determine the computational cost of the derivation of the normalizing constant of the distribution (8.5)
for an $n\times m$ array and $G$ different colors.
\end{exoset}

Just as in Exercise \ref{exo:thecostofpotts}, finding the exact normalizing requires summing over all
possible values of $\bx$, which involves $G^{m\times n}$ terms. And each exponential term involves a sum over
$(m\times n)^2$ terms, even though clever programing of the neighborhood system may reduce the computational
cost down to $m\times n$. Overall, the normalizing constant faces a computing cost of at least
$\text{O}(m\times n\times G^{m\times n})$.

\begin{exoset}\label{exo:PottsisMRF}
Use the Hammersley--Clifford theorem to establish that (8.5) is the joint distribution
associated with the conditionals above. Deduce that the Potts model is an MRF.
\end{exoset}

Similar to the resolution of Exercise \ref{exo:klimt}, using the Hammersley-Clifford representation (8.5) and defining
an arbitrary order on the set $\mathcal{I}$ leads to the joint distribution
\begin{align*}
\pi(\bx) &\propto \frac{\exp\left\{ \beta \sum_{i\in\mathcal{I}}
\sum_{j<i,j\sim i} \mathbb{I}_{x_i=x_j} +\sum_{j>i,j\sim i} \mathbb{I}_{x_i=x_j^\star} \right\}}{
\exp\left\{ \beta \sum_{i\in\mathcal{I}}
\sum_{j<i,j\sim i} \mathbb{I}_{x_i^\star=x_j} +\sum_{j>i,j\sim i} \mathbb{I}_{x_i^\star=x_j^\star} \right\}} \\
&\propto \exp\left\{ \beta \left(\sum_{j\sim i,j<i} \mathbb{I}_{x_i=x_j}
+ \sum_{j\sim i,j>i} \mathbb{I}_{x_i=x_j^\star}
-\sum_{j\sim i, j>i} \mathbb{I}_{x_j^\star=x_i} \right) \right\} \\
&= \exp\left\{ \beta \sum_{j\sim i} \mathbb{I}_{x_i=x_j} \right\}\,. 
\end{align*}
So we indeed recover a joint distribution that is compatible with the initial full conditionals of the
Potts model. The fact that the Potts is a MRF is obvious when considering its conditional distributions.

\begin{exoset}
Derive an alternative to Algorithm 8.17 where the
probabilities in the multinomial proposal are proportional to the
numbers of neighbors $n_{u_\ell,g}$ and compare its performance with
that of Algorithm 8.17.
\end{exoset}

In Step 2 of Algorithm 8.3, another possibility is to select the proposed value of $x_{u_\ell}$ from
a multinomial distribution
$$
\mathcal{M}_G \left( 1;n_1^{(t)}(u_\ell),\ldots,n_G^{(t)}(u_\ell) \right)
$$
where $n_g^{(t)}(u_\ell)$ denotes the number of neighbors of $u_l$ that take the value $g$. This is likely to
be more efficient than a purely random proposal, especially when the value of $\beta$ is high.

\begin{exoset} Show that the Swendsen--Wang improvement given in Exercise \ref{exo:sweswa} also applies
to the simulation of $\pi(\bx|\by,\beta,\sigma^2,\bmu)$.
\end{exoset}

This is kind of obvious when considering that taking into account the values of the $y_i$'s does not
modify the dependence structure of the Potts model. Therefore, if there is a decomposition of the grid
$\mathcal{I}$ into a small number of sub-grids $\mathcal{I}_1,\ldots,\mathcal{I}_k$ such that all the
points in $\mathcal{I}_j$ are independent from one another given the other $\mathcal{I}_\ell$'s, a $k$
step Gibbs sampler can be proposed for the simulation of $\bx$.

\begin{exoset}
Using a piecewise-linear interpolation of $f(\beta)$ based on the values $f(\beta^1),\ldots,f(\beta^M)$, with $0<\beta_1<\ldots
<\beta_M=2$, give the explicit value of the integral
$$
\int_{\alpha_0}^{\alpha_1} \hat{f}(\beta)\,\text{d}\beta
$$
for any pair $0\le\alpha_0<\alpha_1\le 2$.
\end{exoset}

This follows directly from the {\sf R} code in \verb+demo/Chapter.8.R+ as \verb+sumising+, with
$$
\int_{\alpha_0}^{\alpha_1} \hat{f}(\beta)\,\text{d}\beta \approx
\sum_{i,\alpha_0\le \beta_i\le \alpha_1} f(\beta_i) (\beta_{i+1}-\beta_i)\,,
$$
with the appropriate corrections at the boundaries.

\begin{exoset}
Show that the estimators $\widehat\bx$ that minimize the posterior expected losses
$\mathbb{E}^\pi[L_1(\bx,\widehat\bx)|\by)]$  and $\mathbb{E}^\pi[L_2(\bx,\widehat\bx)|\by]$
are $\widehat\bx^{MPM}$ and $\widehat\bx^{MAP}$, respectively.
\end{exoset}

Since
$$
L_1(\bx,\widehat\bx) = \sum_{i\in\mathcal{I}} \mathbb{I}_{x_i\ne\hat x_i}\,,
$$
the estimator $\widehat\bx$ associated with $L_1$ is minimising
$$
\mathbb{E}\left[\sum_{i\in\mathcal{I}} \mathbb{I}_{x_i\ne\hat x_i} \big|\by \right]
$$
and therefore, for every $i\in\mathcal{I}$, $\hat x_i$ minimizes $\mathbb{P}(x_i\ne\hat x_i)$,
which indeed gives the MPM as the solution.
Similarly,
$$
L_2(\bx,\widehat\bx) =  \mathbb{I}_{\bx\ne\widehat\bx}
$$
leads to $\widehat\bx$ as the solution to
$$
\min_{\widehat{\bx}} \mathbb{E}\left[ \mathbb{I}_{\bx\ne\widehat{\bx}} \big| \by \right] =
\min_{\widehat{\bx}} \mathbb{P} \left( \bx\ne\widehat{\bx} \big| \by \right)\,,
$$
which means that $\widehat{\bx}$ is the posterior mode.

\begin{exoset}\label{exo:classX}
Determine the estimators $\widehat\bx$ associated with two loss functions that penalize
differently the classification errors,
$$
L_3(\bx,\widehat\bx) =  \sum_{i,j\in\mathcal{I}}\mathbb{I}_{x_i=x_j}\,\mathbb{I}_{\hat x_i\ne\hat x_j}
\quad
\text{and}
\quad
L_4(\bx,\widehat\bx) =  \sum_{i,j\in\mathcal{I}}\mathbb{I}_{x_i\ne x_j}\,\mathbb{I}_{\hat x_i=\hat x_j}\,.
$$
\end{exoset}

Even though $L_3$ and $L_4$ are very similar, they enjoy completely different properties. In fact, $L_3$ is
basically useless because $\widehat\bx=(1,\cdots,1)$ is always an optimal solution!

If we now look at $L_4$, we first notice that this loss function is invariant by permutation of the classes in $\bx$:
all that matters are the groups of components of $\bx$ taking the same value. Minimizing this loss function then
amounts to finding a clustering algorithm. To achieve this goal, we first look at the difference in the risks when
allocating an arbitrary $\hat x_i$ to the value $a$ and when allocating $\hat x_i$ to the value $b$.
This difference is equal to
$$
\sum_{j, \hat x_j=a} \mathbb{P}(x_i=x_j) - \sum_{j, \hat x_j=b} \mathbb{P}(x_i=x_j)\,.
$$
It is therefore obvious that, for a given configuration of the other $x_j$'s, we should pick the value $a$ that
minimizes
the sum $\sum_{j, \hat x_j=a} \mathbb{P}(x_i=x_j)$. Once $x_i$ is allocated to this value, a new index $\ell$ is to be
chosen for possible reallocation until the scheme has reached a fixed configuration, that is, no $\hat x_i$ need
reallocation.

This scheme produces a smaller risk at each of its steps so it does necessarily converge to a fixed point. What is
less clear is that this produces the global minimum of the risk. An experimental way of checking this is to run the
scheme with different starting points and to compare the final values of the risk.

\begin{exoset}\label{exo:cheapneal}
Since the maximum of $\pi(\bx|\by)$ is the same as that of
$\pi(\bx|\by)^\kappa$ for every $\kappa\in\mathbb{N}$, show that
\begin{equation}\label{eq:samoa}
\pi(\bx|\by)^\kappa = \int \pi(\bx,\theta_1|\by)\,\text{d}\theta_1
\times\cdots\times\int \pi(\bx,\theta_\kappa|\by)\,\text{d}\theta_\kappa\,,
\end{equation}
where $\theta_i=(\beta_i,\bmu_i,\sigma^2_i)$ $(1\le i\le\kappa)$. Deduce from this representation an
optimization scheme that slowly increases $\kappa$ over iterations and that runs a Gibbs sampler for the
integrand of (8.9) at each iteration.
\end{exoset}

The representation (8.10) is obvious since
\begin{align*}
\left( \int \pi(\bx,\theta|\by)\,\text{d}\theta \right)^\kappa &= \int \pi(\bx,\theta|\by)\,\text{d}\theta
\times\cdots\times\int \pi(\bx,\theta|\by)\,\text{d}\theta \\
&= \int \pi(\bx,\theta_1|\by)\,\text{d}\theta_1
\times\cdots\times\int \pi(\bx,\theta_\kappa|\by)\,\text{d}\theta_\kappa
\end{align*}
given that the symbols $\theta_i$ within the integrals are dummies.

This is however the basis for the so-called SAME algorithm of Doucet, Godsill and Robert (2001),
described in detail in Robert and Casella (2004).

\begin{exoset}\label{exo:tinising}
For the Ising model, show that the distribution (8.4) can be also defined as
$$
\pi(\bx)\propto \exp\left(2\beta \sum_{j\sim i} \mathbb{I}_{x_j=x_i=1}\right)
$$
when the number of neighbors is constant.
\end{exoset}

Since
$$
\pi(\bx)\propto \exp\left(\beta \sum_{j\sim i} \mathbb{I}_{x_j=x_i}\right)\,,
$$
we have
\begin{align*}
\pi(\bx) &\propto \exp\left(\beta \sum_{j\sim i} \mathbb{I}_{x_j=x_i=1} + 
\beta \sum_{j\sim i} \mathbb{I}_{x_j=x_i=-1}\right) \\
&= \exp\left(\beta \sum_{j\sim i} \mathbb{I}_{x_j=x_i=1} +
\beta \left[N-\sum_{j\sim i}\mathbb{I}_{x_j=x_i=1}\right]\right) \\
&= \exp\left(2\beta \sum_{j\sim i} \mathbb{I}_{x_j=x_i=1} \right) \exp(N\beta)
\end{align*}
if $N$ denotes the number of connected pairs $i\sim j$.

\begin{exoset}\label{exo:IsingasHC}
Show that the joint distribution (8.4) can be obtained from the full conditionals (8.3)
by virtue of the Hammerseley-Clifford representation (8.1).
\end{exoset}

This is a special case of Exercise \ref{exo:PottsisMRF} since the Ising model is a Potts model with only two modalities.

\begin{exoset}\label{exo:Rverse}
Show that the Ising distribution is symmetric in that inverting the color of all pixels does not change
the probability (8.4).
\end{exoset}

Given the definition of the Ising model as 
$$
\pi(\bx)\propto \exp\left(\beta \sum_{j\sim i} \mathbb{I}_{x_j=x_i}\right)\,,
\eqno{(8.3)}
$$
switching $1$'s and $-1$'s does not modify the right hand side and hence does not change $\pi(\bx)$.

\begin{exoset}\label{exo:tutti}
For the Ising model, run a simulation experiment that should locate the 
limiting value of $\beta$ above which  almost all pixels are of the same color.
Same question for the (negative) limiting value of $\beta$ below which the image
is a perfect checkerboard.
\end{exoset}

A possible approach used in the following code is to resort to simulated annealing, increasing progressively $\beta$
until all sites are of the same color. Opting for a four-neighbour structure, we slightly modify the {\sf R} functions
\begin{verbatim}
xneig4=function(x,a,b,col){
n=dim(x)[1];m=dim(x)[2]
nei=c(x[a-1,b]==col,x[a,b-1]==col)
if (a!=n)
  nei=c(nei,x[a+1,b]==col)
if (b!=m)
  nei=c(nei,x[a,b+1]==col)
sum(nei)
}
\end{verbatim}
and
\begin{verbatim}
isingibbs=function(niter=10^2,n,m=n,beta=1,
   x=matrix(sample(c(-1,1),n*m,rep=TRUE),n,m)){
  for (i in 1:niter){
    sampl1=sample(1:n)
    sampl2=sample(1:m)
    for (k in 1:n){
    for (l in 1:m){
     n0=xneig4(x,sampl1[k],sampl2[l],-1)
     n1=xneig4(x,sampl1[k],sampl2[l],1)
     x[sampl1[k],sampl2[l]]=sample(c(-1,1),1,
                 prob=exp(beta*c(n0,n1)))
     }}}
  x
  }
\end{verbatim}
defined in the book. Then the function
\begin{verbatim}
isinganeal=function(niter=10^3,precis=.1,n,m=n){
  beta=precis
  simu=isingibbs(niter,n,m,beta) 
  while (min(simu)<max(simu)){
   beta=beta+precis
   simu=isingibbs(niter,n,m,beta,x=simu)}
  return(beta)
}
\end{verbatim}
increases the coefficient $\beta$ until all simulated entries are of the same color.

\begin{figure}[h]
\centerline{\includegraphics[width=.7\textwidth]{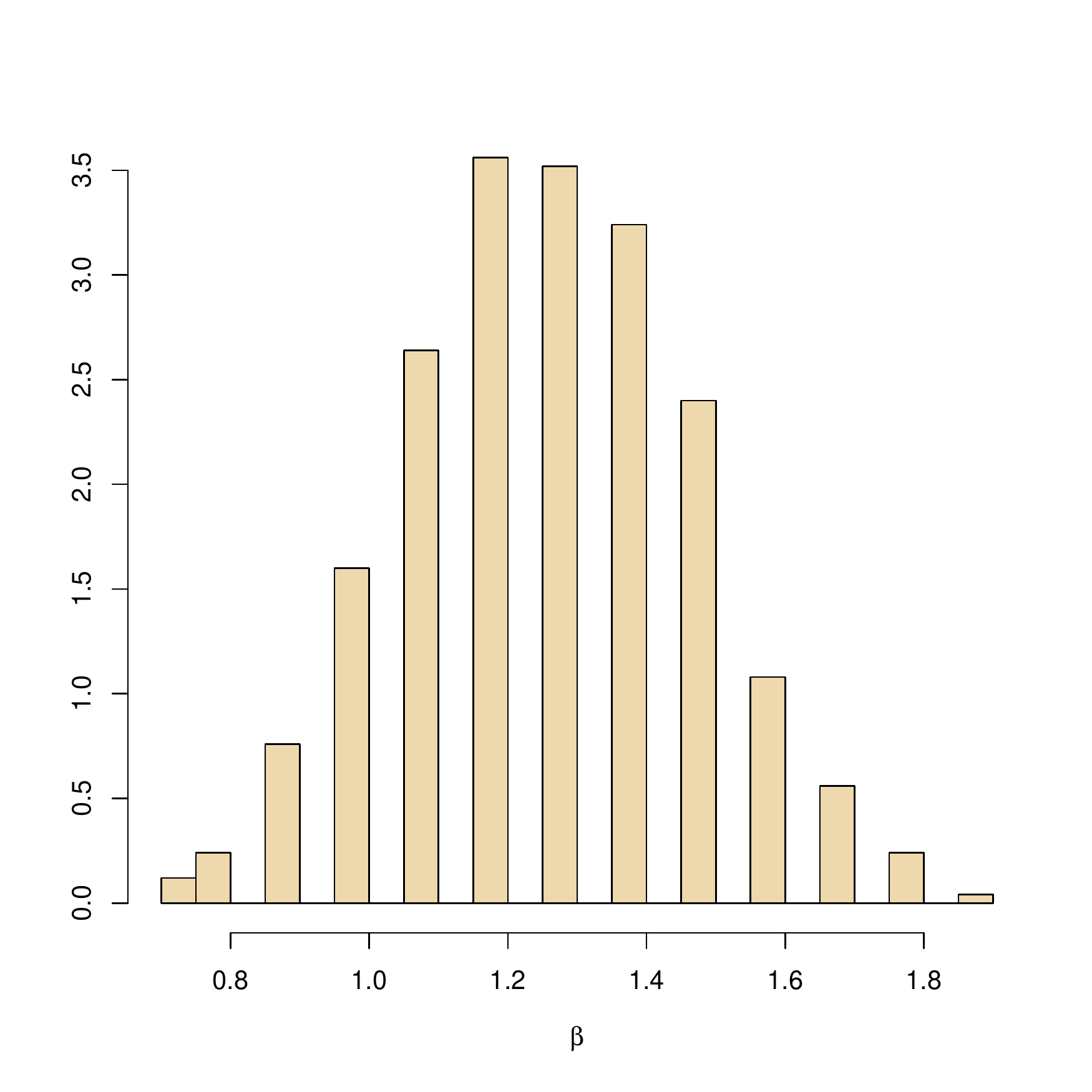}}
\caption{\label{fig:allBlacks}
Empirical distribution of the $\beta$'s leading to a unicolor simulation of the Ising model, for a $(5,5)$ grid,
based on $250$ replications and a precision of $0.1$.}
\end{figure}

Figure \ref{fig:allBlacks} in this manual provides an histogram of the $\beta$'s returned by the above code in the case of a $5\times
5$ grid. It gives indications on the zone to study more precisely the occurence of unicolor grids and the detection of
the cutoff point.

\begin{figure}[h]
\centerline{\includegraphics[width=.7\textwidth]{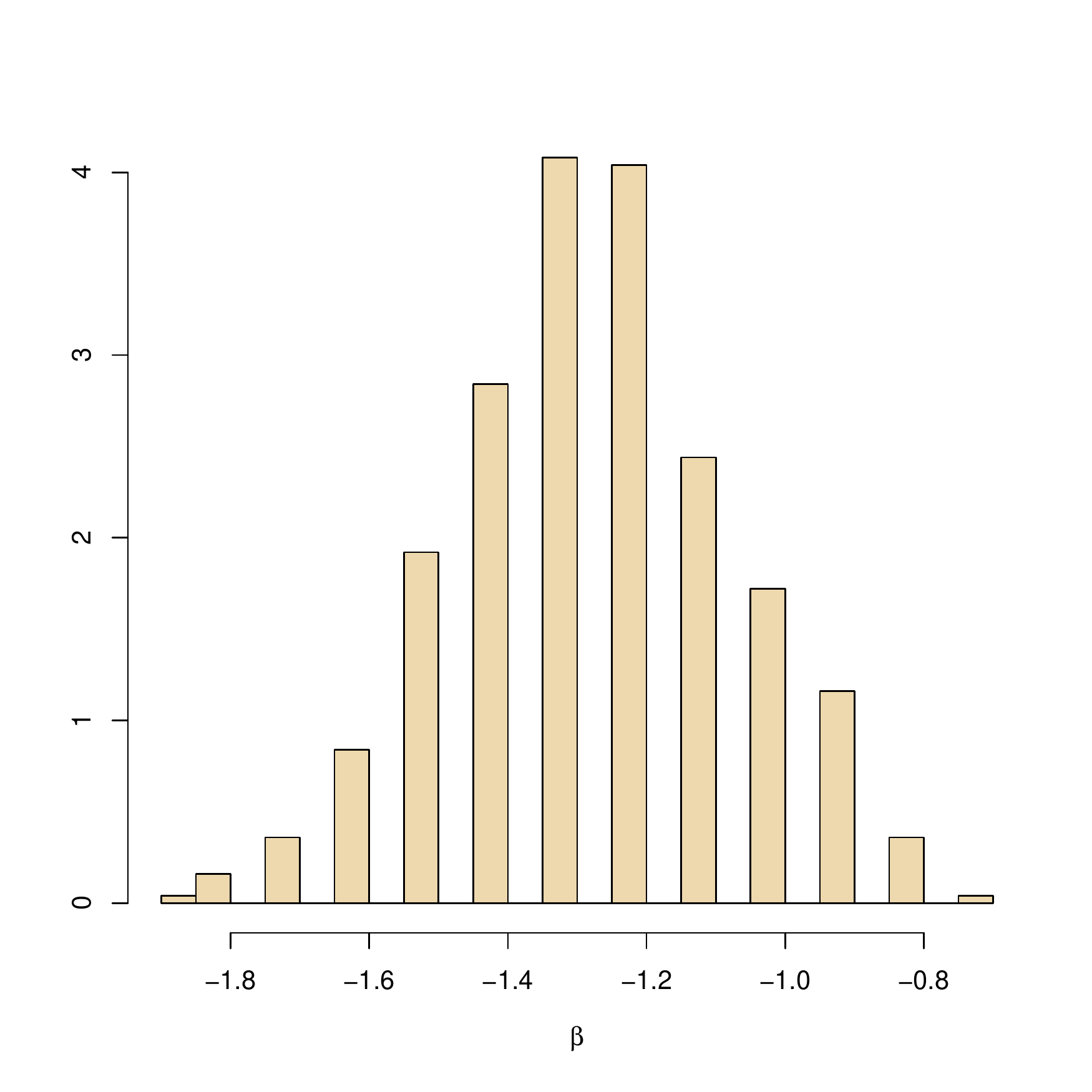}}
\caption{\label{fig:Kasparov}
Empirical distribution of the $\beta$'s leading to a checkerboard simulation of the Ising model, for a $(5,5)$ grid,
based on $250$ replications and a precision of $0.1$.}
\end{figure}

For the opposite case, the coefficient $\beta$ is decreased in \verb+isinganeal+ until
\begin{verbatim}
sum(abs(simu[,-1]+simu[,-m]))+sum(abs(simu[-1,]+simu[-n,]))==0
\end{verbatim}
Figure \ref{fig:Kasparov} in this manual provides an histogram of the $\beta$'s returned by the above code in the case of a $5\times
5$ grid. As for Figure \ref{fig:allBlacks} in this manual, it only provide some indications on the zone of $\beta$'s for producing
checker grids almost surely.

\begin{exoset}\label{exo:!ABC}
Show that the ABC algorithm implemented with $\epsilon=0$ and
a distance between sufficient statistics is not approximate in that 
the output is truly simulated from the posterior distribution $\pi(\theta|\bx)$ $\propto f(\bx|\theta)\pi(\theta)$.
\end{exoset}

When the ABC algorithm is used with a tolerance $\epsilon=0$, the probability of accepting $\theta\sim\pi(\theta)$ in
Algorithm 8.18 is $\mathbb{P}_\theta(S(Y)=S(x))=f^S(S(x)|\theta)$, the probability mass function of the statistic $S(X)$
when $X\sim f(x|\theta)$. Therefore the distribution of the accepted $\theta$'s is 
$$
\pi^\text{ABC}(\theta|x)\propto\pi(\theta)f^S(S(x)|\theta)
$$
which is the {\em exact} posterior distribution of $\theta$ when observing $S(x)$. If $S(\cdot)$ is a sufficient
statistic, this posterior is also equal to the posterior distribution of $\theta$ given the observation $x$. 
Therefore, an ABC simulation of the Potts model posterior in Section 8.3.3 could be rerun with a tolerance of
$\epsilon=0$, albeit at a higher computational cost.

\end{document}